\documentclass[12pt,showpacs,showkeys,amsmath,amssymb]{revtex4}
\usepackage{amsmath,amsfonts,amsthm,amscd,amssymb,latexsym}
\usepackage{longtable}
\usepackage{bm}
\usepackage{dcolumn}
\usepackage{graphicx}
\usepackage{epstopdf}
\usepackage{color}
\usepackage{epsf}
\usepackage{epsfig}
\usepackage{graphicx, epic, eepic, color}

\def\a{\alpha}
\def\b{\beta}

\def\d{\delta}

\def\@{\partial_}

\def\negenspace{\kern-1.1em}

\def\sqr#1#2{{\vcenter{\hrule height.#2pt\hbox{\vrule width.#2pt
height#1pt \kern#1pt \vrule width.#2pt}\hrule height.#2pt}}}

\date{\today}

\begin{document}

\title{Observational Tests of Nonlocal Gravity: Galaxy Rotation Curves and Clusters of Galaxies}

\author{S. Rahvar}
\email{rahvar@sharif.edu}
\affiliation{Perimeter Institute for Theoretical Physics, Waterloo, Ontario N2L 2Y5, Canada \\
and \\
Department of Physics, Sharif University of Technology, P.O. Box 11155-9161, Tehran, Iran}

\author{B. Mashhoon}
\email{mashhoonb@missouri.edu}
\affiliation{Department of Physics and Astronomy,
University of Missouri, Columbia, Missouri 65211, USA}

\begin{abstract}
A classical nonlocal generalization of Einstein's theory of gravitation has recently been developed via the introduction of a scalar causal ``constitutive" kernel that must ultimately be determined from observational data. It turns out that the nonlocal aspect of gravity in this theory can simulate dark matter; indeed, in the Newtonian regime of nonlocal gravity, we recover the phenomenological Tohline-Kuhn approach to modified gravity. A simple generalization of the Kuhn kernel in the context of nonlocal general relativity leads to a two-parameter modified Newtonian force law that involves an additional repulsive Yukawa-type interaction. We determine the parameters of our nonlocal kernel by comparing the predictions of the theory with observational data regarding the rotation curves of spiral galaxies. The best-fitting stellar mass-to-light ratio turns out to be in agreement with astrophysical models; moreover,  our results are consistent with the Tully-Fisher relation for spiral galaxies. Light deflection in nonlocal gravity is consistent with general relativity at  Solar System scales, while beyond galactic scales an enhanced deflection angle is predicted  that is compatible with lensing by the effective ``dark matter". Furthermore, we extend our results to the internal dynamics of rich clusters of galaxies  and show that the dynamical mass of the cluster obtained from nonlocal gravity is consistent with the measured baryonic mass.
\end{abstract}



\maketitle

\section{Introduction}

Lorentz invariance, which is a fundamental symmetry of nature, deals with observations of \emph{ideal inertial} observers in Minkowski spacetime, where the gravitational interaction is assumed to be turned off. However, physical measurements are carried out by observers that are all more or less accelerated. To treat \emph{actual accelerated} observers in Minkowski spacetime, a physical law that relates accelerated and inertial observers is indispensable. The standard theory of special relativity postulates a \emph{pointwise} connection; that is, Lorentz transformations are applied event by event to relate the instantaneous  local inertial rest frame of the accelerated observer with the background global inertial frame. This locality postulate is also essential for the utility of Einstein's heuristic principle of equivalence, as these together establish the \emph{local} Lorentz invariance of the gravitational interaction in general relativity~\cite{Ei}.

The hypothesis of locality in relativistic physics originates from the classical mechanics of Newtonian point particles and rays of radiation. It would be exactly valid if all physical processes could be reduced to  \emph{pointlike coincidences} in Minkowski spacetime. The deviation from locality is expected to be proportional to $\lambda/{\cal L}$, where $\lambda$ is the characteristic wavelength of the phenomenon under observation and ${\cal L}$ is the relevant acceleration length of the observer.  An accelerated observer in Minkowski spacetime is naturally endowed with a local orthonormal tetrad frame that it carries along its world line. The rate of variation of the tetrad frame along the world line is governed by the acceleration tensor of the observer. The components of this tensor consist of the observer's translational acceleration $\mathbf{g}$ and the angular velocity of rotation of its spatial frame $\boldsymbol{\Omega}$ with respect to a locally nonrotating (i.e., Fermi-Walker transported) frame. The typical acceleration lengths of the observer are then given, for instance, by $c^2/|\mathbf{g}|$ and $c/|\boldsymbol{\Omega}|$. For an observer fixed on the Earth, $c^{2}/|\mathbf{g}_{\oplus}|\approx 1$ light year and $c/|\boldsymbol{\Omega}_{\oplus}|\approx 28$ astronomical units; hence, laboratory deviations from locality are expected to be generally rather small in special relativity and this circumstance explains why the locality postulate is an excellent approximation in most situations of practical interest. It is important to recognize that these physical considerations regarding deviations from locality for accelerated observers in Minkowski spacetime cannot be directly extended to the gravitational field, since Einstein's \emph{local} principle of equivalence cannot be applied to situations where locality breaks down.

It has been argued that one must in general go beyond the locality hypothesis of standard special relativity theory and include the past history of the accelerated observer as well~\cite{Mash}. Indeed, Bohr and Rosenfeld pointed out long ago that field determinations cannot be performed instantaneously~\cite{BR1, BR2}. On the basis of these considerations, a \emph{nonlocal special relativity} theory has been developed~\cite{NSR}. In this theory, the field measured by an accelerated observer has in general an additional component that involves an average of the field over the past world line of the observer with a kernel that carries the memory of its past acceleration. The basic scale associated with such acceleration-induced nonlocality is then the acceleration length ${\cal L}$ or, equivalently, the acceleration time ${\cal L}/c$.

The principle of equivalence of inertial and gravitational masses implies a deep connection between inertia and gravitation~\cite{Ei}.  It is therefore natural to try and extend this notion of nonlocality to the gravitational interaction. This can be done via an averaging procedure involving a scalar causal kernel that acts as the weight function for the gravitational memory of past events. Indeed, such a nonlocal generalization of Einstein's theory of gravitation has been devised in which \emph{nonlocality appears to simulate dark matter}~\cite{NG1, NG2, NG3, NG4, NG5, NG6, NG7}. It may appear natural to try and establish a connection between the nonlocal kernels of acceleration-induced nonlocality in Minkowski spacetime and nonlocal gravity. However, such a relation does \emph{not} exist, as it would amount to a nonlocal extension of Einstein's strictly local principle of equivalence. In particular, the lengthscale characteristic of nonlocality in the gravitational case \emph{cannot} be estimated on the basis of the corresponding acceleration length. It is important to emphasize here again that it is not possible to deduce the scalar causal kernel of nonlocal gravity from the kernel of accelerated observers in Minkowski spacetime; indeed, one cannot invoke Einstein's principle of equivalence in this case due to its extreme locality. In the absence of a deeper understanding of the gravitational interaction, we adopt the view that the kernel of nonlocal gravity must be determined from observational data regarding dark matter in astrophysics.

Nonlocal gravity is a tetrad theory, where the tetrad field is locally defined but satisfies integro-differential field equations. To construct a nonlocal generalization of classical general relativity, it is first necessary to extend the Riemannian structure of spacetime by introducing an additional Weitzenb\"ock connection. In the resulting extended framework, known as teleparallelism, general relativity (GR) is expressed in its equivalent teleparallel form, namely, GR$_{||}$. This is formally analogous to electrodynamics and is in fact a  gauge theory of the Abelian group of spacetime translations---see~\cite{1,2,3} and the references cited therein. Nonlocal general relativity has been obtained from GR$_{||}$, the teleparallel equivalent of GR, via a ``constitutive" kernel~\cite{NG1, NG2, NG3, NG4, NG5, NG6, NG7}. A brief account of this theory is provided in this introductory section for the sake of completeness. The gravitational potentials in nonlocal general relativity are given by the tetrad field $e_\mu{}^{\hat \alpha}(x)$, from which one obtains the spacetime metric $g_{\mu \nu}(x)=e_\mu{}^{\hat \alpha}e_\nu{}^{\hat \beta}\eta_{\alpha \beta}$. In our convention,  the Minkowski metric tensor  $\eta_{\alpha \beta}$ is given by diag$(-1,1,1,1)$; moreover, Greek indices run from 0 to 3, while Latin indices run from 1 to 3. The hatted Greek indices ${\hat{\alpha}}$, ${\hat{\beta}}$, etc., refer to \emph{anholonomic} tetrad indices, while $\mu$, $\nu$, etc., refer to \emph{holonomic} spacetime indices. We use units such that $c=1$, unless otherwise specified. The holonomic and anholonomic indices are raised and lowered by means of the metric tensors $g_{\mu \nu}(x)$ and $\eta_{\alpha \beta}$, respectively; furthermore, in order to change a holonomic index of a tensor into an anholonomic index or vice versa, we project the tensor on an appropriate tetrad field.

The arena for nonlocal gravity is the Weitzenb\"ock spacetime that is a parallelizable manifold. That is, the tetrad frame field is globally teleparallel, so that the curvature of the Weitzenb\"ock spacetime vanishes and, just as in the case of flat Minkowski spacetime, it is possible to introduce Cartesian frames for which the corresponding flat connection vanishes as well.  \emph{We will work with such Cartesian tetrad frames throughout this work}. Free test particles and null rays follow respectively timelike and null geodesics of $g_{\mu \nu}$, the metric of the Weitzenb\"ock spacetime.

The field equations of nonlocal gravity are expressed in terms of the \emph{gravitational field strength}  $C_{\mu \nu}{}^{\hat{\alpha}}$, which is a tensor defined by
\begin{equation}\label{I1}
 C_{\mu \nu}{}^{\hat{\alpha}}=\partial_{\mu}e_{\nu}{}^{\hat{\alpha}}-\partial_{\nu}e_{\mu}{}^{\hat{\alpha}}\,.
\end{equation}
This definition is reminiscent of the definition of the electromagnetic field tensor in terms of the vector potential. In our convention, $C_{\mu \nu}{}^{\hat{\alpha}}$ is in fact the torsion of the Weitzenb\"ock spacetime. Moreover, the theory contains two auxiliary field strengths, namely, a modified torsion tensor
\begin{eqnarray}\label{I2}
\mathfrak{C}_{\mu \nu}{}^{\hat{\alpha}} :=\frac 12\,
C_{\mu \nu}{}^{\hat{\alpha}} -C^{\hat{\a}}{}_{[\mu \nu]}+2e_{[\mu}{}^{\hat{\a}} C_{\nu]{\hat{\b}}}{}^{\hat{\b}}\,
\end{eqnarray}
and a tensor density that is linear in the modified torsion tensor
\begin{eqnarray}\label{I3}
{\cal H}^{\mu \nu}{}_{\rho}(x) := \frac{\sqrt{-g(x)}}{\kappa}\Big[\mathfrak{C}^{\mu \nu}{}_{\rho}- \int \Omega^{\mu \mu'} \Omega^{\nu \nu'} \Omega_{\rho \rho'} {\cal K}(x, x')\mathfrak{C}_{\mu' \nu'}{}^{\rho'}(x') \sqrt{-g(x')}~d^4x'\Big]\,,
\end{eqnarray}
where $\kappa=8 \pi G/c^4$,  ${\cal K}$ is the constitutive causal kernel of the theory and $\Omega$ is Synge's \emph{world function}~\cite{Sy}.  In Eq.~\eqref{I3}, we assume that event $x'$ is connected to event $x$ via a unique future directed timelike or null geodesic of $g_{\mu \nu}$; then, 2$\Omega$ is the square of the corresponding proper spacetime distance from $x'$ to $x$. Indices $\mu', \nu', \rho',...$ refer to event $x'$, while indices $\mu, \nu, \rho, ...$ refer to event $x$. Moreover, we define
\begin{equation}\label{I3a}
\Omega_{\mu}(x, x')=\frac{\partial \Omega}{\partial x^{\mu}}, \quad \Omega_{\mu'}(x, x')=\frac{\partial \Omega}{\partial x'^{\mu'}}\,
\end{equation}
and note that covariant derivatives at $x$ and $x'$ commute for any bitensor. Indeed, $\Omega_{\mu \mu'}(x, x')=\Omega_{\mu' \mu}(x, x')$ is a dimensionless bitensor such that
\begin{equation}\label{I3b}
\lim_{x' \to x} \Omega_{\mu \mu'}(x, x')=-g_{\mu \mu'}(x)\,.
\end{equation}
The derivatives of the bitensor $\Omega(x, x')$ appear in Eq.~\eqref{I3} in order to render this constitutive relation covariant under arbitrary transformations of spacetime coordinates; in particular, the integral here is a proper tensor of the third rank in $x$, but what is integrated is a \emph{scalar} function of the integration variable $x'$.
This constitutive relation is nonlocal by virtue of the existence of kernel ${\cal K}$ and is reminiscent of a similar situation in electrodynamics, where the constitutive relations between ($\mathbf{E}$, $\mathbf{B}$) and ($\mathbf{D}$, $\mathbf{H}$) could be nonlocal due to memory effects.
The nature of the causal scalar kernel ${\cal K}$ has been discussed in detail in Refs.~\cite{NG1, NG2, NG3, NG4, NG5, NG6}; as memory fades, the kernel is expected to vanish for events that occurred in the distant past. It is in principle possible that ${\cal K}$ could be derivable from a future deeper theory; however, we assume here that the nonlocal kernel is ultimately determined via observational data~\cite{NG3}. Indeed, to account for the rotation curves of spiral galaxies, we associate nonlocality with a galactic lengthscale $\lambda_0 \sim 1$ kpc.

Let us observe that the field strength $C_{\mu \nu}{}^{\hat{\alpha}}$ and the auxiliary field strengths $\mathfrak{C}_{\mu \nu}{}^{\hat{\alpha}}$ and  ${\cal H}_{\mu \nu}{}^{\hat{\alpha}}$ are all antisymmetric in their first two indices. The field equations of nonlocal gravity are analogous to Maxwell's equations and can be expressed as
\begin{equation}\label{I4}
 \partial_{[\mu} C_{\nu \rho]}{}^{\hat{\alpha}}=0\,,
 \end{equation}
\begin{equation}\label{I5}
  \partial_\nu{\cal H}^{\mu \nu}{}_{\hat{\alpha}} =\sqrt{-g}~(T_{\hat{\alpha}}{}^\mu + E_{\hat{\alpha}}{}^\mu)\,.
\end{equation}
Here, $T_{\hat{\alpha}}{}^\mu$ is the matter energy-momentum tensor and $E_{\hat{\alpha}}{}^\mu$ is the energy-momentum tensor of the gravitational field defined by
\begin{equation}\label{I6}
\sqrt{-g}~ E_{\hat{\alpha}}{}^\mu:=-\frac 14  e^\mu{}_{\hat{\alpha}}(C_{ \nu \rho}{}^{\hat{\beta}}
{\cal H}^{\nu \rho}{}_{\hat{\beta}}) + C_{{\hat{\alpha}} \nu}{}^{\hat{\beta}} {\cal H}^{\mu \nu}{}_{\hat{\beta}}\,.
\end{equation}
We note that $E_{\mu \nu}$ is traceless just as in electromagnetism. The relationship
 between this \emph{tensor} and the energy-momentum \emph{pseudotensor} of the gravitational field in GR has been discussed in detail in Chap. 10 of Ref.~\cite{2}. It follows from Eq.~\eqref{I5} that
\begin{equation}\label{I7}
\partial_{\mu}\Big[\sqrt{-g}~(T_{\hat{\alpha}}{}^\mu + E_{\hat{\alpha}}{}^\mu)\Big] = 0\,,
\end{equation}
which expresses the energy-momentum conservation law in nonlocal gravity.

It is interesting to remark that, for our conventional choice of Cartesian tetrad frame, Eq.~\eqref{I4} is automatically satisfied. This is reminiscent of the source-free part of Maxwell's equations expressed in terms of the vector potential. Hence, the main field equation of nonlocal gravity is Eq.~\eqref{I5}, which is a nonlinear integro-differential equation for the tetrad field $e_{\mu}{}^{\hat{\alpha}}$. If the scalar kernel ${\cal K}$ vanishes, the theory reduces to GR$_{||}$, the teleparallel equivalent of GR. It must be emphasized that nonlocal gravity's basic ansatz, namely, the \emph{linear} constitutive relation that introduces  the \emph{scalar} kernel in Eq.~\eqref{I3}, is hardly unique; indeed, the present approach appears to be the simplest possible way of formulating a nonlocal generalization of classical general relativity.

To clarify further the nature of this nonlocal gravity theory, let us express Maxwell's equations for the electrodynamics of media in the form
\begin{equation}\label{I8}
\partial_{[\mu}F_{\nu \rho]}=0\,
\end{equation}
and
\begin{equation}\label{I9}
\partial_{\nu}H^{\mu \nu}=\frac{4 \pi}{c} j^\mu\,
\end{equation}
that are analogous to Eqs.~\eqref{I4} and~\eqref{I5}, respectively. Here $(\mathbf{E}, \mathbf{B}) \mapsto F_{\mu \nu}$, where $F_{\mu \nu}=\partial_{\mu} A_{\nu} -\partial_{\nu} A_{\mu}$ is the electromagnetic field tensor and $A_{\mu}$ is the vector potential, while $(\mathbf{D}, \mathbf{H}) \mapsto H_{\mu \nu}$, which is the auxiliary field tensor, and $j^\mu$ is the current of free charges. It is well known that the constitutive relation between $H_{\mu \nu}$ and $F_{\mu \nu}$ is in general nonlocal. Thus in the treatment of such media, Maxwell's equations~\eqref{I8} and~\eqref{I9} do \emph{not} formally change, only the \emph{constitutive} law connecting the two field strengths involves a nonlocal kernel. In much the same way, Einstein's gravitational field equations~\eqref{I4} and~\eqref{I5}, as expressed in the teleparallel equivalent of general relativity, GR$_{||}$, require a constitutive relation between ${\cal H}_{\mu \nu \rho}$ and $\mathfrak{C}_{\mu \nu \rho}$---or, equivalently, $C_{\mu \nu \rho}$. In GR$_{||}$ proper, we have the local relation ${\cal H}_{\mu \nu \rho}=(\sqrt{-g}/\kappa) \mathfrak{C}_{\mu \nu \rho}$; however, this local  constitutive relation could be made nonlocal via a constitutive kernel as in Eq.~\eqref{I3}. Thus we do \emph{not} formally change the gravitational field equations of GR$_{||}$; indeed, only the local constitutive law is made nonlocal in this theory by the introduction of a causal scalar kernel. This is then the genesis of the nonlocal gravity theory.

No exact solution of the field equation of nonlocal gravity is known at present; therefore, we have resorted to the general linear approximation and its Newtonian limit~\cite{NG1, NG2, NG3, NG4, NG5, NG6, NG7}. Thus we limit our considerations to the weak-field regime, since the nonlinear strong-field regime of nonlocal gravity has not yet been investigated. In particular, exact cosmological models and  the nature of black holes in nonlocal gravity are beyond the scope of our treatment. In connection with gravitational radiation,  detailed investigations reveal that the implications of nonlocal gravity theory for linearized gravitational waves are essentially the same as in GR~\cite{ NG6, NG7}; that is, the orbital decay of relativistic binary systems due to gravitational radiation damping as well as the standard GR treatment of linearized gravitational radiation of frequency $\gtrsim 10^{-8}$ Hz is expected to be consistent with nonlocal gravity.   Indeed, the galactic nonlocality lengthscale of $\lambda_0 \sim 1$ kpc is much larger than the orbital radius of a relativistic binary pulsar, so that nonlocal effects are expected to be negligibly small, just as they are in the Solar System~\cite{NG3}; moreover, the frequency associated with linearized gravitational waves of wavelength $\lambda_0 \sim 1$ kpc is $\sim 10^{-11}$ Hz, which is much smaller than the frequency of radiation that could be detectable with present methods. For the relatively high-frequency  gravitational radiation that is of current observational interest ($\gtrsim 10^{-8}$ Hz), nonlocal effects essentially average out and can therefore be safely ignored in practice~\cite{NG6}.

The main purpose of the present work is to discuss further the Newtonian regime within the framework of nonlocal gravity in connection with the problem of dark matter and compare the predictions of the theory with observational data regarding galaxy rotation curves as well as the dynamics of clusters of galaxies. We show in the next section that nonlocality leads to a simple modification of the Newtonian inverse-square force law that depends on two free parameters $\alpha$ and $\mu$. We determine these parameters in section III by comparing the predictions of nonlocal gravity theory with observational data regarding the rotation curves of spiral galaxies. We show that nonlocal gravity is consistent with the  relevant astrophysical models and the empirical Tully-Fisher relation. After fixing the parameters of our nonlocal gravity model, we then turn our attention in section IV to the implications of nonlocal gravity for the gravitational physics of the Solar System. We show in section V that the dynamics of rich clusters of galaxies is consistent with our nonlocal gravity model. Section VI contains a brief discussion of our main results.

\section{Nonlocal Gravity: Newtonian Regime}

General relativity reduces to the Newtonian gravitation theory in the correspondence limit, in which we formally let $c \to \infty$. The gravitational field equations of GR then reduce to the Poisson equation
\begin{equation}\label{II1A}
\nabla^2\phi=4\pi G\rho\,,
\end{equation}
where $\phi$ is the Newtonian gravitational potential and $\rho$ is the density of matter. In nonlocal general relativity, Poisson's equation is modified such that the source on right-hand side of Eq.~\eqref{II1A} acquires an additional nonlocal component that can be interpreted as the density of ``dark" matter.

Consider matter of density $\rho$ confined to a finite region of space in Minkowski spacetime such that the resulting gravitational field is everywhere weak and vanishes infinitely far from the source. Treating the gravitational perturbation of Minkowski spacetime to linear order within the framework of nonlocal gravity, Eqs.~\eqref{I3} and~\eqref{I5} take the form
\begin{equation}\label{II1B}
\frac{\partial}{\partial x^\sigma}\Big[\mathfrak{C}_{\mu}{}^{\sigma}{}_{\nu}(x)+\int{\cal K}(x, y) \mathfrak{C}_{\mu}{}^{\sigma}{}_{\nu}(y)~d^4y\Big] = \kappa~ T_{\mu \nu}\,,
\end{equation}
since $E_{\mu \nu}$ can be neglected at the linear order.
In the linear weak-field approximation of nonlocal gravity, the preferred frame field $e_\mu{}^{\hat \alpha}(x)$ is replaced by $\delta_\mu^\alpha$ plus small perturbations that constitute the gravitational potentials of linearized nonlocal gravity. More specifically, we let
\begin{equation}\label{II1C}
 e_\mu{}^{\hat{\alpha}}={\d}_\mu ^{\alpha}+\psi^{\alpha}{}_\mu\,, \quad  e^\mu{}_{\hat{\alpha}}=\d^\mu _{\alpha} -\psi^\mu{}_{\alpha}\,,
\end{equation}
where the perturbation  $\psi_{\mu \nu}$ is treated to first order and the distinction between holonomic and anholonomic indices disappears at this level of approximation. The sixteen components of
$\psi_{\mu \nu}$ can be decomposed into its symmetric and antisymmetric parts, namely, $\psi_{\mu \nu}=\psi_{(\mu \nu)}+\psi_{[\mu \nu]}$. It follows from the orthonormality of the preferred frame field that the spacetime metric in the linear regime is given by $g_{\mu \nu}=\eta_{\mu \nu} +h_{\mu \nu}$, where $h_{\mu \nu}=2\psi_{(\mu \nu)}$. It is then possible to show that to linear order the Einstein tensor is given by
\begin{equation}\label{II1D}
G_{\mu \nu}=\partial_\sigma \mathfrak{C}_{\mu}{}^{\sigma}{}_{\nu}\,.
\end{equation}
Moreover, it follows from a detailed theoretical investigation regarding the nature of the nonlocal scalar kernel ${\cal K}(x, y)$ that in the linear approximation, we must have a \emph{universal} function of $x-y$, namely,
\begin{equation}\label{II1E}
{\cal K}(x, y) = K(x-y)\,.
\end{equation}

In the Newtonian regime, there are no retardation effects, as $c \to \infty$; therefore, we can assume that
\begin{equation}\label{II1F}
 K(x-y)=\delta(x^0-y^0)~ k(\mathbf{x}-\mathbf{y})\,.
\end{equation}
Moreover, the only relevant component of Eq.~\eqref{II1B} in the Newtonian limit is the $\mu=\nu=0$ one with $G_{00}=(2/c^2)\nabla^2\phi$, $\mathfrak{C}_{0i0}=(2/c^2)\partial_i \phi$ and $T_{00}=\rho c^2$; then, using the convolution property of kernel $k$, an integration by parts and Gauss's theorem, we finally arrive at the modified Poisson equation. We note that the only significant part of $\psi_{\mu \nu}$ that survives in the transition to the Newtonian regime is its symmetric part given by $2\psi_{\mu \nu}=h_{\mu \nu}$, where, just as in GR,
\begin{equation}\label{II1G}
h_{\mu \nu}=-\frac{2\phi}{c^2}~\rm{diag}(1,1,1,1)\,.
\end{equation}
In this way, Eq.~\eqref{II1B} reduces in the Newtonian regime to a nonlocally modified Poisson equation that can be expressed as the Fredholm integral equation
\begin{equation}\label{II1}
\Psi (\mathbf{x})+\int k (\mathbf{x}-\mathbf{y})\Psi (\mathbf{y})d^3y= \psi (\mathbf{x})\,,
\end{equation}
where $\Psi = \nabla^2\phi$ and $\psi=4\pi G\rho$.
The nonlocal memory term in Eq.~\eqref{I3} reduces here to an instantaneous average over all space as retardation effects vanish in the Newtonian regime; in fact, the nonlocal Newtonian regime may be formally considered to be the limit of nonlocal gravity as $c\to \infty$. This transition---briefly outlined above---has been discussed in detail in Refs.~\cite{NG2, NG3, NG4, NG5, NG6}; in particular, the consequences of nonlocality for the gravitational physics of the Solar System have been explored in Ref.~\cite{NG3} and shown to be negligible at present based on a galactic nonlocality scale length $\lambda_0$ of  order 1 kpc. This is essentially because the dimension of the Solar System is very small compared to $\lambda_0 \sim 1$ kpc, the nonlocality scale that is necessary to provide a satisfactory explanation for the rotation curves of spiral galaxies~\cite{NG3}. We revisit this issue in section IV, once we determine that $\lambda_0 \approx 3 \pm 2$ kpc from the comparison of Eq.~\eqref{II1} with the rotation curves of spiral galaxies in section III.

It is not known whether nonlocal gravity in its general form can be derived from an action principle. There are in general problems with action principles for nonlocal theories if the
kernel is not symmetric---and causal kernels cannot be symmetric in time; in fact, this issue has been discussed in detail in Ref.~\cite{NG2}. On the other hand, it is possible to derive Eq.~\eqref{II1} from a variational principle if we assume that $k (\mathbf{x}-\mathbf{y})$ is only a function of $|\mathbf{x}-\mathbf{y}|$ and therefore symmetric. Indeed, it turns out that $k$ is invariant under the exchange of $\mathbf{x}$ and $\mathbf{y}$ for all nonlocal Newtonian kernels of interest in this paper.  In this case, the variation of $S$,
\begin{equation}\label{II1a}
 S = \int {\mathfrak L}~ d^3x
\end{equation}
with
\begin{equation}\label{II1b}
{\mathfrak L} = \frac{1}{8 \pi G} \Big[(\boldsymbol{\nabla}_{\mathbf{x}}~\phi)^2+\int  k (\mathbf{x}-\mathbf{y})(\boldsymbol{\nabla}_{\mathbf{x}}~\phi) \cdot (\boldsymbol{\nabla}_{\mathbf{y}}~\phi) ~d^3y\Big]+\rho ~ \phi\,
\end{equation}
results in Eq.~\eqref{II1}.

In Eq.~\eqref{II1}, which is a Fredholm integral equation of the second kind~\cite{Tr}, the density of matter in effect determines the sum of $\nabla^2\phi$ and its convolution with kernel $k$. We can formally solve this equation via the Liouville-Neumann method of successive substitutions. That is, we modify Eq.~\eqref{II1} by moving the integral term to the right-hand side; then, we replace $\Psi (\mathbf{y})$ in the integrand by its value given by the modified Eq.~\eqref{II1}. By repeating this procedure, we eventually obtain an infinite (Neumann) series that may or may not converge. If the Neumann series converges uniformly, we obtain a unique solution of the form
\begin{equation}\label{II2}
\Psi (\mathbf{x})=\psi (\mathbf{x})+\int q (\mathbf{x}-\mathbf{y})\psi (\mathbf{y})d^3y\,,
\end{equation}
where $q$ is the \emph{reciprocal} kernel; indeed, $k$ and $q$ are reciprocal of each
other~\cite{Tr}. This result can be written as
\begin{equation}\label{II3}
\nabla^2\phi=4\pi G(\rho+\rho_D), \quad \rho_D (\mathbf{x})=\int q(\mathbf{x}-\mathbf{y})\rho (\mathbf{y})d^3y\,,
\end{equation}
where $\rho_D$ has the interpretation of the density of ``dark matter". That is, the nonlocal aspect of gravity appears in Eq.~\eqref{II3} as an extra ``dark" matter source whose density is the convolution of the reciprocal kernel $q$ with the density of matter $\rho$. In this sense, nonlocality simulates dark matter. Moreover, no such dark matter exists in the complete absence of matter; i.e., $\rho_D=0$ if $\rho=0$.

In principle, we can determine the reciprocal convolution kernel $q$ from the comparison of the nonlocal theory with observational data regarding the ``flat" rotation curves of spiral galaxies~\cite{RF, RW, SR}. For any continuous  function $f$ that is absolutely integrable as well as square integrable over all space, let ${\hat f}$ denote its spatial Fourier integral transform; then, it follows from the definition of $\rho_D$ in Eq.~\eqref{II3} and the convolution theorem that ${\hat q}={\hat \rho_D}/{\hat \rho}$ in the Fourier domain. Moreover, under similar conditions, Eqs.~\eqref{II1} and~\eqref{II2} imply that $(1+{\hat q})(1+{\hat k})=1$. It can be shown that if $(1+{\hat q})\ne 0$, then one can obtain  kernel $k$ from the knowledge of the reciprocal kernel $q$~\cite{NG5}. Of the reciprocal kernels $k$ and $q$, if one is symmetric and hence only a function of $|\mathbf{x}-\mathbf{y}|$, then so is the other one.

It is interesting to recall here briefly the phenomenological Tohline-Kuhn modified-gravity approach to the problem of dark matter in galaxies and clusters of galaxies. An excellent review of this topic has been given by Bekenstein~\cite{B}. Imagine the circular motion of stars in the disk of a spiral galaxy. A ``flat" rotation curve implies that at any radius $r$ outside the central core of the spiral galaxy, all of the stars rotate with the same \emph{constant} circular speed $v_c$. Thus the centripetal acceleration of each star is $v_c^2/r$. Assuming Newton's fundamental laws of motion, the radial gravitational force of attraction experienced by a star must be equal to its mass multiplied by $v_c^2/r$. It follows that the main component of the radial gravitational force in the disk of a spiral galaxy must vary as $1/r$ with radial distance $r$ away from the center of the galaxy. Following this line of thought, Tohline~\cite{T} assumed that the gravitational force varies with distance as $r^{-2} + r^{-1}/\lambda_0$, where $\lambda_0 \sim 1$ kpc is a constant length. Thus for $r \gg \lambda_0$, the force of gravity varies as $1/r$ on the scale of galaxies. Tohline showed that this modified force law leads to the stability of a spinning galactic disk~\cite{T}. Therefore, in Tohline's approach, the Newtonian gravitational potential for a point mass $M$ would be modified by a logarithmic term, namely,
\begin{equation}\label{II4}
\phi_T(\mathbf{x})=-\frac{GM}{|\mathbf{x}|}
+\frac{GM}{\lambda_0}\ln\left(\frac{|\mathbf{x}|}{\lambda_0}\right)\,,
\end{equation}
where $\lambda_0$ is a constant galactic length of order 1\,kpc. Tohline's suggestion was generalized by Kuhn and his collaborators~\cite{B, K}. Indeed, Kuhn proposed a \emph{nonlocal} modification of Poisson's equation of the form~\eqref{II3} with the kernel
\begin{equation}\label{II5}
q_K(\mathbf{x}-\mathbf{y})=\frac{1}{4\pi\lambda_0}
\frac{1}{|\mathbf{x}-\mathbf{y}|^2}\,,
\end{equation}
such that $\phi_T$ is a solution of Eq.~\eqref{II3} with Kuhn's kernel $q_K$ when $\rho(\mathbf{x})=M \delta(\mathbf{x})$. The Tohline-Kuhn approach may appear to be in conflict with the empirical Tully-Fisher law~\cite{TF}. However, we take the view that for a fair comparison with the Tully-Fisher relation~\cite{TF}, one should include the electromagnetic radiation aspects of the issue as well~\cite{NG5}. We will return to this important topic at the end of section III.

The ``flat" rotation curve of a spiral galaxy extends out to the edge of the galaxy. On the other hand, the gravitational interaction must be defined over all space. Within the framework of nonlocal gravity, we need to generalize Kuhn's kernel in order to ensure that the total effective ``dark mass" is finite for a realistic distribution of matter; moreover, the corresponding Neumann series must be uniformly convergent to ensure the existence of  kernel $k$. We therefore introduce two new parameters in order to modify the behavior of Kuhn's kernel near $r=0$ and $r=\infty$. Two explicit examples were worked out in detail in Ref.~\cite{NG5} and further discussed in Ref.~\cite{NG7}, namely,
\begin{equation}\label{II6}
q_1=\frac{1}{4\pi \lambda_0}~ \frac{1+\mu (a_0+r)}{(a_0+r)^2}~e^{-\mu r}\,,
\end{equation}
\begin{equation}\label{II7}
q_2=\frac{1}{4\pi \lambda_0}~ \frac{1+\mu (a_0+r)}{r(a_0+r)}~e^{-\mu r}\,,
\end{equation}
where $r=|\mathbf{x}-\mathbf{y}|$ and $\lambda_0$ is, as before, a parameter that is expected to be of the order of 1~kpc. Parameters $\mu$ and $a_0$ are such that  $0<\mu a_0 \ll1$; in fact, we assume that $0<\mu \lambda_0<1$ and $0< a_0/\lambda_0 \ll 1$. The reciprocal kernels $q_1$ and $q_2$ are real positive functions that are integrable as well as square integrable over all space. Moreover, $q_1$ is finite everywhere and its Fourier integral transform is a real positive function if $a_0/\lambda_0$ is sufficiently small compared to unity, while $q_2$ diverges only at $r=0$ and its Fourier integral transform is always real  and positive. It has been shown in Ref.~\cite{NG5}, using the Fourier transform method, that it is then possible to infer the existence of the corresponding symmetric kernels $k_1$ and $k_2$, respectively. Furthermore, $k_1$ and $k_2$ have been numerically determined for $ \lambda_0=10$~kpc, $\mu^{-1}=10~\lambda_0$ and $a_0=10^{-3} \lambda_0$ in Ref.~\cite{NG5}; indeed, in this case $-k_1$ and $-k_2$ turn out to be positive functions of $r$ that fall off very fast with increasing $r$ and are effectively zero beyond $2.5~ \lambda_0$.

The behavior of the reciprocal kernel as $r \to 0$ accounts for the main difference between $q_1$ and $q_2$. Indeed, as $a_0 / \lambda_0 \to 0$,  $q_1$ and $q_2$ both become equal to $q$,
\begin{equation}\label{II8}
 q=\frac{1}{4 \pi \lambda_0}\frac{(1+\mu r)}{r^2}e^{-\mu r}\,.
\end{equation}
It is demonstrated in Appendix A that for the purposes of the present work, it is permissible to ignore $a_0$ in practice. Henceforth,  we adopt the two-parameter reciprocal kernel~\eqref{II8} for the sake of simplicity. It is important to emphasize, however, that more complicated kernels can also be considered by suitable generalizations of Eqs.~\eqref{II6} and~\eqref{II7}. Indeed, in nonlocal gravity, $q(\mathbf{x})$ is a \emph{universal} function of $\mathbf{x}$ that could depend on any number of constant parameters.

Adopting kernel~\eqref{II8}, we can solve for the modified force law of the Newtonian regime of nonlocal gravity and study the implications of this theory for the motion of particles and light rays.

\subsection{Modified force law}

In nonlocal gravity, the test particle follows a geodesic of the metric tensor $g_{\mu \nu}$; therefore,  in the Newtonian regime, we recover the usual force law
\begin{equation}\label{II9}
\frac{d^2\mathbf{x}}{dt^2}=-\boldsymbol{\nabla} \phi(\mathbf{x})\,,
\end{equation}
where $\phi$ is a solution of the modified Poisson equation~\eqref{II3} with the kernel $q$ given by Eq.~\eqref{II8} in the present context.

Let us note that in Eq.~\eqref{II3}, the nonlocal relation between the potential $\phi$ and matter density $\rho$ is \emph{linear}; therefore, it is possible to write
\begin{equation}\label{II10}
\phi(\mathbf{x}) =  G \int \xi (\mathbf{x}-\mathbf{y}) \rho(\mathbf{y})d^3y\,,
\end{equation}
where the \emph{Green} function $\xi$ is given by
\begin{equation}\label{II11}
  \nabla^2\xi (\mathbf{x}) = 4\pi [\delta (\mathbf{x})+  q(\mathbf{x})]\,.
\end{equation}
Thus $G \xi(\mathbf{x})$ is the gravitational potential at $\mathbf{x}$ due to a point particle of unit mass located at the origin of spatial coordinates.
It follows that the force on a test particle can also be expressed as the vector sum of the forces over the source, namely,
\begin{equation}\label{II12}
- \boldsymbol{ \nabla}_{\mathbf{x}}~ \phi (\mathbf{x}) = G \int [\boldsymbol{\nabla}_{\mathbf{y}}~\xi(\mathbf{x}-\mathbf{y})] ~\rho(\mathbf{y})d^3y\,.
\end{equation}
We conclude that, in computing the influence of an extended distribution of matter, it does not in the end matter whether one sums over the potential or the force, just as in Newtonian gravity. The test particle in the gravitational potential moves in such a way that $v^2+2\phi$ is conserved.

Let $F(r)$ be the magnitude of the radial attractive force of gravity between point masses $m_1$ and $m_2$ that are a distance $r$ apart; then, $F=Gm_1m_2f(r)$, where $f(r)=~ \partial \xi/\partial r$. It follows from Eq.~\eqref{II11} that
\begin{equation}\label{II13}
\frac{1}{r^2}\frac{d}{dr}(r^2f) = 4\pi \Big[\delta (\mathbf{x})+  q(\mathbf{x})\Big]\,.
\end{equation}
The solution of this equation is
\begin{equation}\label{II14}
f(r) =\frac{1}{r^2}\Big[1+ \alpha - \alpha(1+\frac{1}{2} \mu r)e^{-\mu r}\Big]\,,
\end{equation}
where we have introduced a new parameter $\alpha$ defined by
\begin{equation}\label{II15}
 \alpha  \mu \lambda_0=2\,
\end{equation}
and we have chosen the integration constant in Eq.~\eqref{II14} to be $\alpha$ in order to ensure that we recover the Newtonian inverse-square force law for $\mu r \to 0$. Here, $\lambda_0$ is, as before, a constant galactic lengthscale characteristic of nonlocality and $2/(\mu \lambda_0)= \alpha$, where $\alpha >2$ is a dimensionless parameter. In the explicit numerical examples worked out in Ref.~\cite{NG5}, for instance, $\alpha=20$.

We conclude that in the Newtonian regime of nonlocal gravity, the Newtonian gravitational force is modified such that it behaves like a Yukawa-type interaction. That is, for two test particles $m_1$ and $m_2$, their mutual gravitational attraction can be essentially characterized by a universal Yukawa-type force given by
\begin{equation}\label{II16}
 F(r)=\frac{Gm_1m_2}{r^2}\left \{1+\alpha \Big[1-(1+\frac{1}{2}\mu r)e^{-\mu r}\Big]\right \}\,.
\end{equation}
It is useful to express $F(r)$ in the form
\begin{equation}\label{II16a}
 F(r)=\frac{Gm_1m_2}{r^2}+ \frac{Gm_1m_2}{\lambda_0 r}~U(\mu r)\,,
\end{equation}
where $U(x)$ is given by
\begin{equation}\label{II16b}
 U(x)=\frac{2 e^{-x}}{x}\Big(e^x-1-\frac{1}{2}x\Big)\,.
\end{equation}
It is clear from Eq.~\eqref{II16a} that the basic nonlocality lengthscale is $\lambda_0$, since for $\lambda_0 \to \infty$, we recover the Newtonian inverse-square force law; indeed, for $\lambda_0 \to \infty$, the reciprocal kernel vanishes in Eq.~\eqref{II8} and thus Eq.~\eqref{II3} reduces to the Poisson equation.

For $x: 0 \to \infty$, $U(x)$ is a positive function ($U\ge 0$) that starts from $U=1$ at $x=0$ and then decreases monotonically as $x$ increases and tends to zero as $x \to \infty$ such that $x\,U(x) \to 2$. For $0 < x \ll 1$, we can write
\begin{equation}\label{II16c}
 U(x)=1-\frac{1}{6}~ x^2 + \frac{1}{12} ~x^3 - \frac{1}{40}~ x^4 + {\cal O}(x^5)\,.
\end{equation}
Thus for $\mu r < 1$, neglecting terms of order $(\mu r)^2$ in the expansion of $U(\mu r)$, we find that
Eq.~\eqref{II16a} reduces to the Tohline-Kuhn force~\cite{T,K}
\begin{equation}\label{II17}
F_{TK}(r)=\frac{Gm_1m_2}{r^2}+ \frac{Gm_1m_2}{\lambda_0 r}\,,
\end{equation}
which leads to flat rotation curves of spiral galaxies, but also involves a logarithmic potential that diverges at infinity. It turns out that the corresponding effective amount of dark matter is then infinite. To correct this situation in our nonlocal framework, parameter $\mu \ne 0$ is indispensable. For instance, it follows from Eq.~\eqref{II16} that
\begin{equation}\label{II18}
F(r \to \infty)\to \frac{Gm_1m_2}{r^2}(1+\alpha)\,.
\end{equation}
From the viewpoint of the test particle of mass $m_1$, say, $m_2$ has effective mass $(1+\alpha) m_2$, consisting of $m_2$ and its ``dark" component $\alpha m_2$, and vice versa. Alternatively, we may say that on the largest (cosmological) scales, the effective gravitational constant is $G(1+\alpha)$.

We should note here the remarkable similarity of Eq.~\eqref{II16} with the corresponding force law in the weak-field regime of  the MOdified Gravity (``MOG") theory~\cite{MR}. In fact, the corresponding MOG result can be obtained from Eq.~\eqref{II16} by replacing the factor of $\frac{1}{2}$ in front of $\mu r$ with   unity. Within the MOG framework, the extra factor of $1+\alpha$ is interpreted as leading to an effective gravitational constant given by $G(1+\alpha)$.

It is useful to calculate here the gravitational potential $\phi (r)$ due to a point source of mass $M$ such that $\phi \to 0 $ when $r \to \infty$, as expected.  From
\begin{equation}\label{II19}
\phi(r)= GM \int_\infty^r f(r')dr' \,,
\end{equation}
we find
\begin{equation}\label{II20}
 \phi(r)=-\frac{GM}{r}\Big[1+\alpha (1- e^{-\mu r})\Big]+\frac{GM}{\lambda_0}\int_{\infty}^{r}\frac{e^{-\mu r'}}{r'}dr'\,.
\end{equation}
The last integral can be expressed as $-E_1(\mu r)$, where $E_1(u)$ for $u>0$ is the \emph{exponential integral function}~\cite{A+S}
\begin{equation}\label{II21}
E_1(u):=\int_{u}^{\infty}\frac{e^{-t}}{t}dt\,.
\end{equation}
This positive function is such that for $u>0$,
\begin{equation}\label{II22}
 \frac{e^{-u}}{u+1} < E_1(u) \le \frac{e^{-u}}{u}\,,
\end{equation}
see formula 5.1.19 in Ref.~\cite{A+S}.  Moreover, let us note that
\begin{equation}\label{II23}
E_1(x)=-C-\ln x -\sum_{n=1}^{\infty}\frac{(-x)^n}{n~ n!}\,,
\end{equation}
where $C=0.577...$ is Euler's constant. In fact, $E_1(u)$ is a monotonically decreasing function that behaves like $-\ln u$ near $u=0$ and vanishes exponentially as $u \to \infty$.

We now turn to the motion of rays of radiation in the gravitational potential $\phi$.

\subsection{Light deflection}

The Yukawa-type force law leads to changes in the dynamics of galaxies and clusters of galaxies. Moreover,  rays of radiation follow null geodesics of the metric $g_{\mu \nu}$; therefore, light rays are also affected by the Yukawa-type potential $\phi(r)$. Indeed, as in general relativity, $g_{\mu \nu}=\eta_{\mu \nu}+h_{\mu \nu}$, where $h_{\mu \nu}=-(2\phi/c^2)$ diag(1, 1, 1, 1) and $\phi$ is the modified Newtonian potential given by Eq.~\eqref{II3}.

As is well known, in the first post-Newtonian approximation of general relativity, the net deflection angle of a light ray due to a ``Newtonian" potential $\phi$ is given by twice the Newtonian expectation. For instance, let ${\cal F}(r)$ be the attractive gravitational force on a test point particle of unit mass located outside of a spherical source of mass $M$; then, the net deflection angle $\Delta$ of a light ray propagating outside the source is given by
\begin{equation}\label{III1}
 \Delta=\frac{4\zeta}{c^2}\int_0^{\frac{\pi}{2}}{\cal F}\Big(\frac{\zeta}{\sin \theta}\Big)~\frac{d\theta}{\sin \theta}\,,
\end{equation}
where $\zeta=r \sin \theta$ is the impact parameter of the light ray and the scattering angle $\theta: 0 \to \pi$, as the unperturbed light ray is bent by the gravitational attraction of the spherical mass.

To illustrate formula~\eqref{III1}, let the spherical mass be a point source of mass $M$ such that ${\cal F}(r)=GMf(r)$. The calculation of the deflection angle is then straightforward and the result is that the net deflection angle $\Delta$ is given by
\begin{equation}\label{III2}
 \Delta=\frac{4GM}{c^2 \zeta}\Big \{1+\alpha [1- {\cal I}(\mu \zeta)] \Big \}\,,
\end{equation}
where
\begin{equation}\label{III3}
 {\cal I}(\mu \zeta)=\int_0^{\frac{\pi}{2}}(\sin \theta +\frac{1}{2}\, \mu \zeta)e^{-\frac{\mu \zeta}{\sin \theta}}~d\theta\,
\end{equation}
is such that for $0<\mu \zeta \ll 1$, ${\cal I} \approx 1- \pi \mu \zeta/4$.  It follows that in this case $\Delta \approx \Delta_E + 2 \pi G M /(c^2 \lambda_0)$, so that the Einstein deflection angle $\Delta_E=4GM/(c^2 \zeta)$ increases by a \emph{constant} amount in proportion to the mass of the source. This result is the same as that given in Ref.~\cite{NG3} based on the Tohline-Kuhn force~\eqref{II17}. Moreover, for impact parameters $\zeta \gg \mu^{-1}$, we note that ${\cal I}(\mu \zeta)$ vanishes exponentially as $\mu \zeta \to \infty$, so that $\Delta \to \Delta_E (1+\alpha)$, where the extra constant factor of $\alpha$ takes due account of the effective ``dark matter" associated with mass $M$. The implications of these results for gravitational lensing in general, and the Bullet Cluster~\cite{BC1, BC2} in particular, are beyond the scope of this paper and will be discussed elsewhere.

Consider next a homogeneous sphere of constant density $\rho$ and fixed radius $r_0$ as the source. What is the attractive force ${\cal F}$ due to the gravitational influence of the sphere's mass $M=(4\pi/3) r_0^3\,\rho$ on a test particle of unit mass held at a distance $r$, $r > r_0$, from the center of the sphere? To calculate ${\cal F}$, one must first compute the force  that would be directed toward the center of the sphere due to each spherical shell of radius $r'$ and then integrate over $r'$ from $0 \to r_0$. It follows that the net force of attraction pointing toward the center of the sphere is given by
\begin{eqnarray}\label{III4}
\nonumber {\cal F}(r>r_0)=\frac{GM}{r^2}(1+\alpha)-\frac{\pi \alpha G  \rho}{2r^2\mu^3}\Big \{ [3(\mu r+1)W_1(\mu r_0)+W_2(\mu r_0)]~e^{-\mu r}\\
 -\int_{\mu (r-r_0)}^{\mu (r+r_0)}u(u-\mu r)(u-2\mu r)E_1(u)du\Big\}\,,
\end{eqnarray}
where
\begin{equation}\label{III5}
 W_1(t)=(t-1)e^{t}+(t+1)e^{-t}\,,
\end{equation}
\begin{equation}\label{III6}
 W_2(t)=(t^2-2t+2)e^{t}-(t^2+2t+2)e^{-t}\,.
\end{equation}
The substitution of Eq.~\eqref{III4} in Eq.~\eqref{III1} leads to the net deflection angle of a light ray in this case.

For $\mu r \to \infty$, it is possible to show that the integral in Eq.~\eqref{III4} behaves as $\exp{(-\mu r)}$, so that very far from the source, the force of attraction is simply given by the Newtonian force augmented by the ``dark matter" factor $(1+\alpha)$. This result is ultimately based on the relations
\begin{equation}\label{III7}
 \int u^nE_1(u)du=\frac{u^{n+1}}{n+1}E_1(u)+\frac{1}{n+1}\int u^ne^{-u}du\,,
\end{equation}
\begin{equation}\label{III8}
 \int u^n e^{-u}du=-\Big[u^n+\sum_{k=1}^{n}n(n-1)...(n-k+1)u^{n-k}\Big]e^{-u}\,
\end{equation}
as well as the following asymptotic expansion
\begin{equation}\label{III9}
 \int_z^{\infty}\frac{e^{-u}}{u^n}du\sim \frac{e^{-z}}{z^n}\Big\{1-\frac{n}{z}+\frac{n(n+1)}{z^2}-\frac{n(n+1)(n+2)}{z^3}+...\Big\}\,,
\end{equation}
all for $n=1,2,3,...$, and based on the method of integration by parts. It is interesting to note that for $n=0, 1, 2, ...,$
\begin{equation}\label{III10}
\int_0^{\infty}u^nE_1(u) du=\frac{n!}{n+1}\,.
\end{equation}

It is clear that the formula for the deflection angle given by  Eq.~\eqref{III1} must be modified if the light ray passes through the spherical mass. One must consider separately the exterior regions with $r>r_0$ and the interior region with $r < r_0$. For the interior region, one finds
\begin{eqnarray}\label{III11}
\nonumber {\cal F}(r < r_0)=\frac{GMr}{r_0^3}(1+\alpha)-\frac{\pi \alpha G \rho}{r^2\mu^3}\Big \{-2\mu r+3\mu r \int_0^{\mu (r_0+r)} u^2E_1(u)du\\
\nonumber +[\mu^2r_0^2\sinh(\mu r)+(\mu r_0+1)Q(\mu r)]e^{-\mu r_0}\\
 -\frac{1}{2}\int_{\mu (r_0-r)}^{\mu (r_0+r)}u(u+\mu r)(u+2\mu r)E_1(u)du\Big\}\,,
\end{eqnarray}
where
\begin{equation}\label{III12}
Q(x):=3x\cosh x- \sinh x\,.
\end{equation}
One can easily check that for $r=r_0$, Eqs.~\eqref{III4} and~\eqref{III11} give the same result.

Finally, we consider in Eq.~\eqref{III4} the special case where $0 < \mu r \ll 1$, as, for instance, in the gravitational physics of the Solar System if the Sun is assumed to be a sphere of uniform density.  In this case, Eq.~\eqref{II16} reduces to the Tohline-Kuhn force~\eqref{II17} and a detailed calculation reveals that
\begin{equation}\label{III13}
 {\cal F}(r>r_0)=\frac{GM}{r^2}+ \frac{GM}{r \lambda_0} \Big[1-\frac{1}{5}\, \frac{r_0^2}{r^2} +{\cal O}\Big(\frac{r_0^4}{r^4}\Big)\Big]\,,
\end{equation}
in agreement with Eq.~(53) of Ref.~\cite{NG3}. It follows that for $r \gg r_0$, one may neglect the small corrections in Eq.~\eqref{III13} to the Tohline-Kuhn force~\eqref{II17}, which can therefore be used for estimating the effects of nonlocality in the Solar System as in section IV.

The rest of this paper is devoted to the confrontation of the nonlocal gravity theory with observational data.

\section{Observational Tests: Rotation Curves of Spiral Galaxies}

To investigate the rotation curves of spiral galaxies within the framework of nonlocal gravity, we assume that there is no actual dark matter; therefore, such a galaxy essentially consists of baryonic matter in the form of stars and interstellar gas. We ignore dust in our analysis, as the mass of the dust is at most a few percent of the mass of the interstellar matter. Using Eq.~\eqref{II16}, the effective gravitational acceleration of a test particle due to an extended distribution of matter with density $\rho(\mathbf{x})$ can be written in the form
\begin{equation}
\mathbf{a}(\mathbf x) = - G
\int\frac{\rho(\mathbf x')(\mathbf{x}-\mathbf{x'})}{|{\mathbf x}-{\mathbf x'}|^3}\Big[1+\alpha
-\alpha e^{-\mu|{\mathbf x}-{\mathbf x'}|}(1+\frac{\mu}{2}|{\mathbf x}-{\mathbf x'}|)\Big]~d^3x'\,. \label{acceleration}
\end{equation}
Assuming cylindrical symmetry for the galactic disk and introducing cylindrical polar coordinates $\hat r$, $\theta$ and $z$ with $\mathbf{\hat r}=(\hat r, \theta)$, the radial component of the acceleration is given approximately by
\begin{equation}
a_{\hat r}(\hat r) = G\sum_{{\hat r}'=~0}^\infty \sum_{\theta'=~
0}^{2\pi}\frac{\Sigma({\hat r}')}{|{\mathbf {\hat r}}- {\mathbf {\hat r}'}|^3}(-\hat r +
\hat r' \cos \theta')\Big(1+\alpha
-\alpha e^{-\mu|{\mathbf {\hat r}}-{ \mathbf {\hat r'}}|} -\frac{1}{2}\mu\alpha|{\mathbf {\hat r}}-{\mathbf {\hat r'} }|e^{-\mu |{\mathbf {\hat r}}-{\mathbf{\hat r'}}|}\Big)\hat r' \Delta
\hat r' \Delta\theta'\,,
\label{cal}
\end{equation}
where we have divided the vertically compressed disk into a large, but finite, number of discrete elements. Here, $\Sigma(\hat r)$ represents the isotropic column density of a spiral galaxy and
 $|\mathbf{\hat r}-\mathbf{\hat r'}| = \sqrt{{\hat r^2} + {\hat r'^2} - 2{\hat r} {\hat r'} \cos\theta'}$, where we have set $\theta=0$ with no loss in generality.
For the function $\Sigma(\hat r)$ in Eq.~\eqref{cal}, one can either use the observed column density of baryonic matter directly or employ a fitting function. Adopting the latter alternative in our work, we use~\cite{fathi10}
\begin{equation}
\Sigma(\hat r) = \Sigma_0 \exp\left(-\frac{\hat r}{\hat R}\right)\,, \label{sigma}
\end{equation}
where the central column density $\Sigma_0$ and $\hat R$ must be determined for stars and gas separately. Adding the integrated mass of the disk, $2 \pi \Sigma_0 \hat R^2$, for each component, we obtain the total mass of the disk $M_{disk}$.

What is actually measured via astronomical observation is the surface density of the luminosity of a galaxy, which decreases from the center towards the edge of the galaxy. From the intrinsic galactic luminosity $L$, we can obtain the total mass of stars in the galaxy, $M_{\star}$, from the stellar mass-to-light ratio
$M_\star/L = \Upsilon_\star$. Moreover, the mass of the gaseous component of the galaxy can be obtained from the total mass of the hydrogen gas, $M_{\rm H}$, via $M_{gas} = \frac{4}{3}M_{\rm H}$, due to the hydrogen to helium abundance in Big Bang nucleosynthesis, while $M_{\rm H}$ can be determined from the 21 cm radiation of the atomic hydrogen. In our analysis of the rotation curves of galaxies, we will henceforth follow the same general approach as has already been used in Ref.~\cite{MR} for testing the weak-field approximation of MOG model. For the first part of our analysis, we choose a subsample of galaxies in The HI Nearby Galaxy Survey (THINGS) catalog with precise  velocity and gas profiles \cite{things}.  The rotation curves have been measured by the Doppler effect using  $21$ cm radiation of neutral hydrogen gas.  In this wavelength, moreover,  we use the observed surface density of the brightness of these galaxies given in Ref.~\cite{Blok2008}. For the stellar contribution, we use the luminosity of stars in the $3.6~\mu$m band. In shorter wavelengths, there would be more absorption and scattering of radiation by dust particles; therefore, it is advantageous to employ longer wavelengths for measuring the surface brightness of galaxies. In addition, longer wavelengths probe old stellar populations, while shorter wavelengths would be more sensitive to populations of young stars. In the present study, we take the stellar  luminosity distributions for this sample of galaxies in the $3.6~\mu$m band images from the SINGS (SIRTF Nearby Galaxies Survey, now known as Spitzer Infrared Nearby Galaxies Survey) catalog~\cite{kenn}.

Our first aim in this section is to fit the rotation curves of twelve galaxies from the THINGS catalog and find the best-fitting values for the parameters  $\alpha$ and $\mu$  of the reciprocal kernel of nonlocal gravity (NLG) in the Newtonian regime, as well as $\Upsilon_\star^{3.6}$, the stellar mass-to-light ratio in the $3.6~\mu$m band. We include $\Upsilon_\star^{3.6}$ as a parameter, since its value  depends on the astrophysical model employed and is rather uncertain in practice due to various factors including the initial mass function of stars, the presence of dust in the galaxy and the inclination angle of the galaxy with respect to the observer. Indeed, $\Upsilon_\star$ has not been measured for any galaxy. However, it has been measured for the stars in our cosmic neighborhood via the \emph{Hipparcos} satellite~\cite{flynn}; in this way, the local average value of $\Upsilon_\star$ in the $B$ band has been found to be about 1.4 $M_\odot/L_\odot$. Table~\ref{tab1} contains the list of twelve galaxies in the THINGS catalog consisting of six  low surface brightness (LSB) and six  high surface brightness (HSB) galaxies with their corresponding parameters and the best-fitting values for  the three
parameters of our model, namely, $\alpha$, $\mu$ and $\Upsilon_\star^{3.6}$, within $1\sigma$ error. Appendix B contains a brief account of our model-fitting scheme. Figure~\ref{fig1} shows the rotation curves of  the twelve galaxies with the corresponding best fits.  Combining the observational data for the twelve galaxies along the lines mentioned in Appendix B, we find $\alpha = 10.94\pm2.56$ and $\mu = 0.059\pm0.028~{\rm kpc^{-1}}$, which are thus the ``average" values and the corresponding error estimates for the parameters of nonlocal gravity (NLG) theory. Summing the reduced chi-squared values given in Table I  for the twelve THINGS galaxies and dividing the result by twelve results in the \emph{average} value of reduced $\chi^2$ given by $\overline{\chi^2} = 1.13$.

In the rest of this section, we fix the parameters of nonlocal gravity on the basis of the THINGS data, namely, we set
\begin{equation}\label{parameters1+2}
\alpha = 10.94\pm2.56\,, \qquad  \mu = 0.059\pm0.028~{\rm kpc^{-1}}\,.
\end{equation}
We have verified that there is no degeneracy here; that is, in all of our numerical work, we have always found one global minimum in the $(\alpha, \mu, \Upsilon_\star^{3.6})$ space.
Moreover, it follows from Eq.~\eqref{II15} that $\lambda_0=2/(\alpha \mu)$, so that in our model
\begin{equation}\label{parameter3}
\lambda_0 \approx 3 \pm 2~ {\rm kpc}\,.
\end{equation}

\begin{table*}
\begin{center}
\caption{\label{tab1} Galaxies from the THINGS
catalog~\cite{things} with the best-fit parameter values obtained from fitting the
observed rotation curves to the theoretical nonlocal gravity rotation curves.
The columns contain (1) the name of the
galaxy, (2) the type of galaxy, (3) the distance of the galaxy, (4) the
color of galaxy in the $J-K$ band, (5) the best-fitting value of $\alpha$, (6)
the best-fitting value of $\mu$, (7) the stellar mass-to-light
ratio for each galaxy $\Upsilon_\star^{3.6}$ in the $3.6~\mu$m band, and (8) the $\chi^2$ per degree of freedom from fitting.
The error estimates are obtained from the likelihood functions given in Figure 1. The observational data for the rotation curves of spiral galaxies are taken from the THINGS catalog~\cite{things}; moreover, the stellar luminosities of the galaxies in the $3.6~\mu$m band are taken from the SINGS catalog~\cite{kenn}.}

\begin{tabular}{|c|c|c|c|c|c|c|c|}
\hline\hline
Galaxy  & Type & Distance & $J-K$ & $\alpha$& $\mu$ & $\Upsilon_\star^{3.6}(\rm{NLG})$ &$\chi^2/N_{\rm{d.o.f.}}$ \\
 &  & (Mpc) & &  &  $({\rm kpc}^{-1})$&$(M_\odot/L_\odot)$ &  \\
 (1)   & (2)  & (3)   & (4)            & (5) &(6)                &(7)                  &(8)        \\
\hline
NGC 3198 & HSB & $13.8$  &  $0.940^{+0.051}_{-0.051}$ & $6.59^{+1.31}_{-0.84}$ & $0.024^{+0.006}_{-0.005}$ & $0.89^{+0.10}_{-0.09}$ & $0.50$ \\
NGC 2903 & HSB & $8.9$ &   $0.915^{+0.024}_{-0.024} $  & $6.51^{+0.75}_{-0.98} $ &  $0.070^{+0.007}_{-0.008}$& $1.29^{+0.10}_{-0.10}$ &  1.6\\
NGC 3521 & HSB & $10.7$ & $0.953^{+0.027}_{-0.027}$   & $7.06^{+1.92}_{-1.32}$   & $0.022^{+0.007}_{-0.011}$ & $0.85^{+0.06}_{-0.05}$ &1.50  \\
NGC 3621 & HSB & $6.6$ &  $0.860^{+0.042}_{-0.042}$ & $14.67^{+0.91}_{-1.29}$   & $0.024^{+0.007}_{-0.006}$ & $0.52^{+0.10}_{-0.04}$ & 1.30\\
NGC 5055 & HSB & $10.1$ & $0.961^{+0.027}_{-0.027}$   & $7.28^{+1.08}_{-1.01}$ & $0.038^{+0.008}_{-0.008}$ & $0.52^{+0.08}_{-0.08}$ &0.37\\
NGC 7331 & HSB & $14.7$  &      $1.03^{+0.024}_{-0.024}$   & $4.67^{+0.84}_{-1.01}  $ & $ 0.034^{+0.005}_{-0.006}                            $ &   $0.44^{+0.06}_{-0.04}$ & 0.79\\
NGC 2403 & LSB & $3.2$ & $0.790^{+0.031}_{-0.031}$ &  $17.68^{+1.00}_{-0.70}$ & $ 0.024^{+0.007}_{-0.006}$& $0.67^{+0.08}_{-0.07}$ &  3.04\\
DDO 154 &LSB  &4.3      & ---  & $20.01^{+0.64}_{-0.43}$& $0.227^{+0.010}_{-0.008} $ & $1.66^{+0.07}_{-0.09}$ &1.49 \\
IC 2574  &LSB  &4.0  &$0.766^{+0.115}_{-0.115}$    & $ 13.48^{+1.48}_{-1.66}$ & $0.058^{+0.014}_{-0.007}$ & $0.36^{+0.03}_{-0.10}$  & 0.36\\
NGC 0925 &LSB  &9.2      &$0.867^{+0.063}_{-0.063}$    & $14.67^{+0.77}_{-0.84}$ & $0.089^{+0.008}_{-0.009}$  & $0.25^{+0.04}_{-0.05}$  &  0.93\\
NGC 2366 & LSB &3.4       &       $0.667^{+0.146}_{-0.146}$    &  $12.43^{+1.07}_{-1.06}$ & $0.067^{+0.036}_{-0.008}$  &$1.24^{+0.10}_{-0.25}$   &0.062    \\
NGC 2976  &  LSB &3.6   &       $0.821^{+0.036}_{-0.036}$   & $6.57^{+0.98}_{-1.61}$  &  $0.028^{+0.036}_{-0.007}$ &$1.11^{+0.84}_{-0.09}$  &  1.67\\
\hline
\end{tabular}
\end{center}
\end{table*}

\setcounter{figure}{0}
\begin{figure*}
\begin{center}
\begin{tabular}{cc}
\includegraphics[width=85mm]{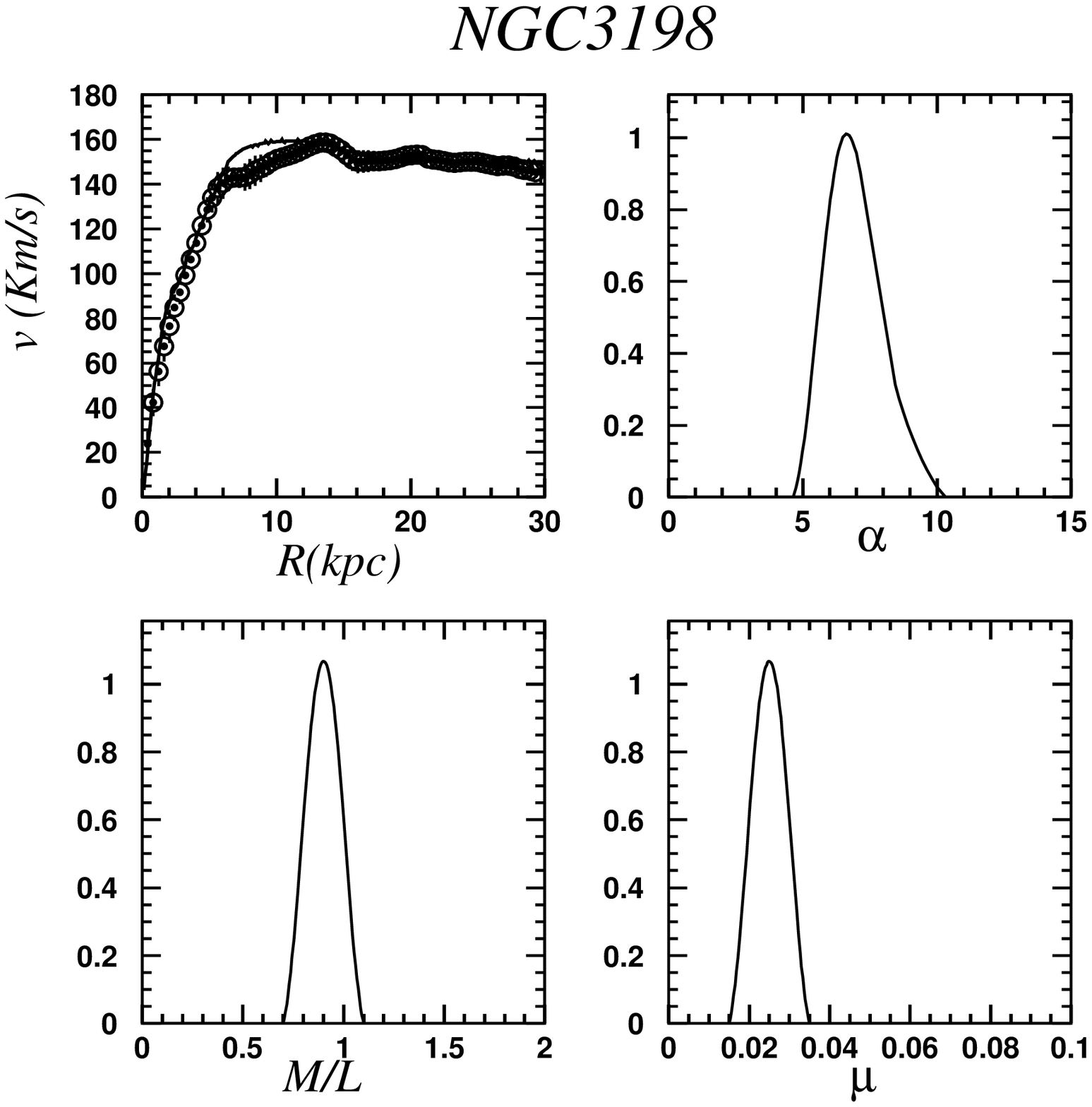} &
\includegraphics[width=85mm]{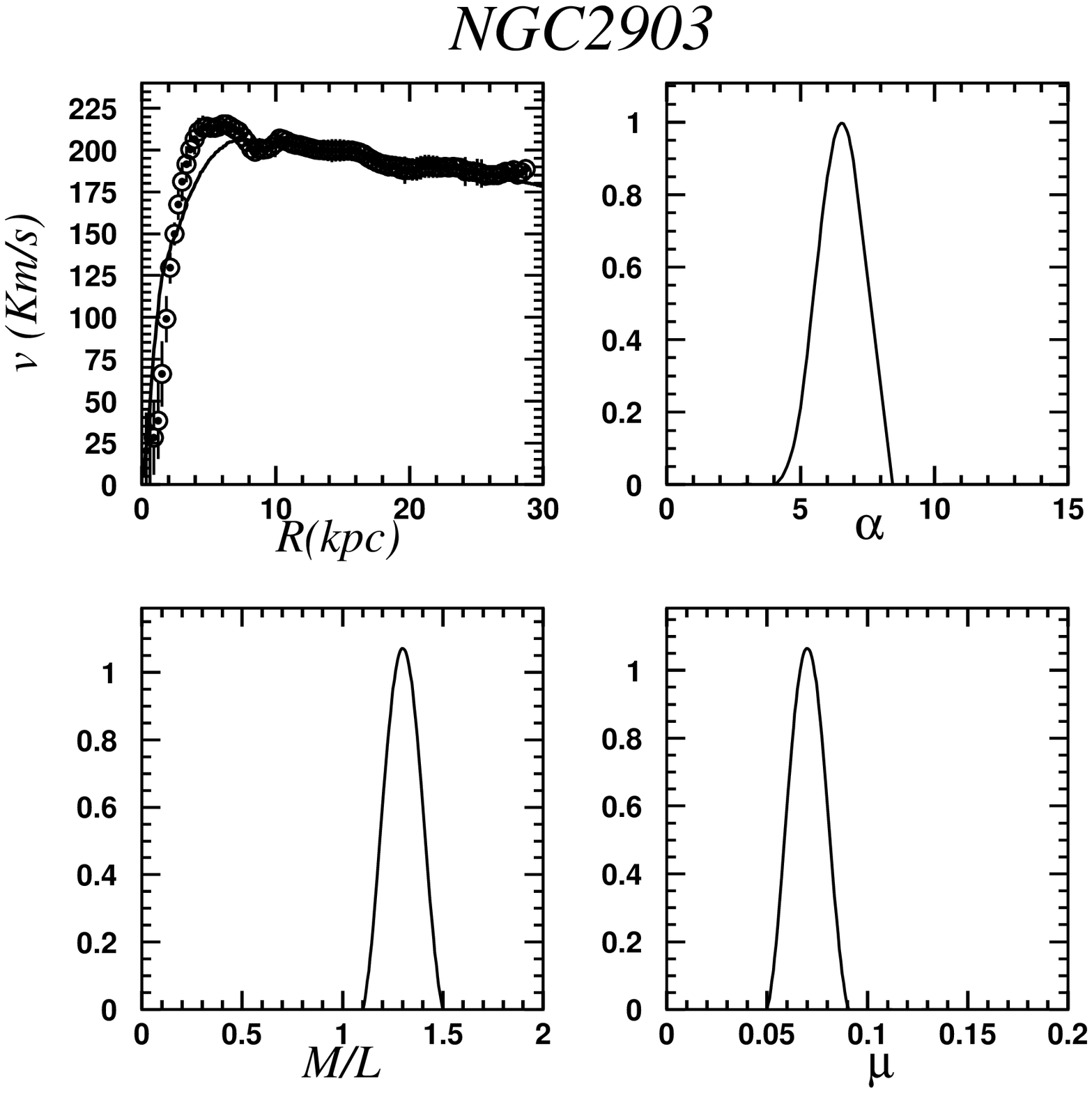}\\
\includegraphics[width=85mm]{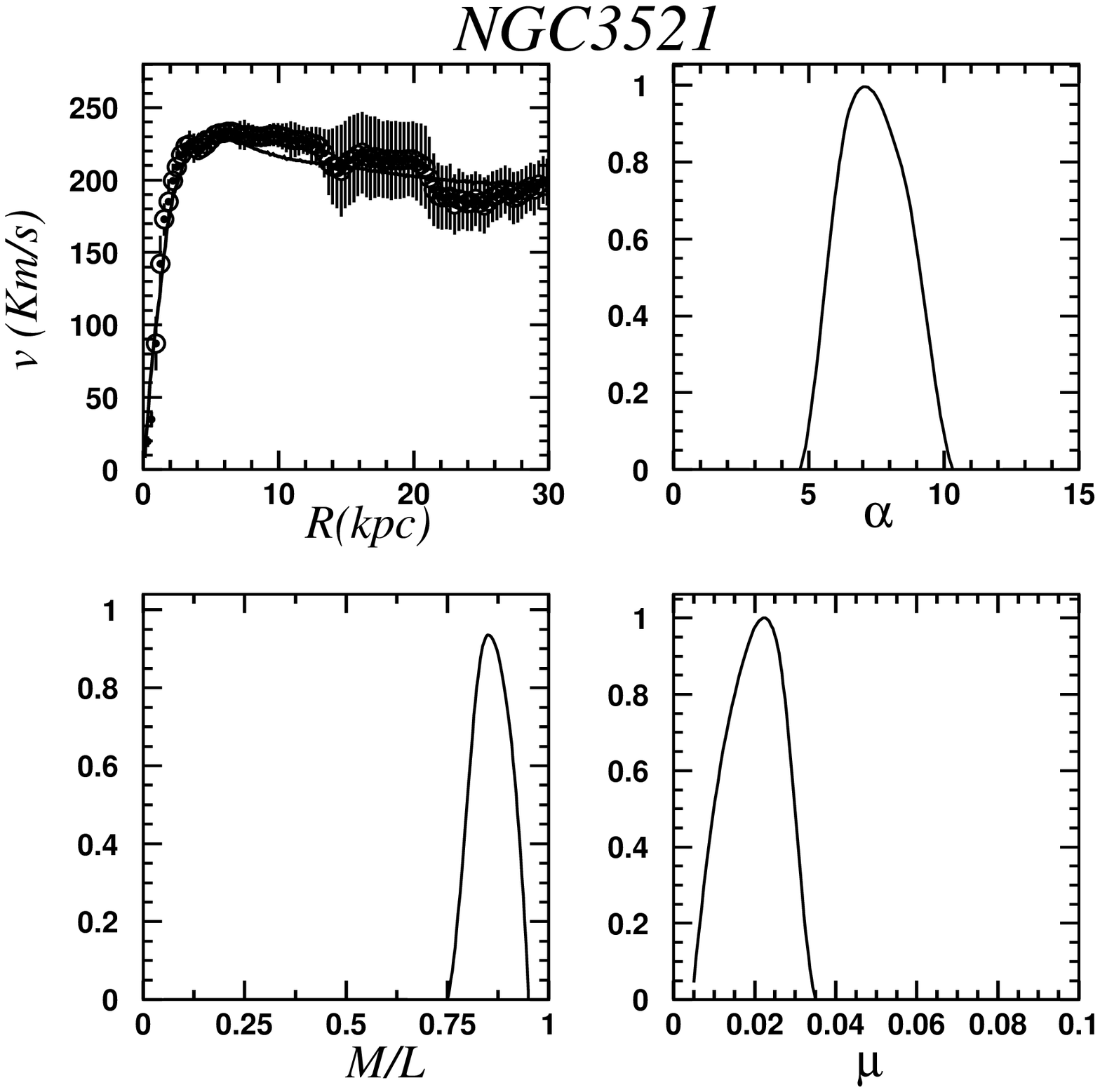}&
\includegraphics[width=85mm]{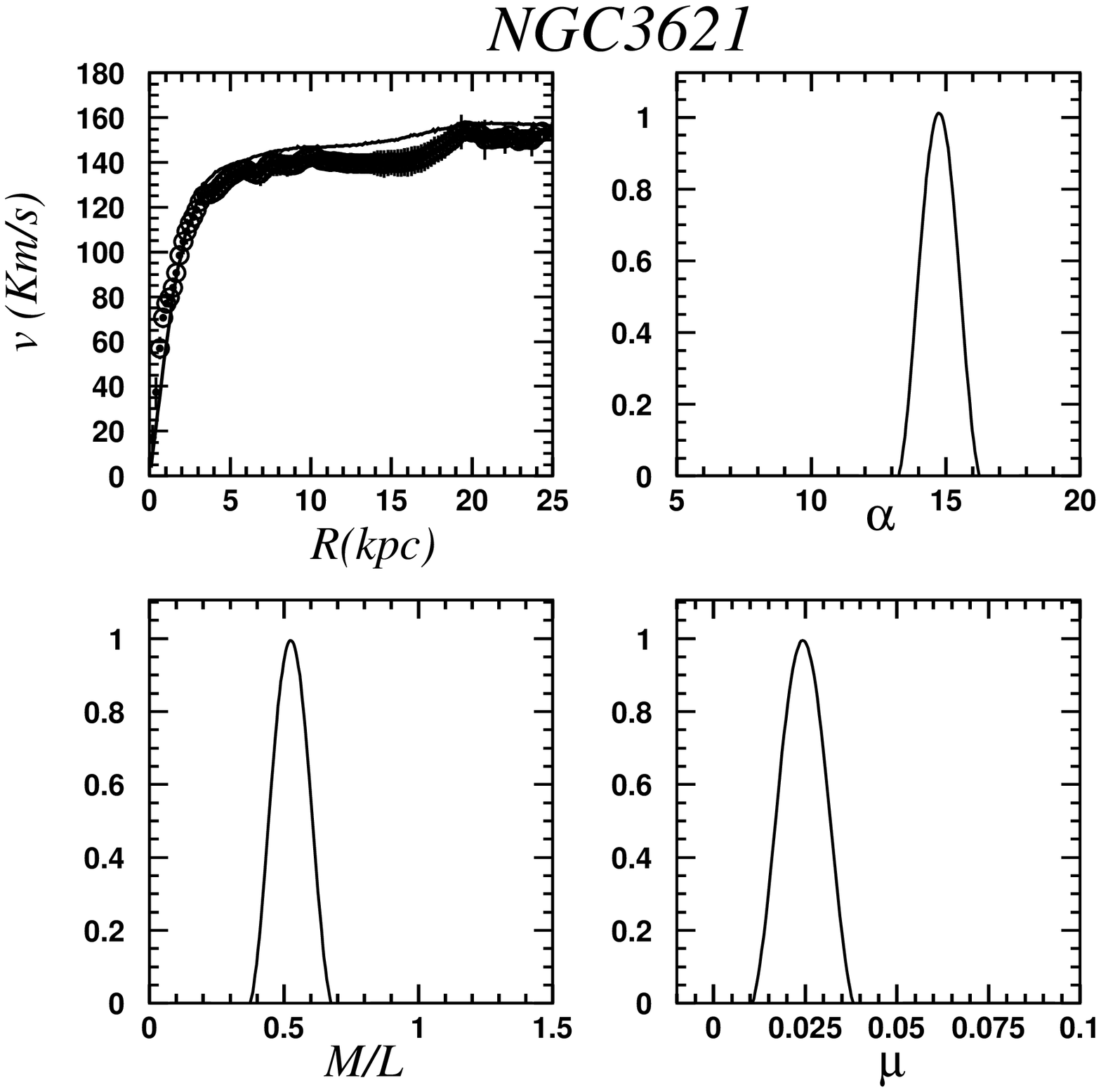} \\
\end{tabular}
\end{center}
\caption {The best fit for the THINGS galaxies listed in Table~\ref{tab1}, with
the corresponding marginalized likelihood functions of $\alpha$, $\mu$ and $\Upsilon_\star^{3.6}$. Both HSB and LSB galaxies are represented here.
Table~\ref{tab1} contains the best-fit values of the parameters with the
corresponding error estimates. \label{fig1}}
\label{fig1}
\end{figure*}

\setcounter{figure}{0}
\begin{figure*}
\begin{center}
\begin{tabular}{cc}
\includegraphics[width=85mm]{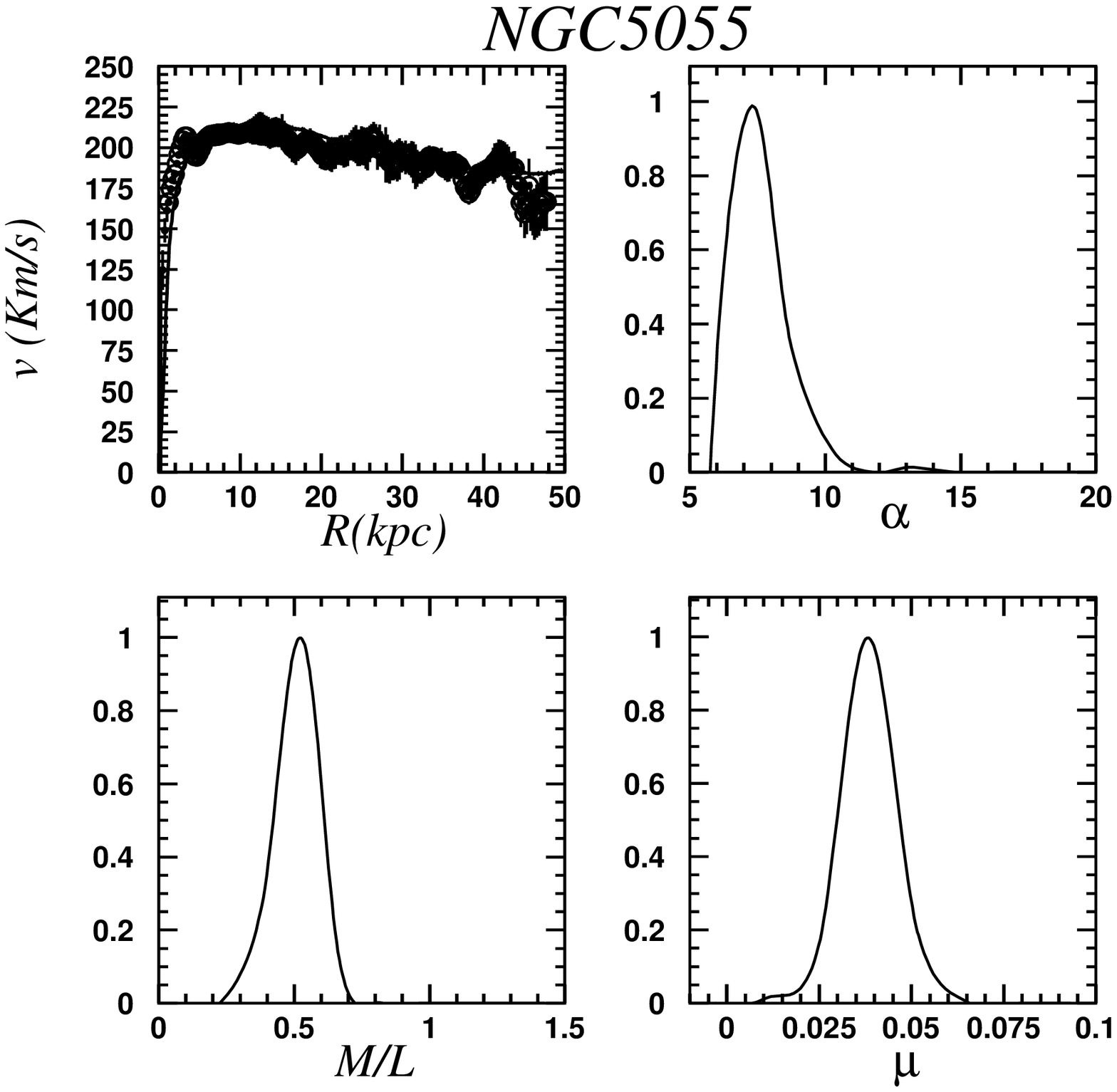} &
\includegraphics[width=85mm]{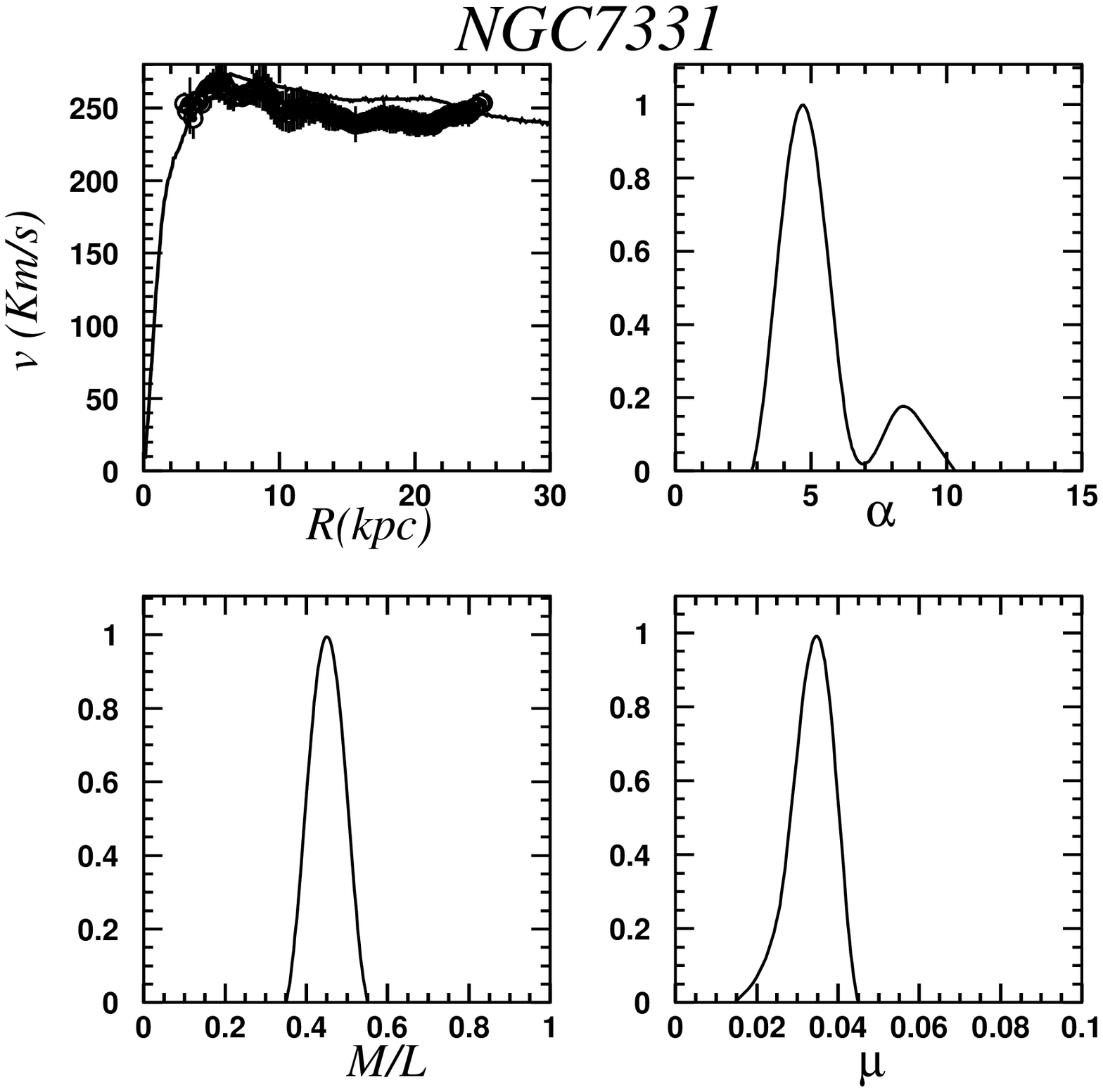} \\
\includegraphics[width=85mm]{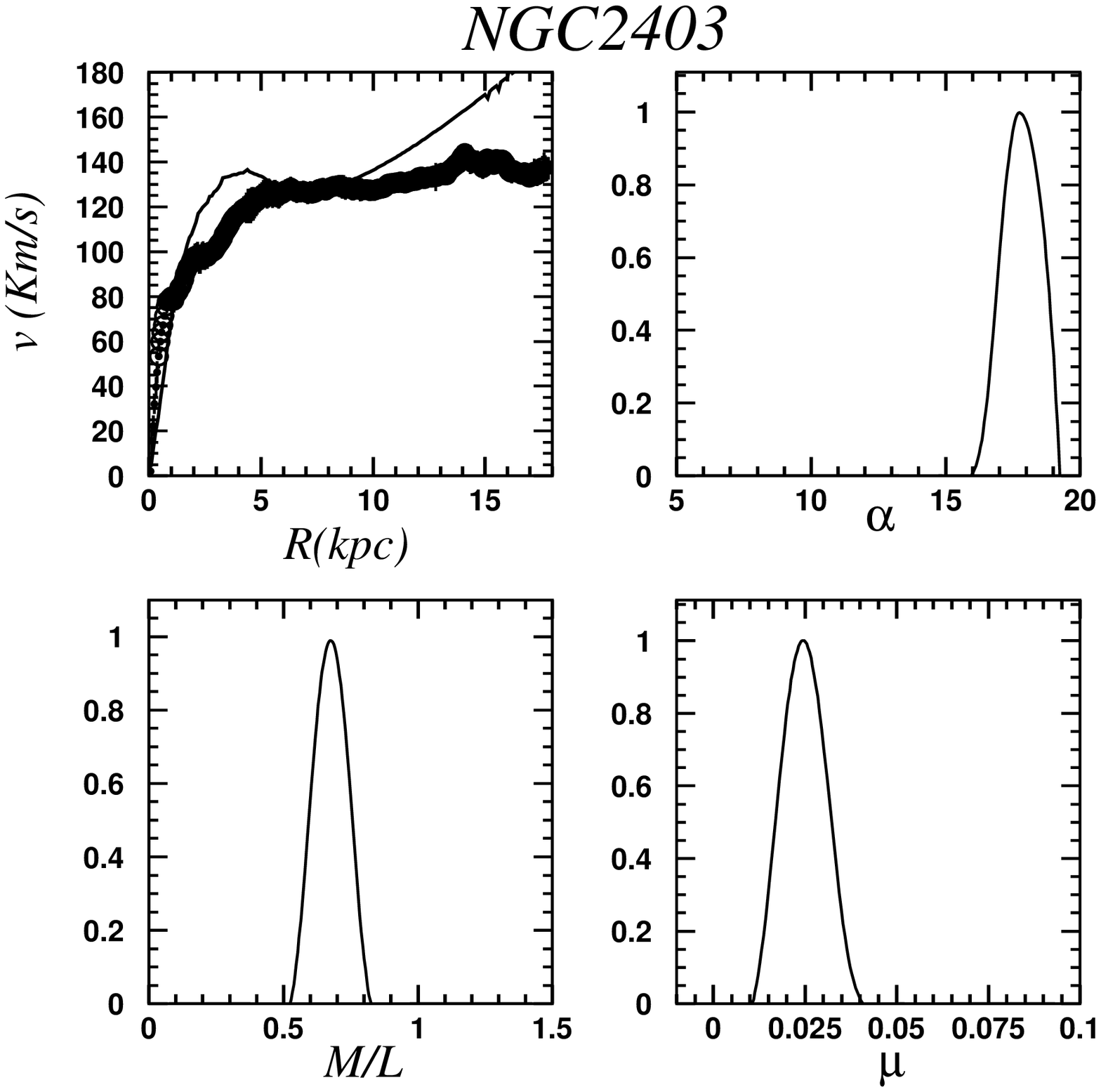}&
\includegraphics[width=85mm]{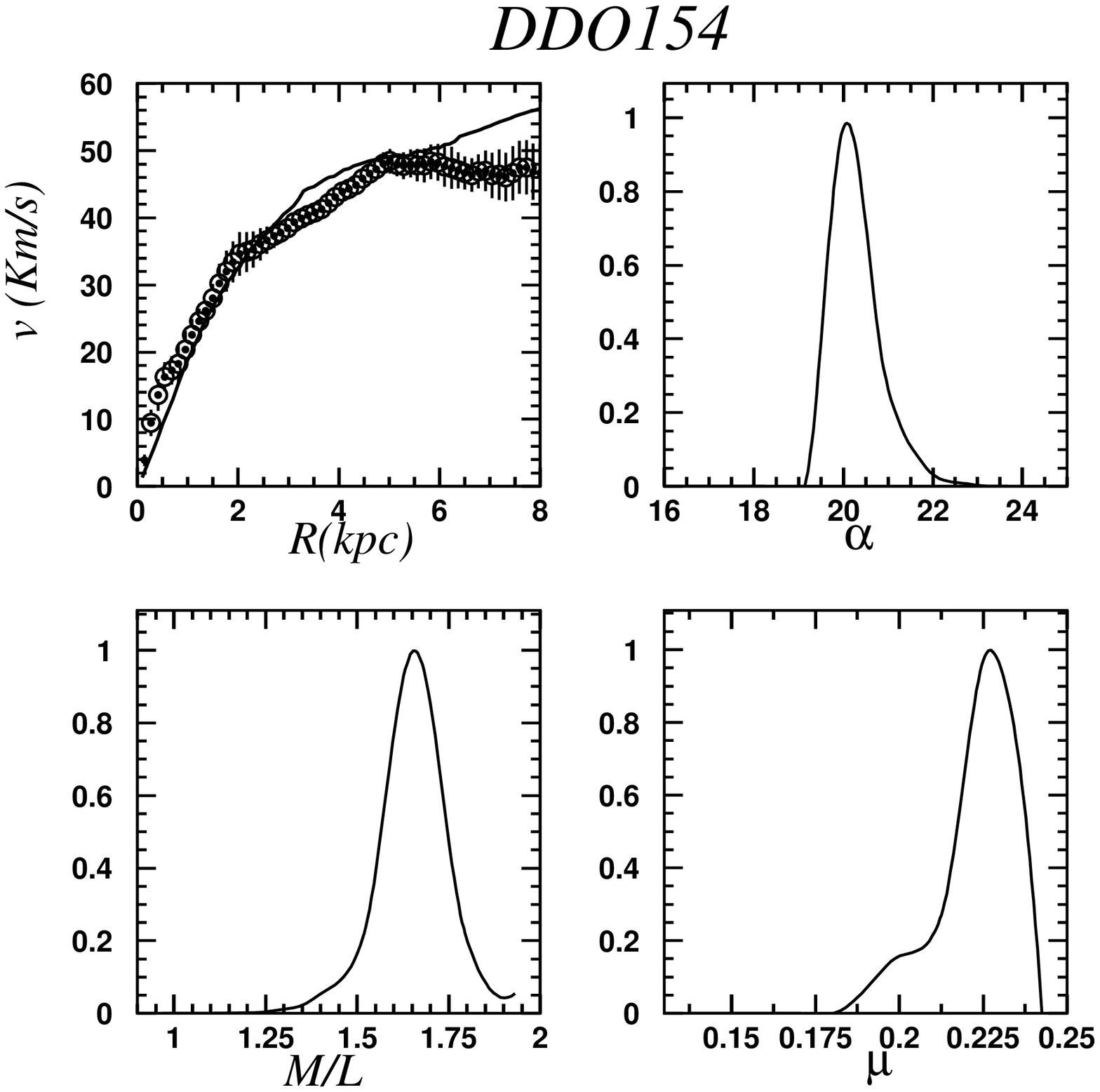} \\
\end{tabular}
\end{center}
\caption {Continued ... }
\end{figure*}

\setcounter{figure}{0}
\begin{figure*}
\begin{center}
\begin{tabular}{cc}
\includegraphics[width=85mm]{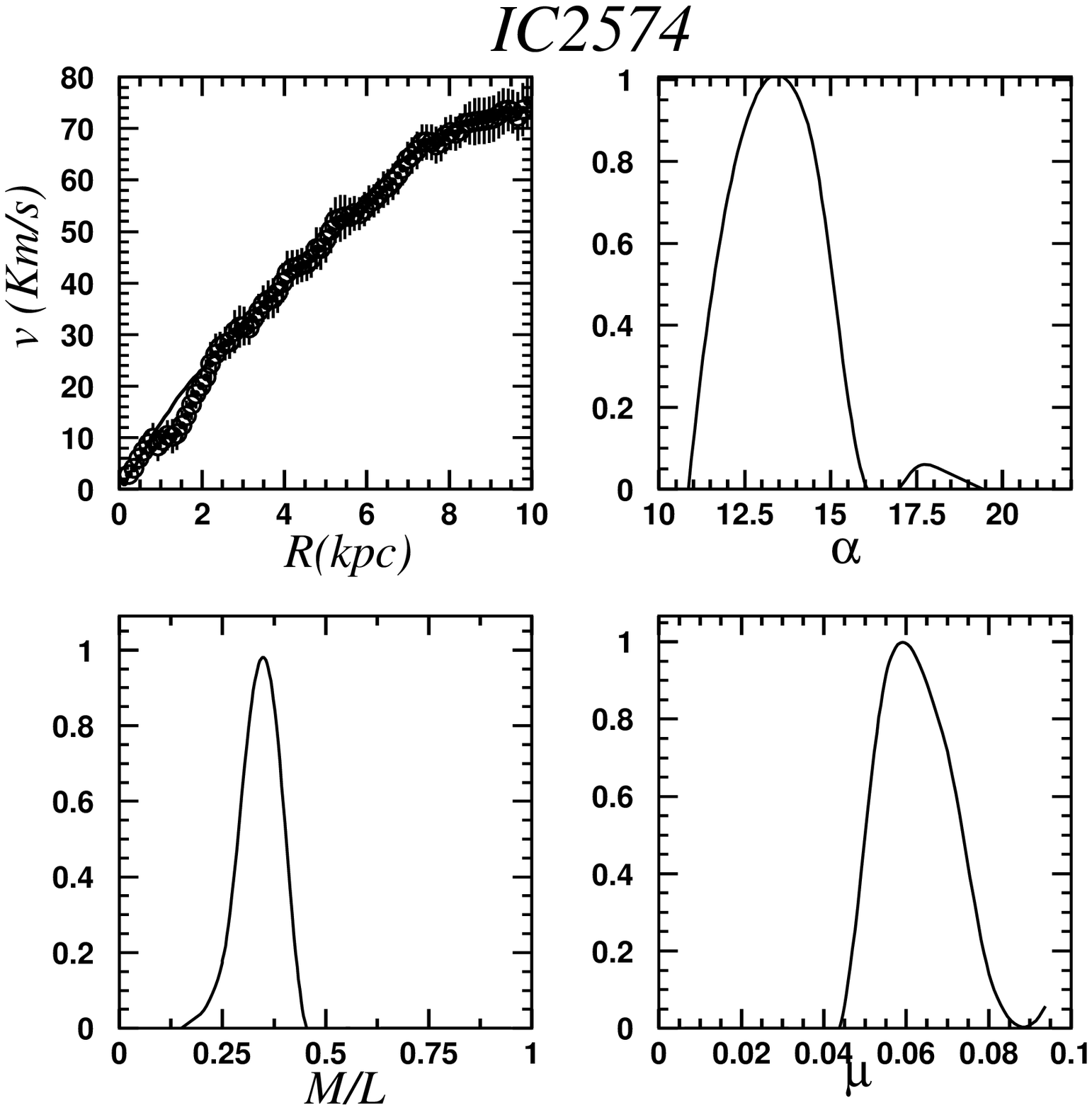}&
\includegraphics[width=85mm]{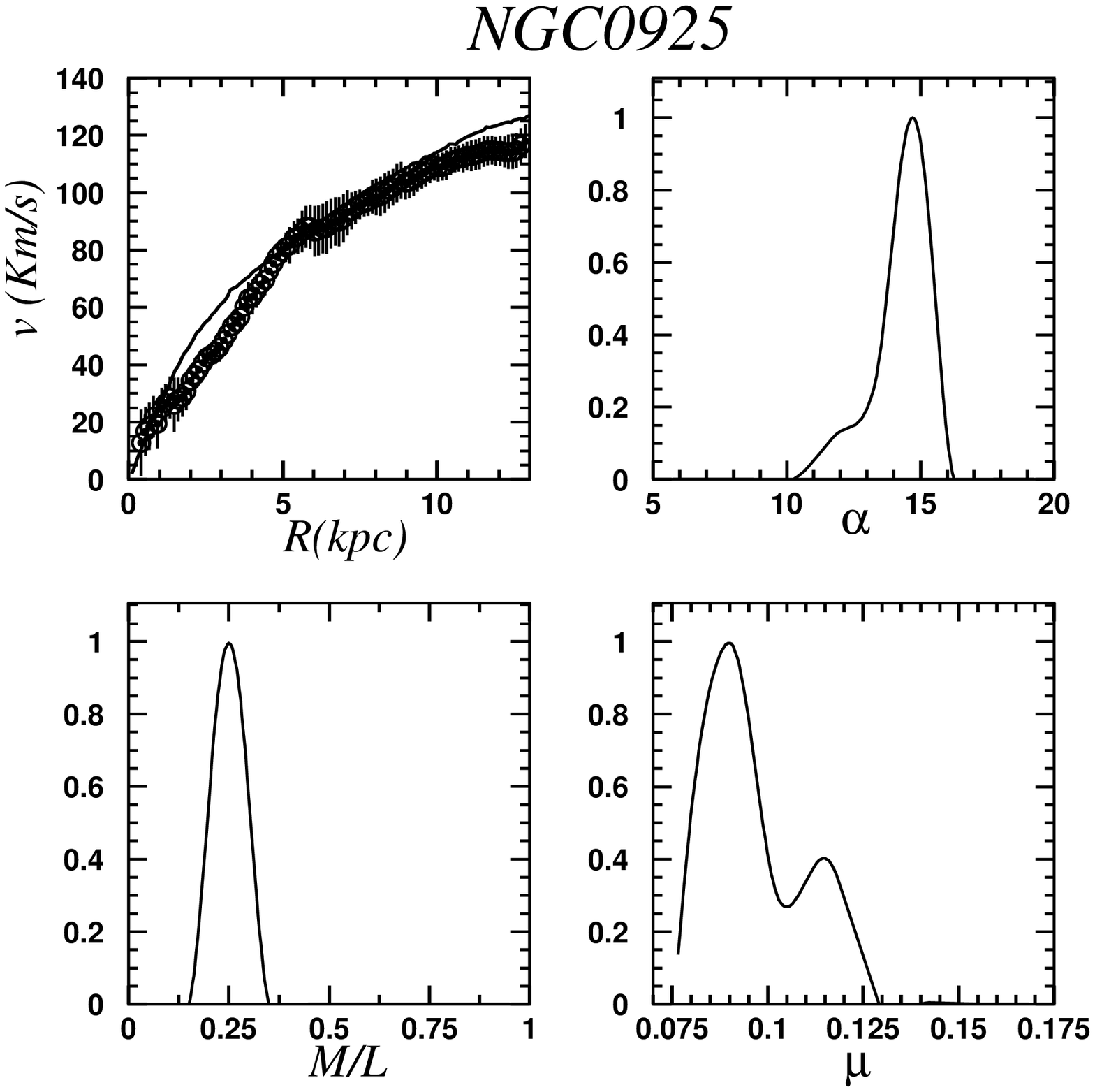} \\
\includegraphics[width=85mm]{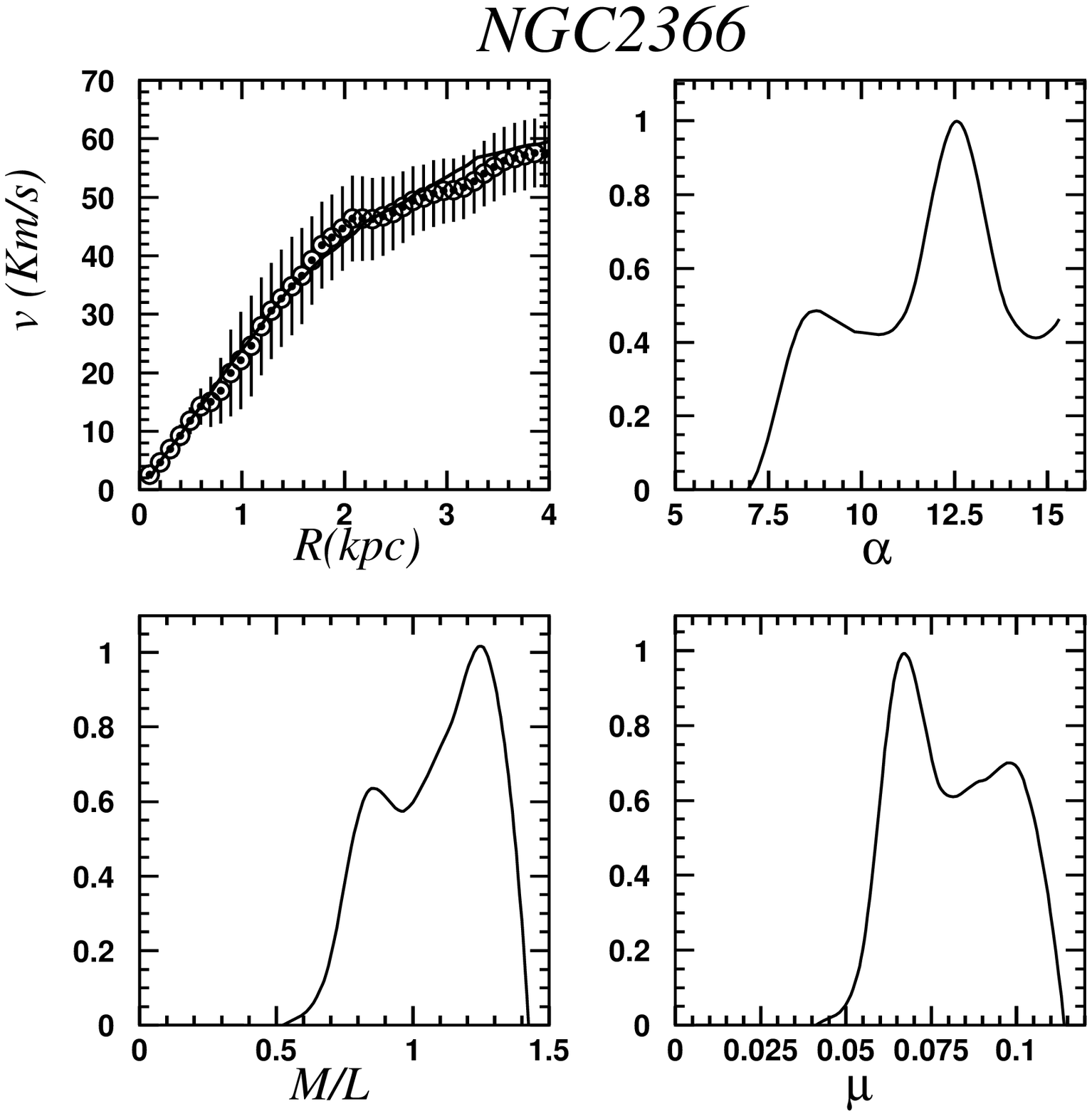}&
\includegraphics[width=85mm]{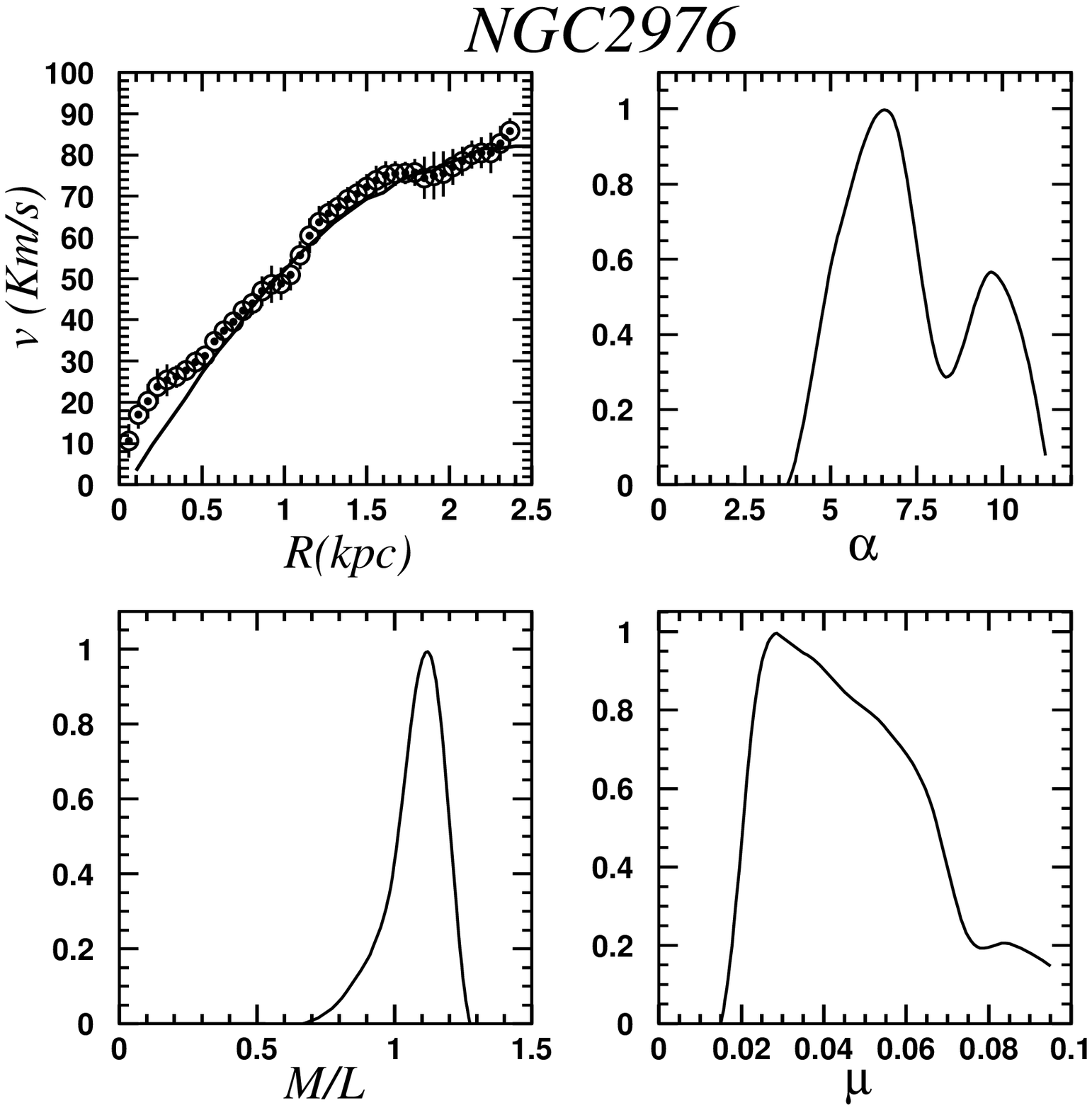} \\
\end{tabular}
\end{center}
\caption {Continued ...}
\end{figure*}

\setcounter{figure}{1}
\begin{figure*}
\begin{center}
\begin{tabular}{ccc}
\includegraphics[width=60mm]{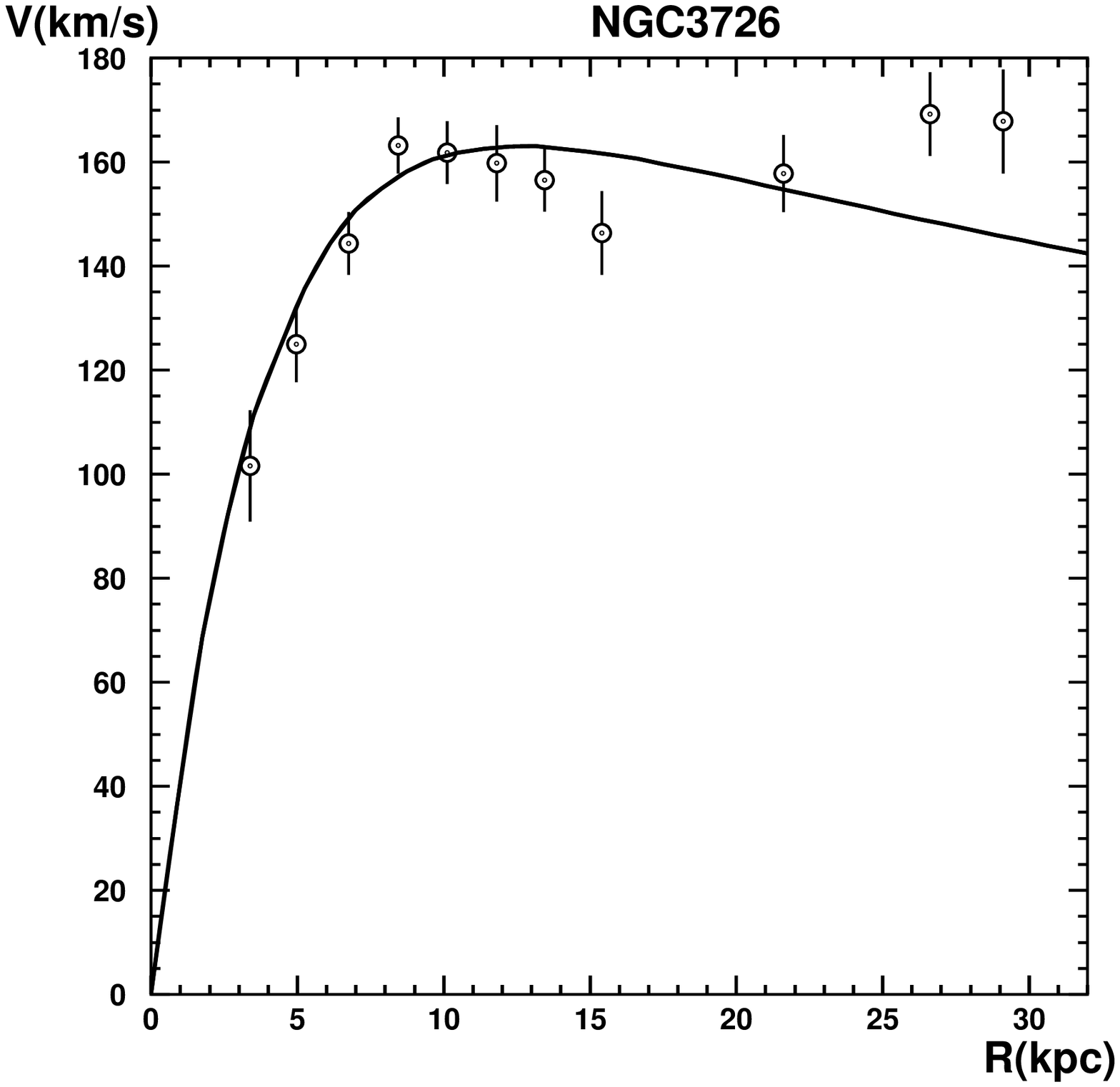}&
\includegraphics[width=60mm]{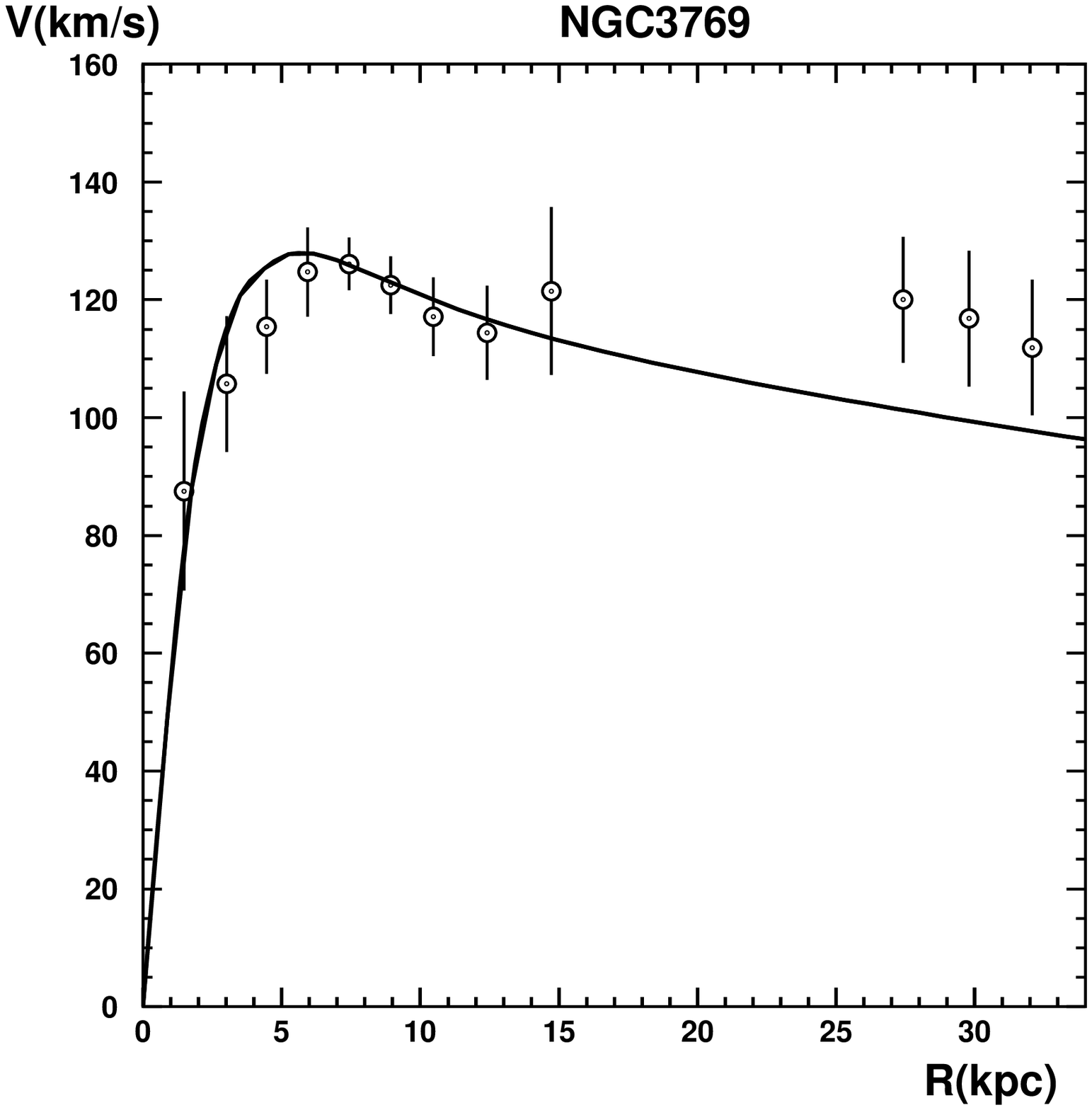}&
\includegraphics[width=60mm]{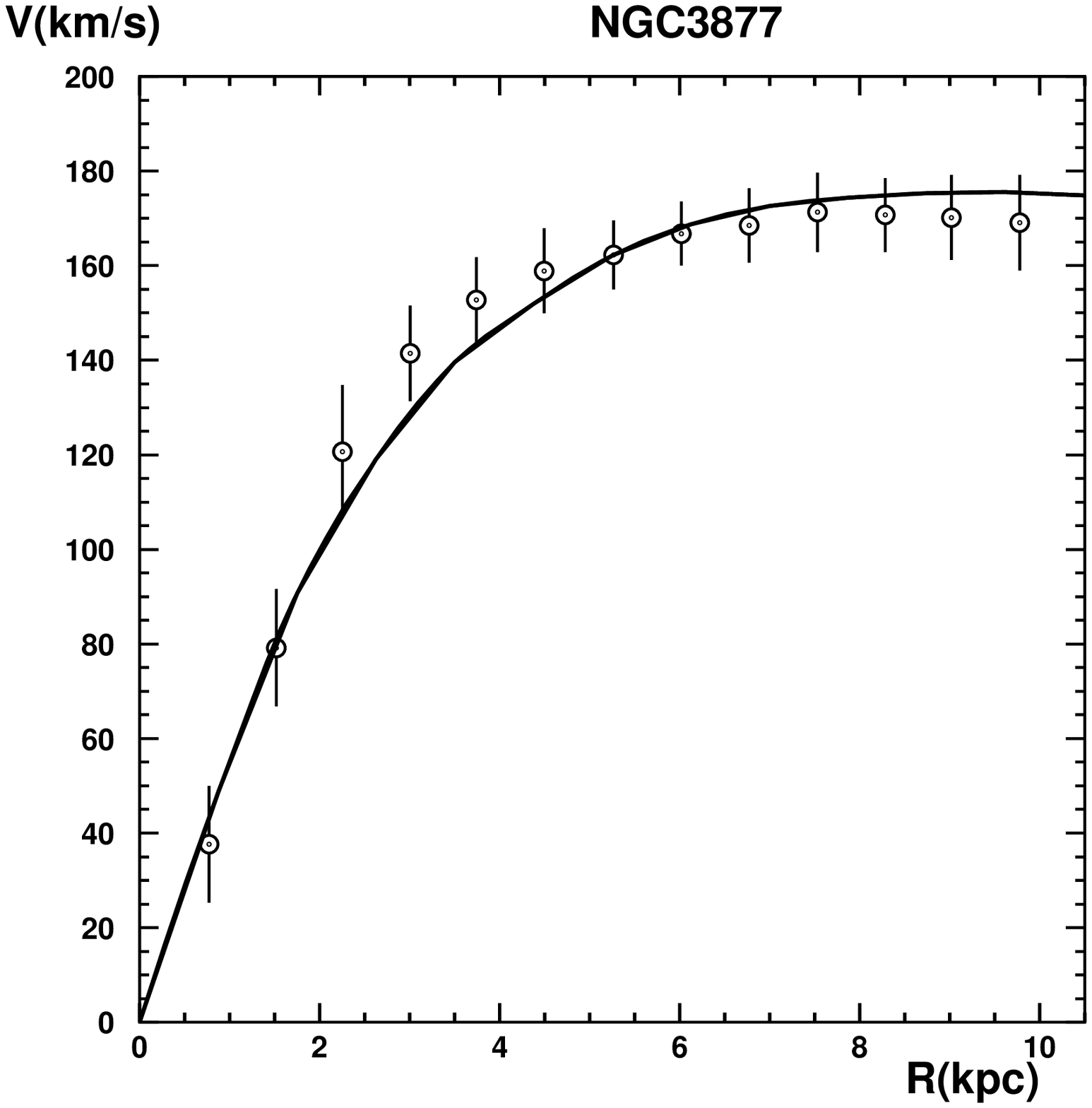} \\
\includegraphics[width=60mm]{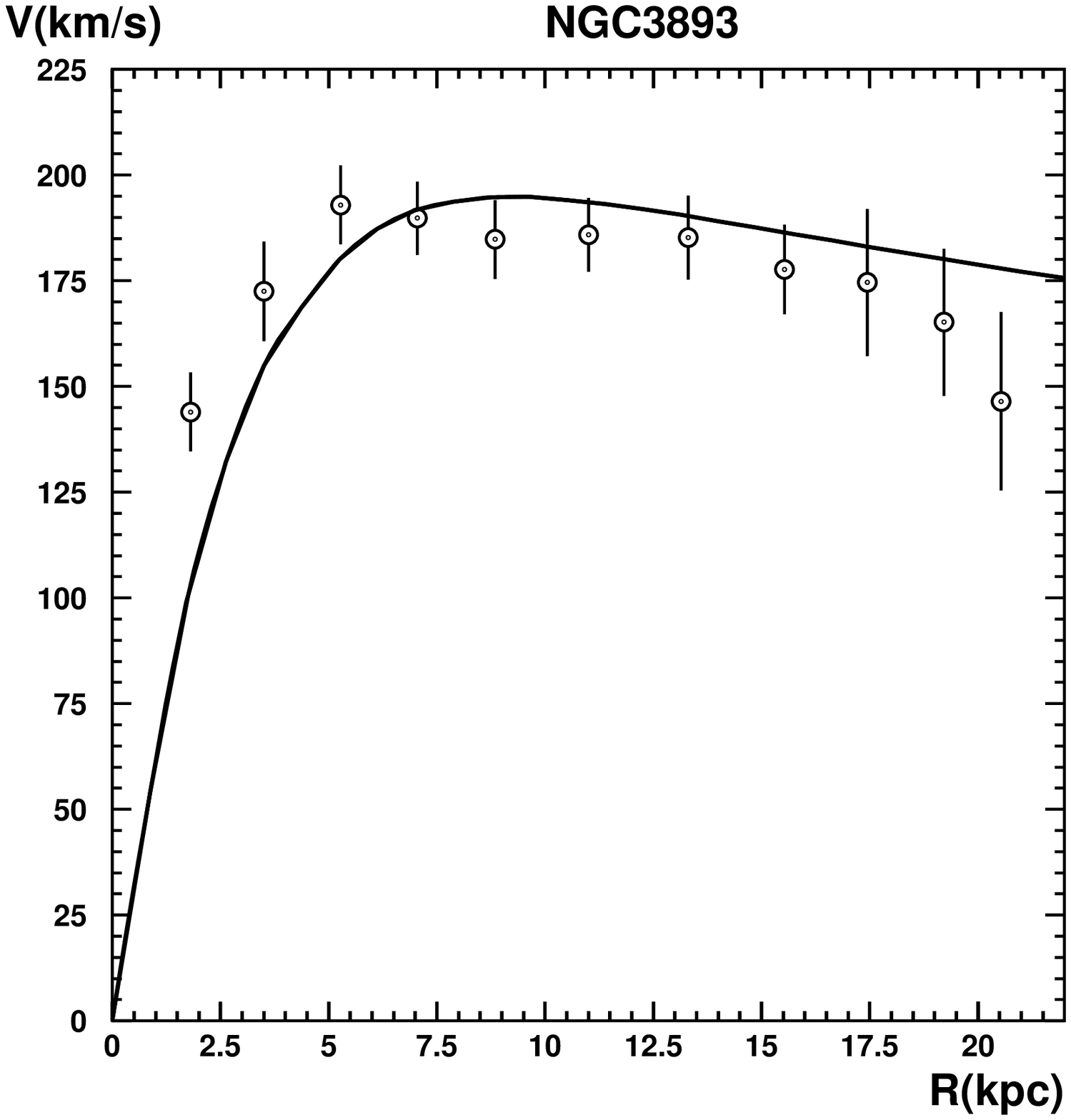} &
\includegraphics[width=60mm]{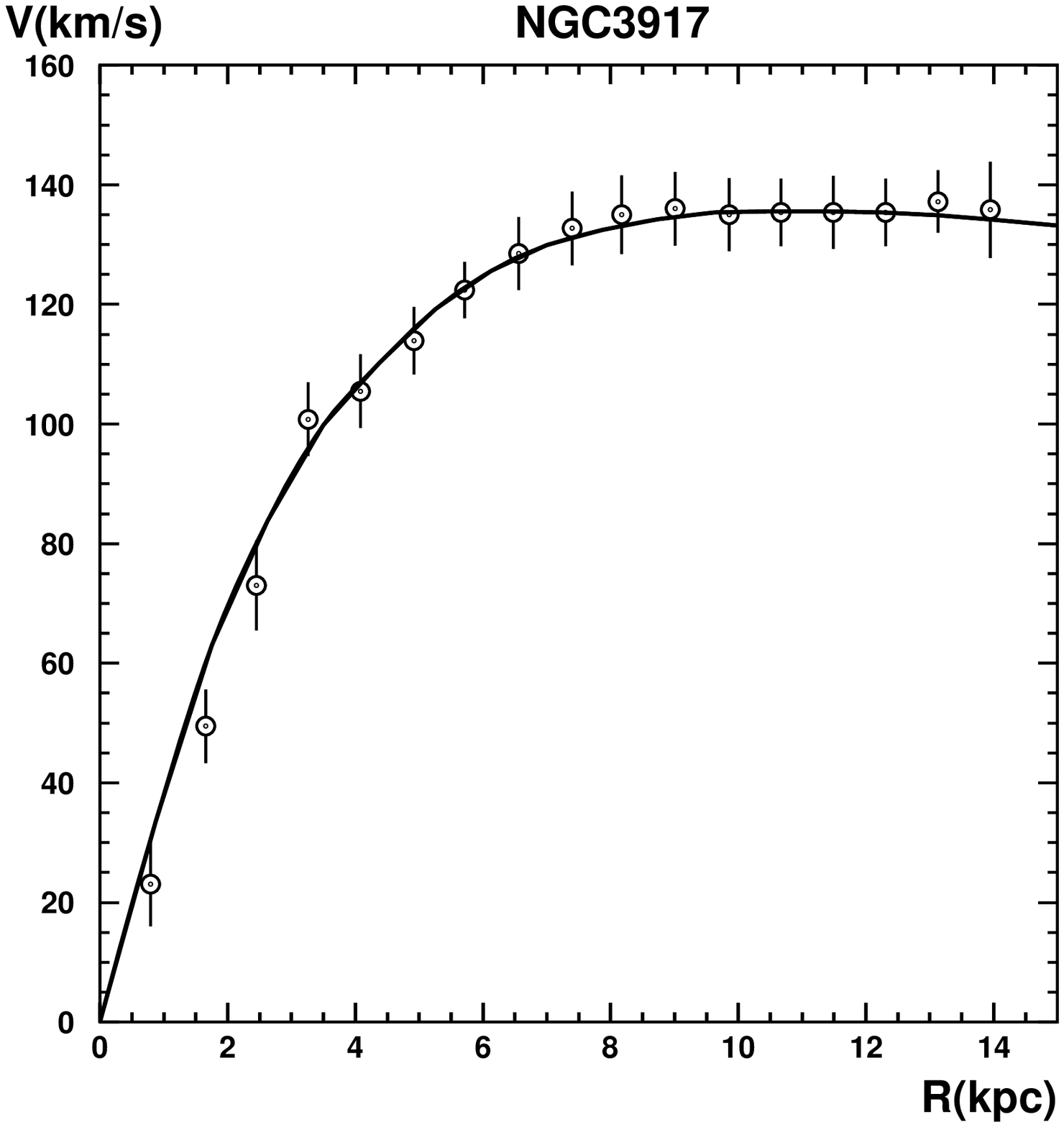}&
\includegraphics[width=60mm]{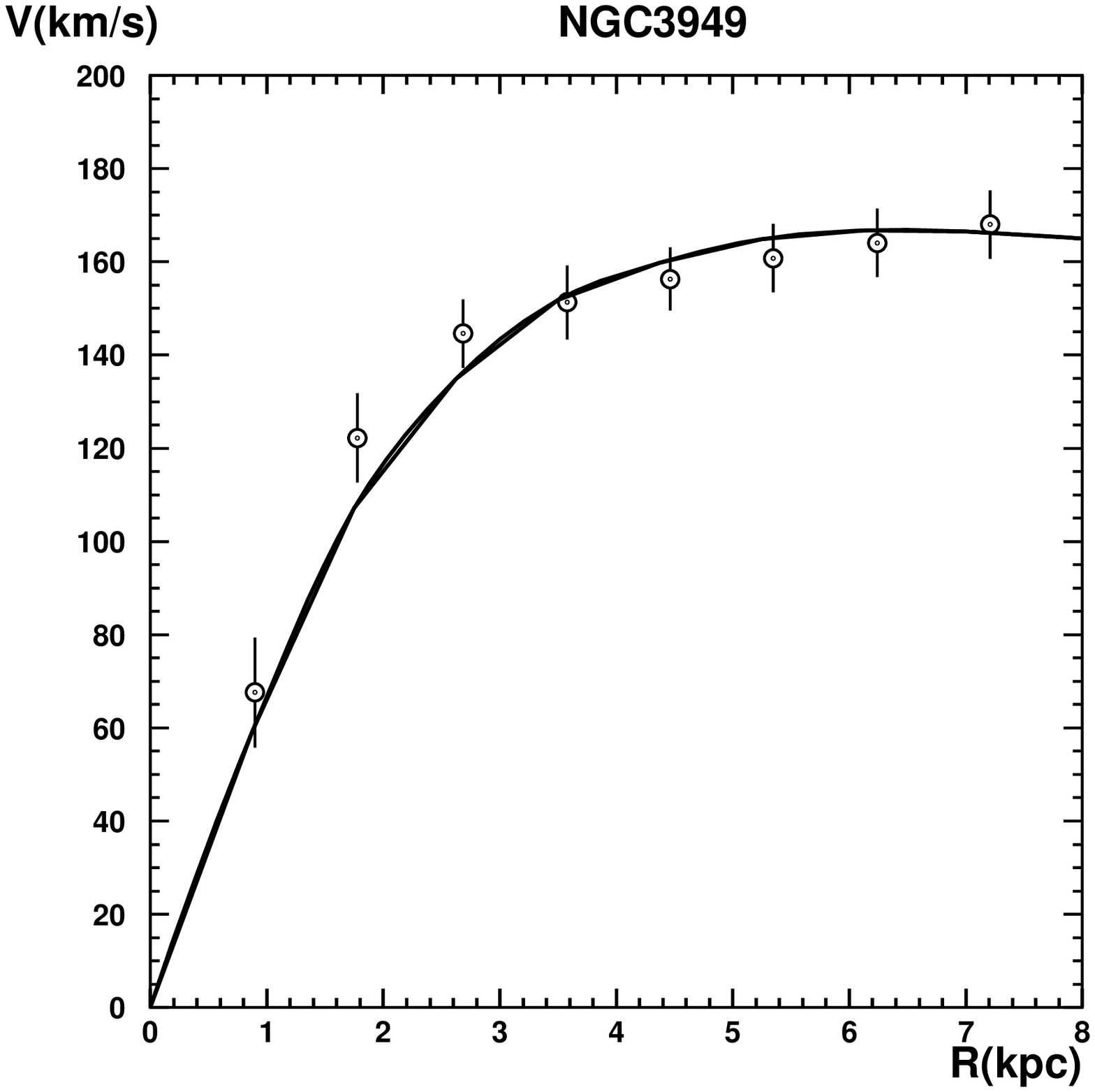}\\
\includegraphics[width=60mm]{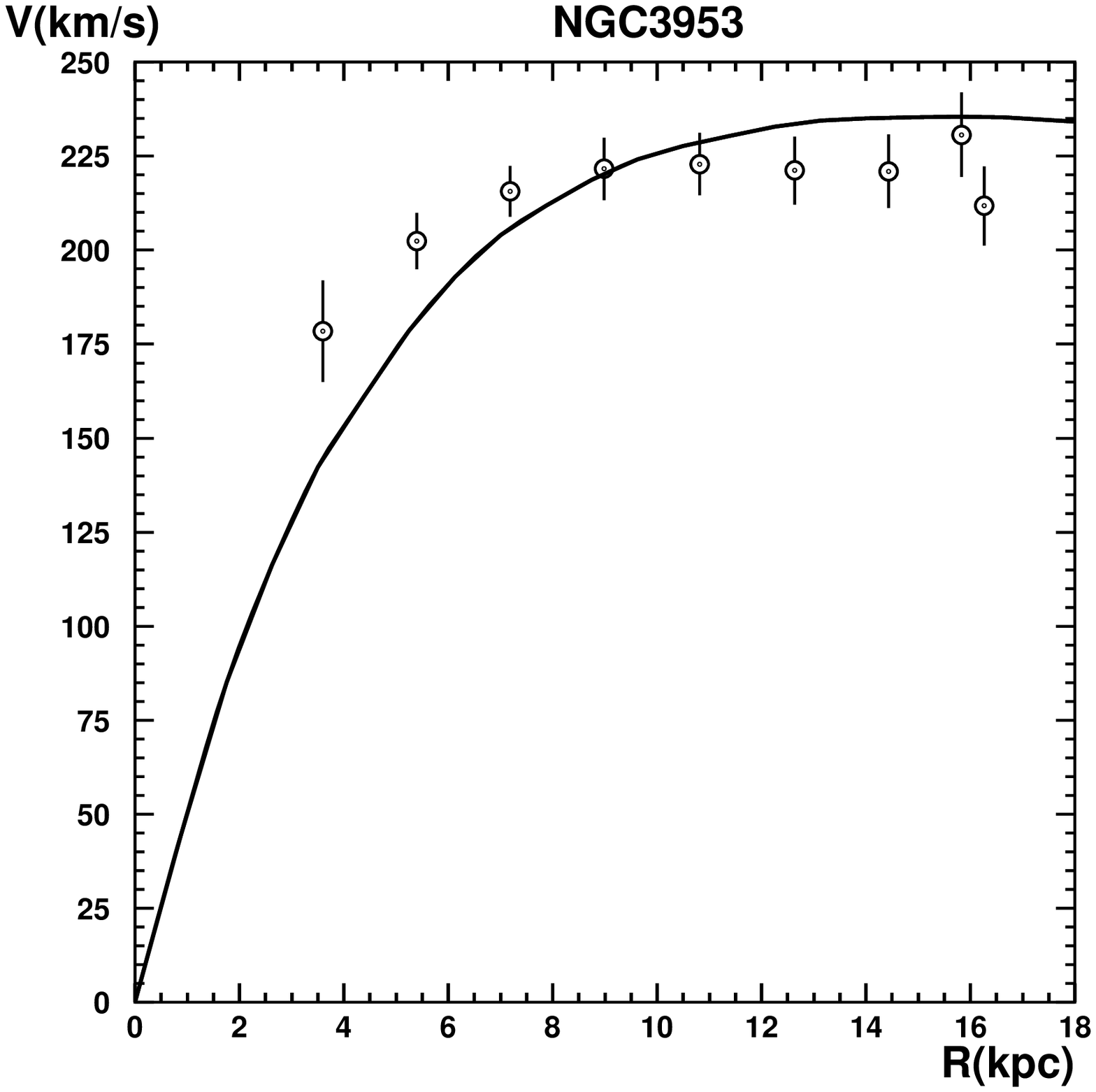}&
\includegraphics[width=60mm]{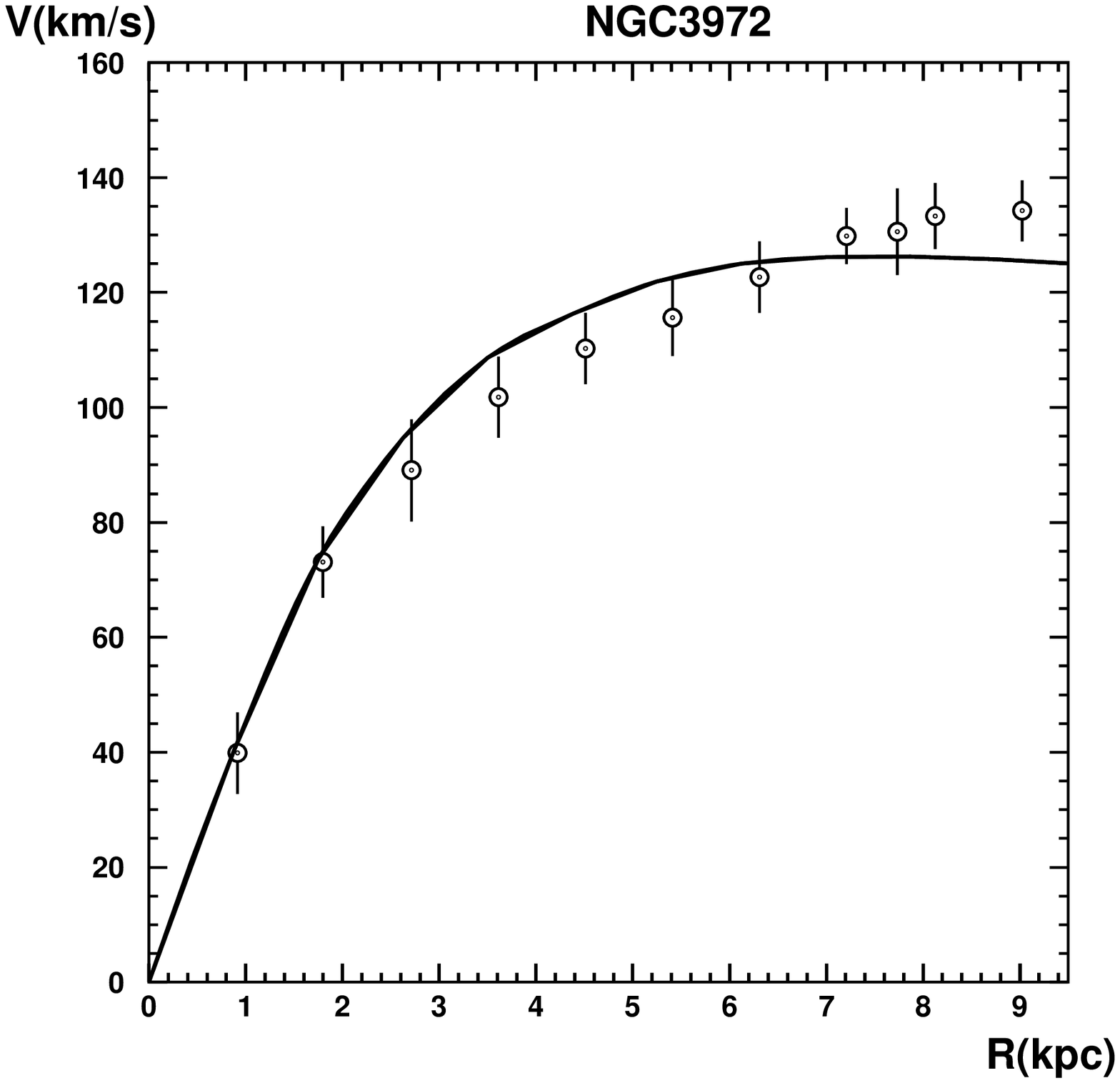}&
\includegraphics[width=60mm]{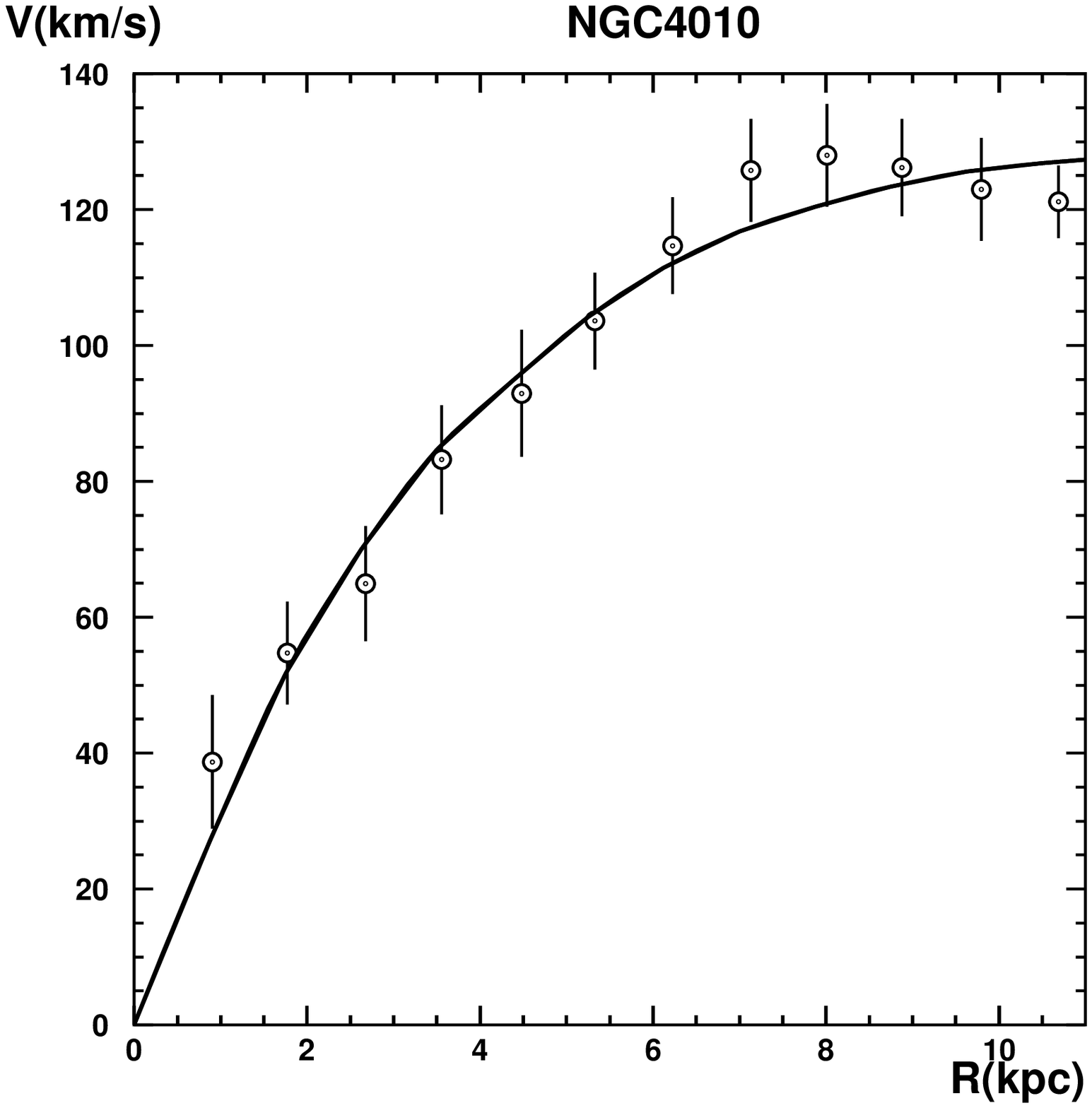}\\
\end{tabular}
\end{center}
\caption {The best fit to the rotation velocity curves of the sample of
Ursa Major galaxies. We fix $\alpha =10.94$ and $\mu = 0.059~{\rm kpc}^{-1}$
from the fits to the THINGS catalog, and we take the stellar
mass-to-light ratio $\Upsilon_\star$ as the free degree of freedom.
\label{fig2} }
\newpage
\end{figure*}
%\clearpage

\setcounter{figure}{1}
\begin{figure*}
\begin{center}
\begin{tabular}{ccc}
\includegraphics[width=60mm]{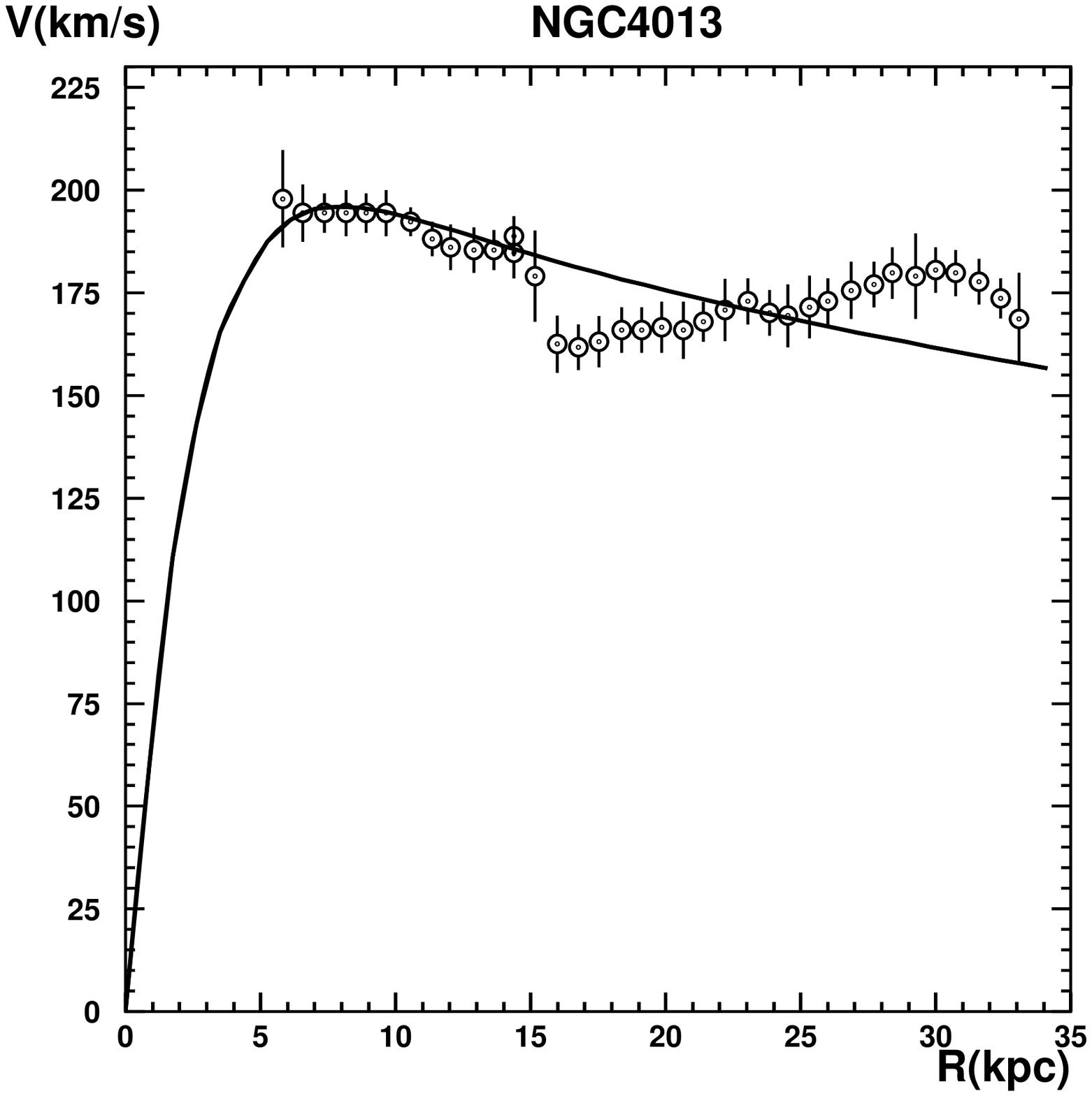} &
\includegraphics[width=60mm]{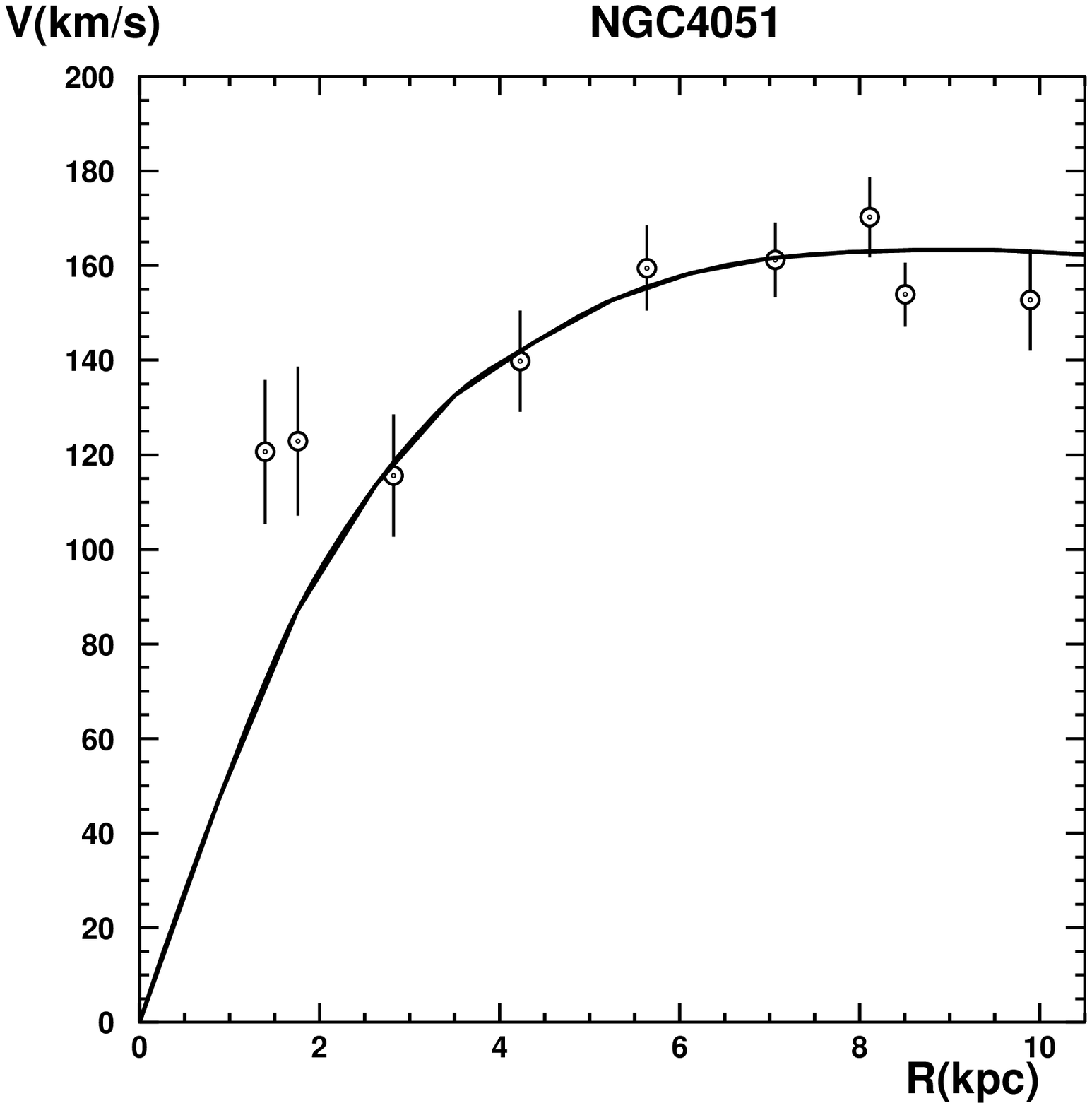}&
\includegraphics[width=60mm]{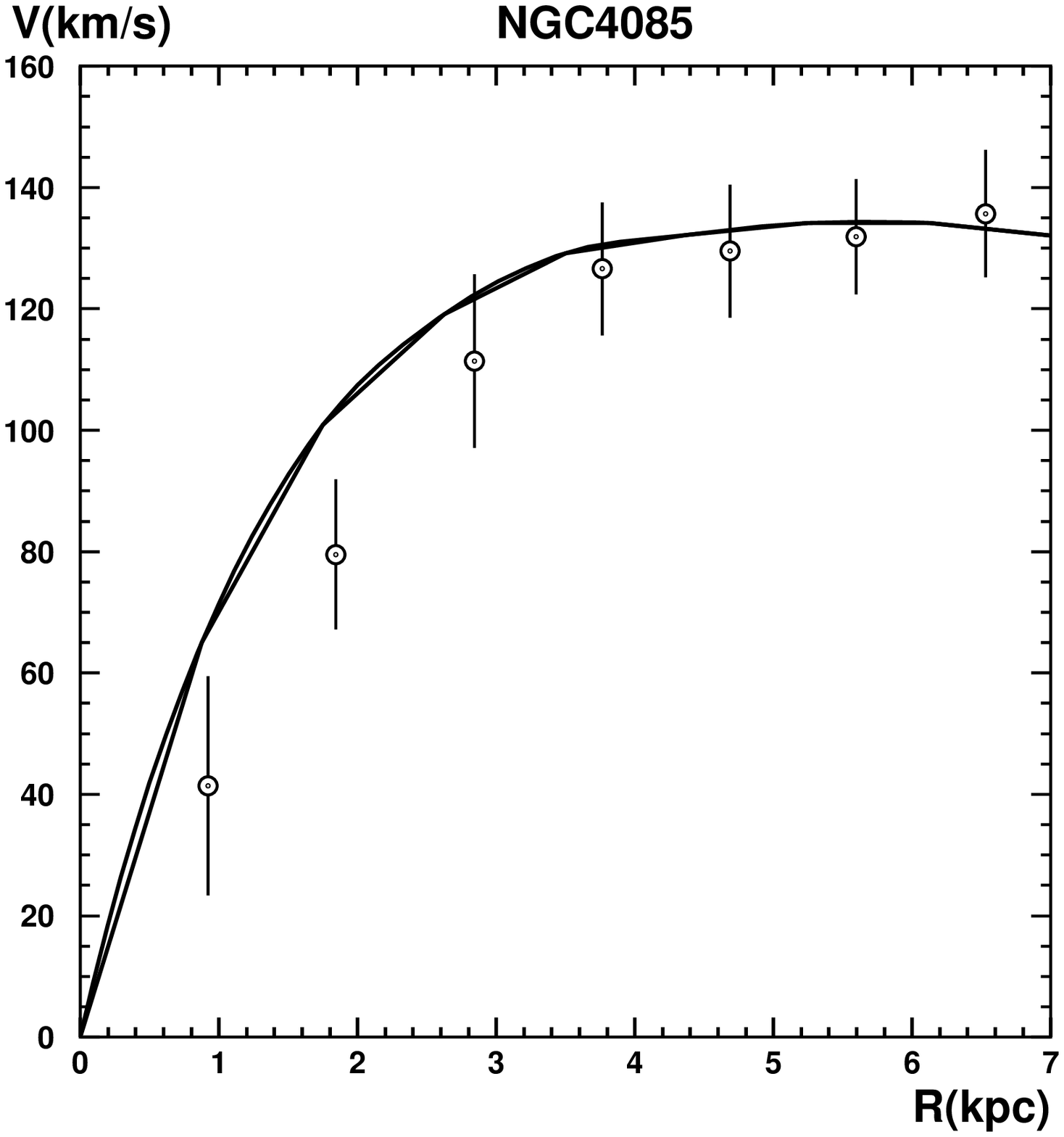} \\
\includegraphics[width=60mm]{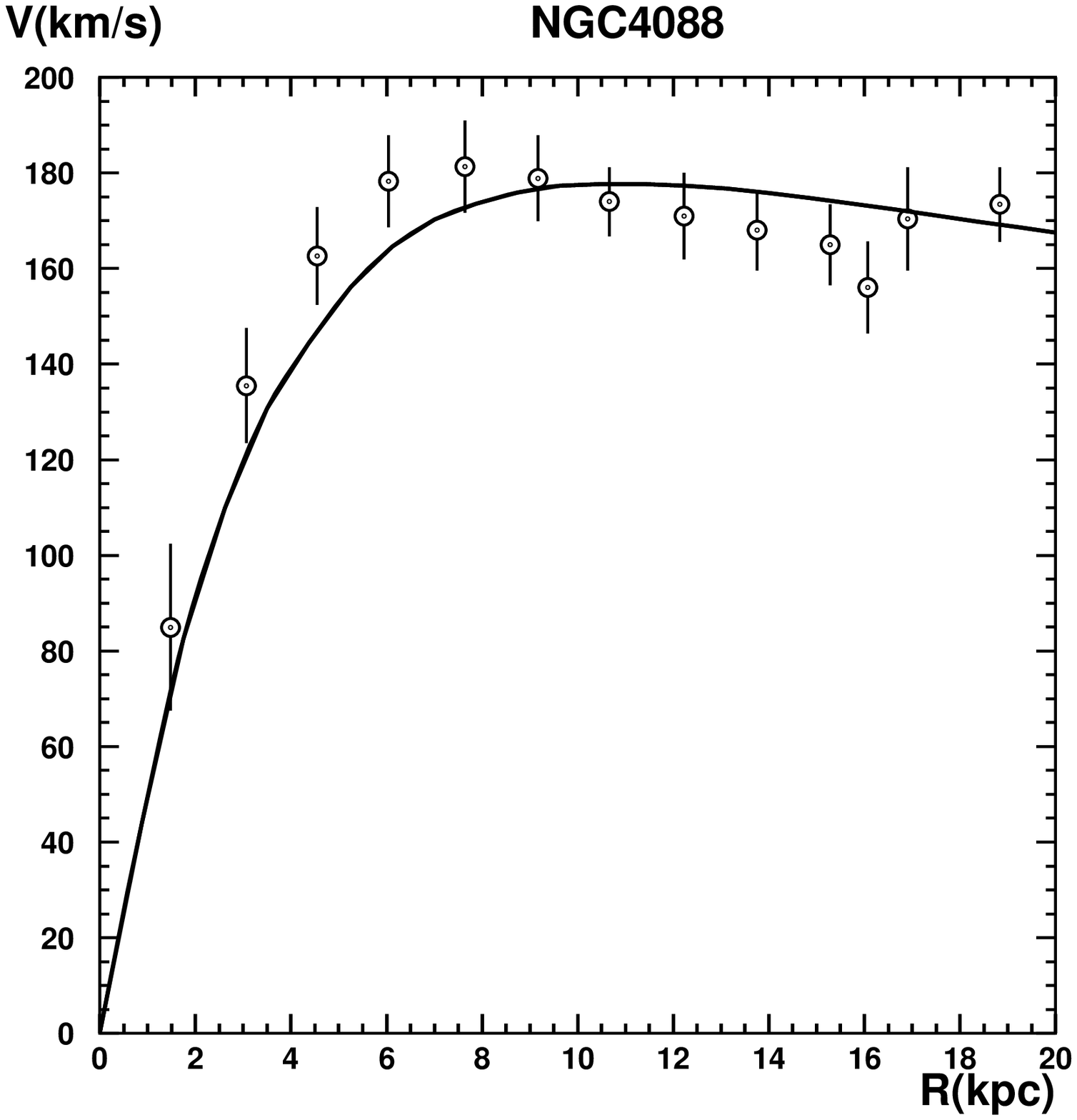} &
\includegraphics[width=60mm]{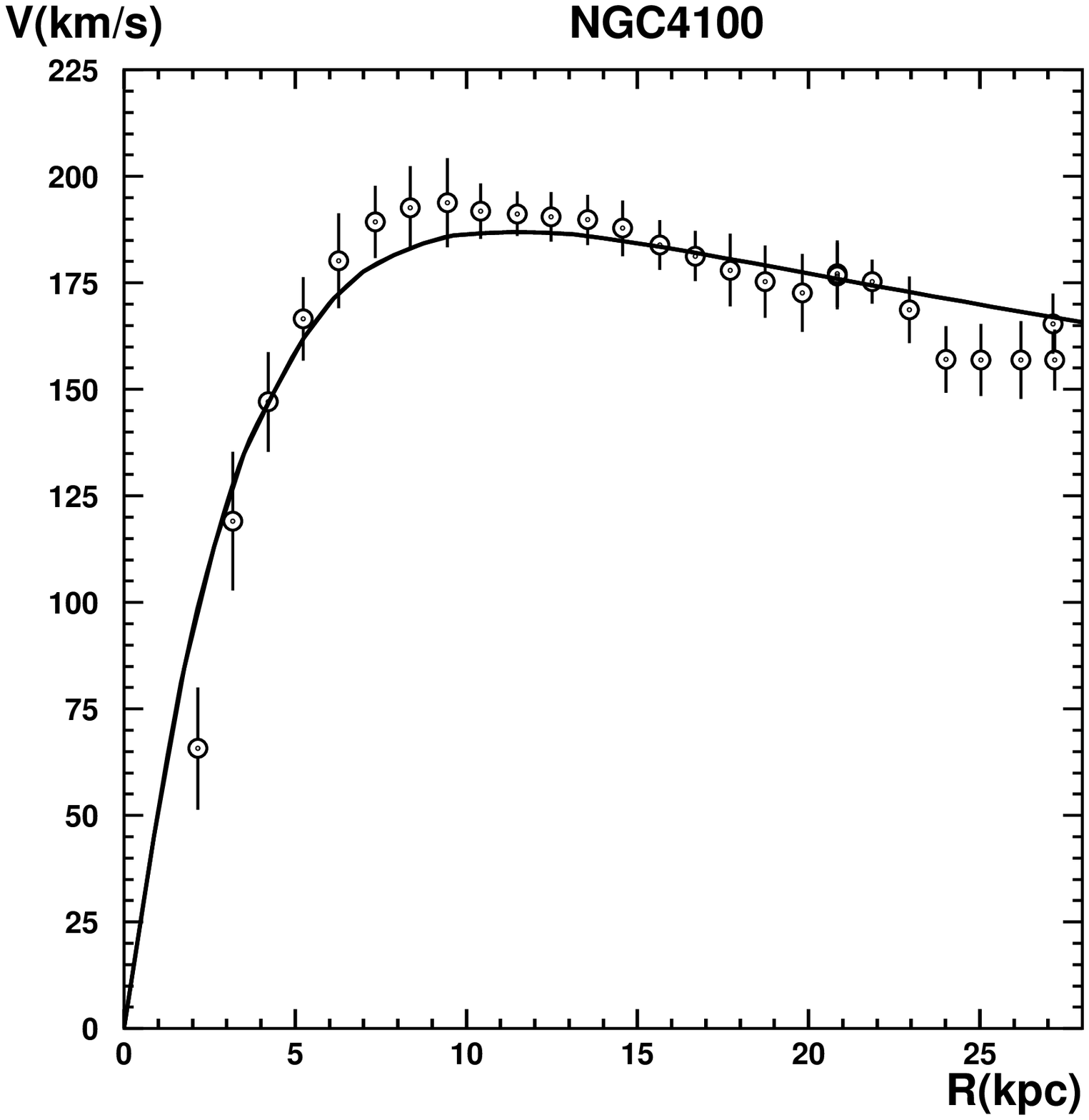}&
\includegraphics[width=60mm]{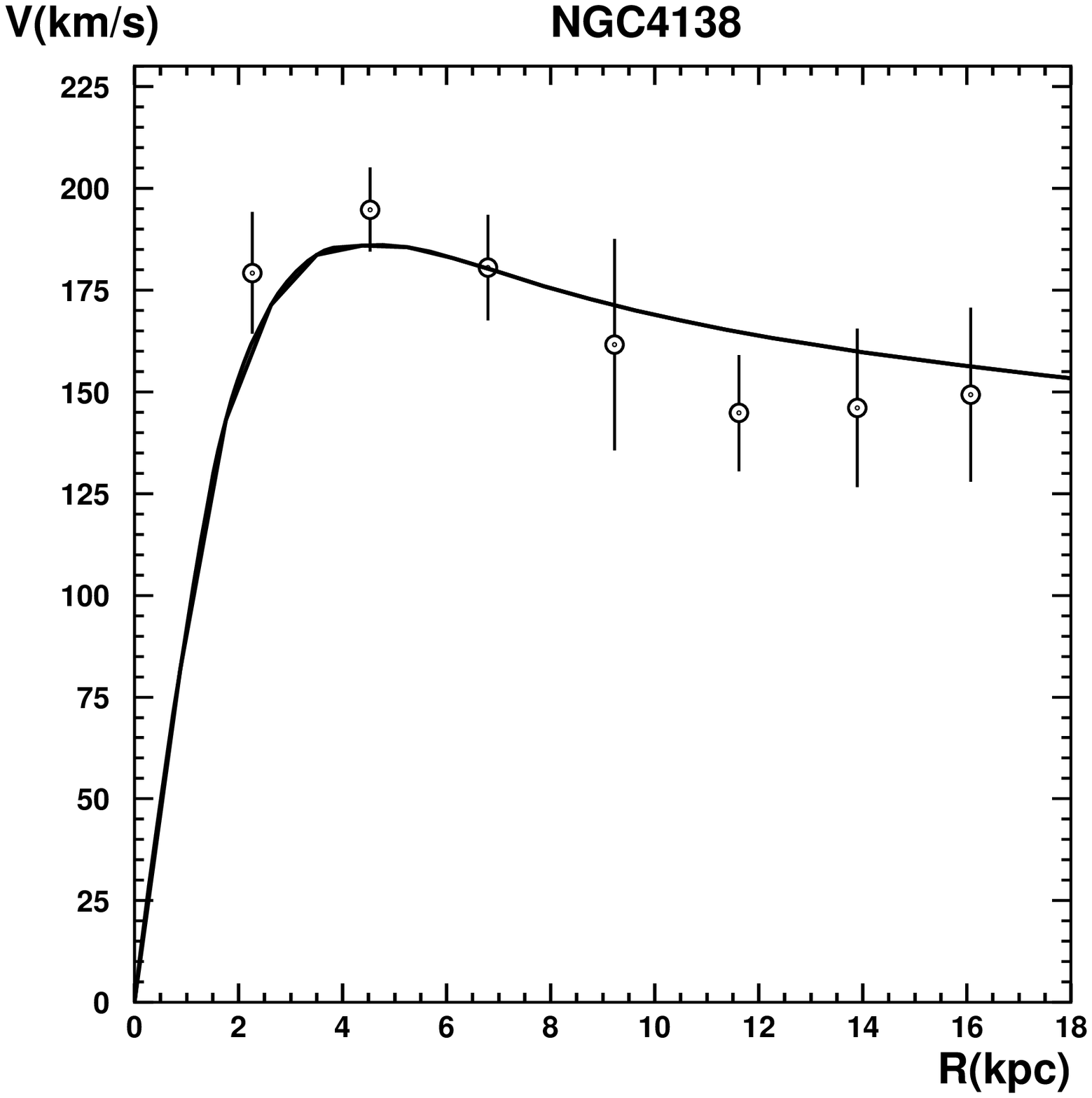}\\
\includegraphics[width=60mm]{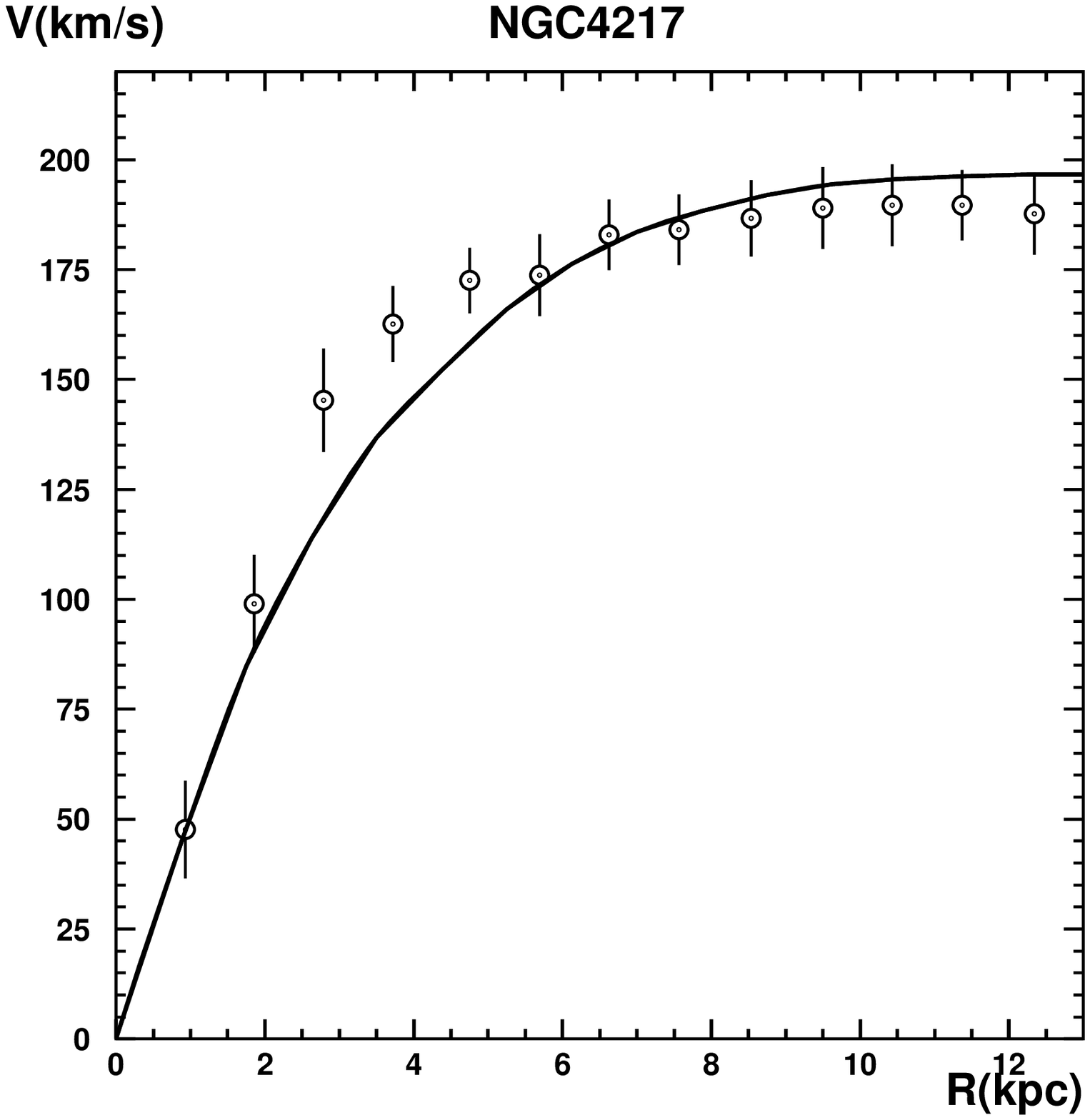}&
\includegraphics[width=60mm]{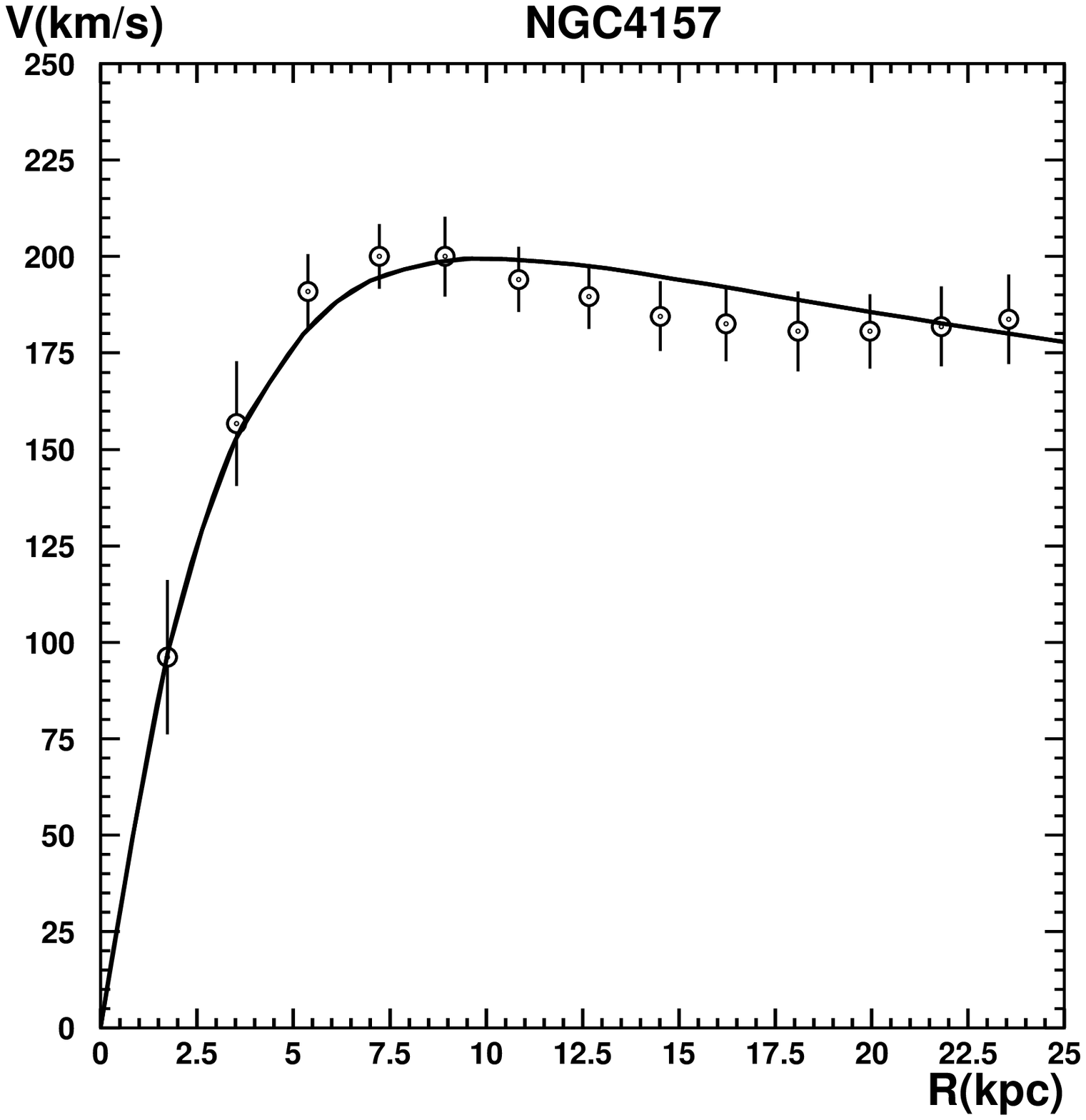}&
\includegraphics[width=60mm]{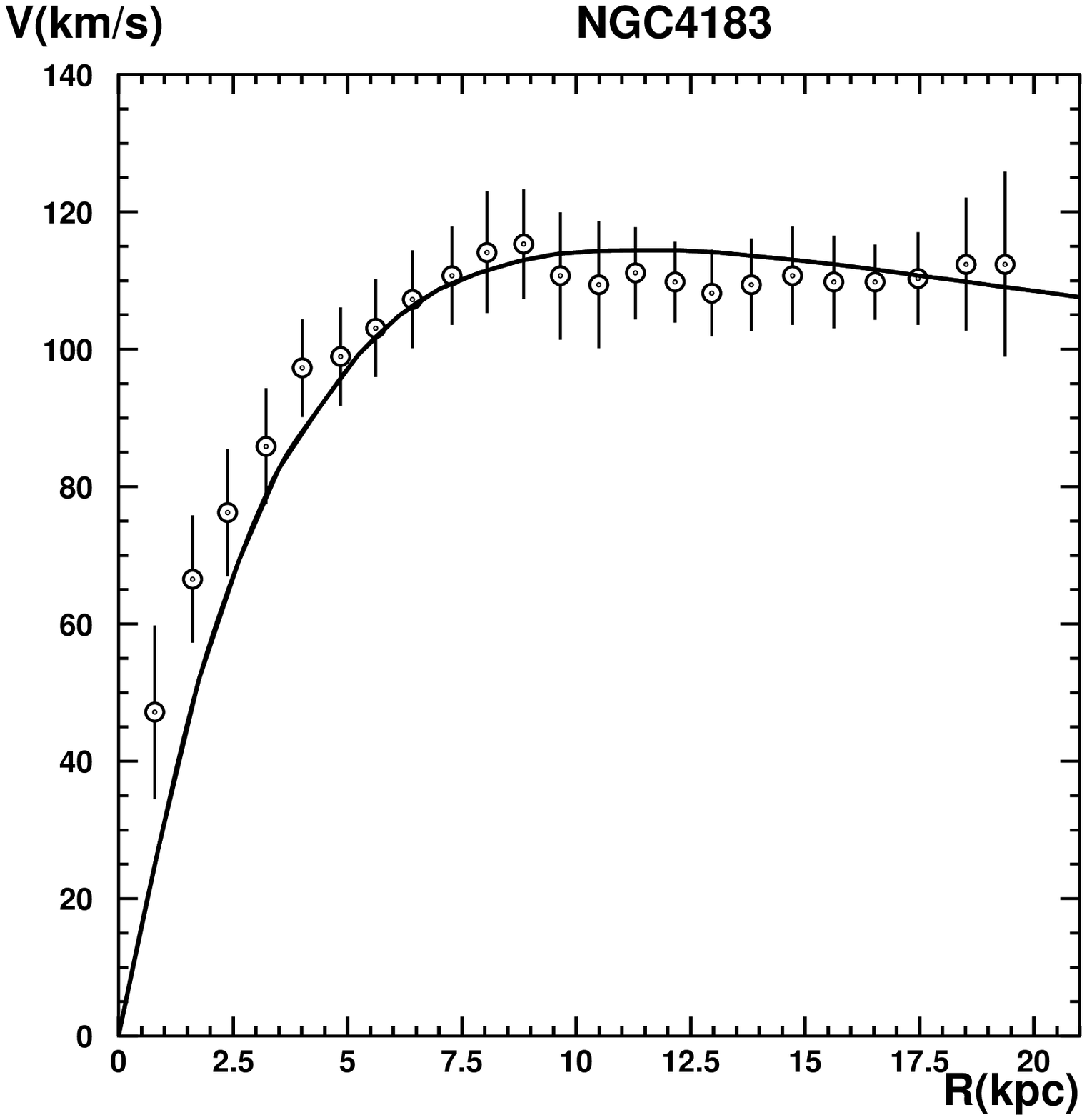}\\
\end{tabular}
\end{center}
\caption {Continued ...}
\newpage
\end{figure*}

\setcounter{figure}{1}
\begin{figure*}
\begin{center}
\begin{tabular}{ccc}
\includegraphics[width=60mm]{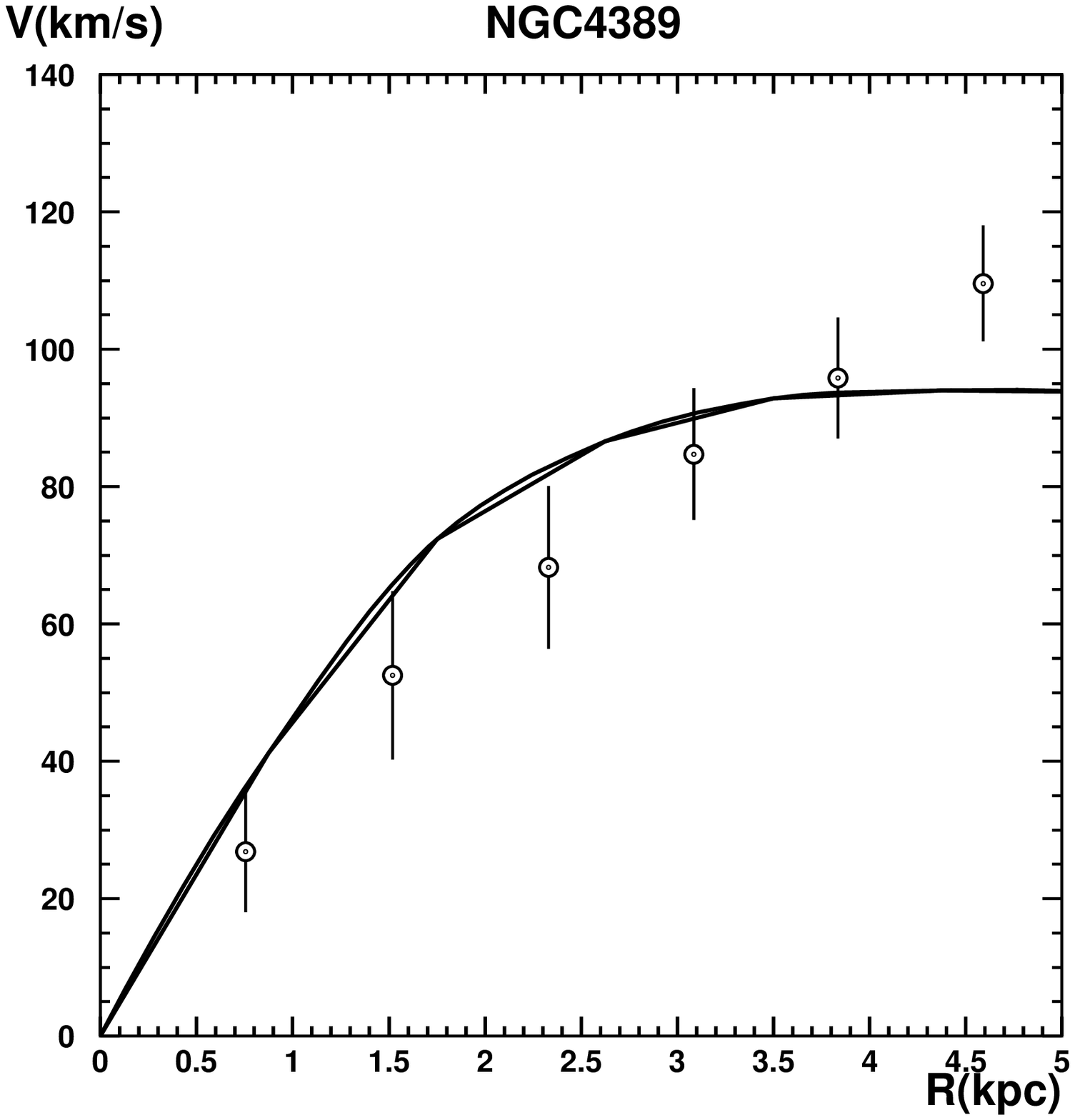} &
\includegraphics[width=60mm]{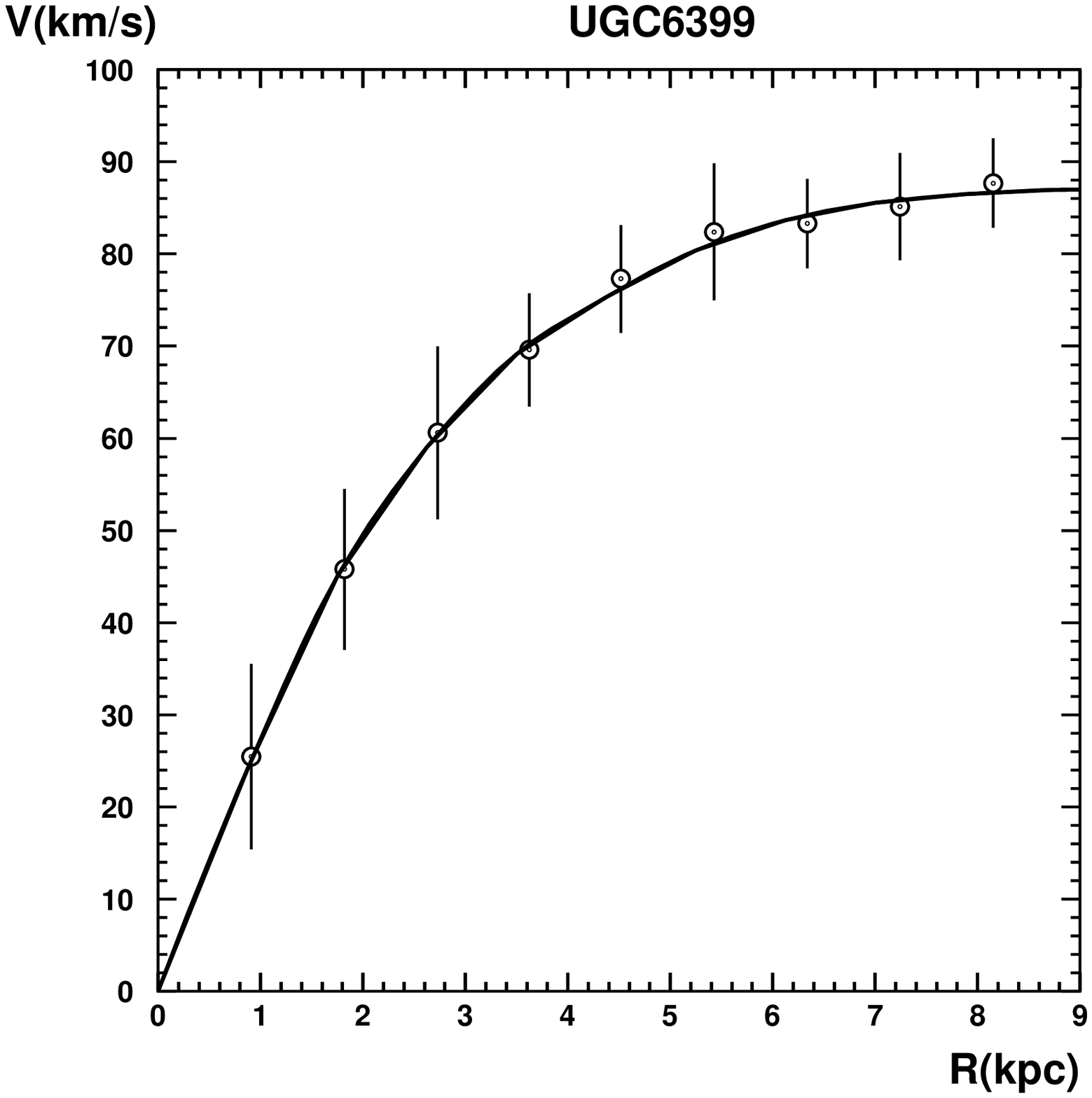}&
\includegraphics[width=60mm]{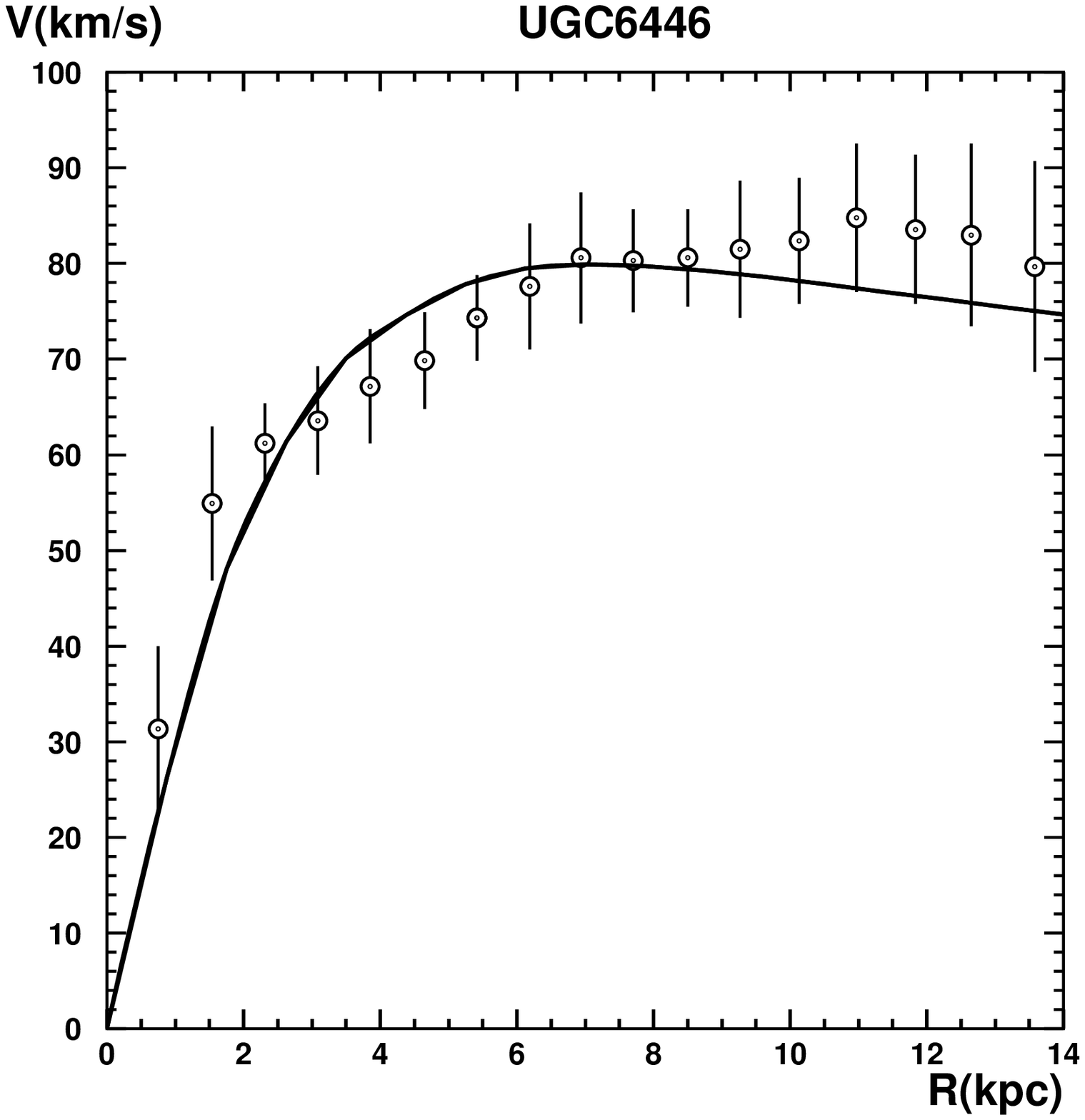} \\
\includegraphics[width=60mm]{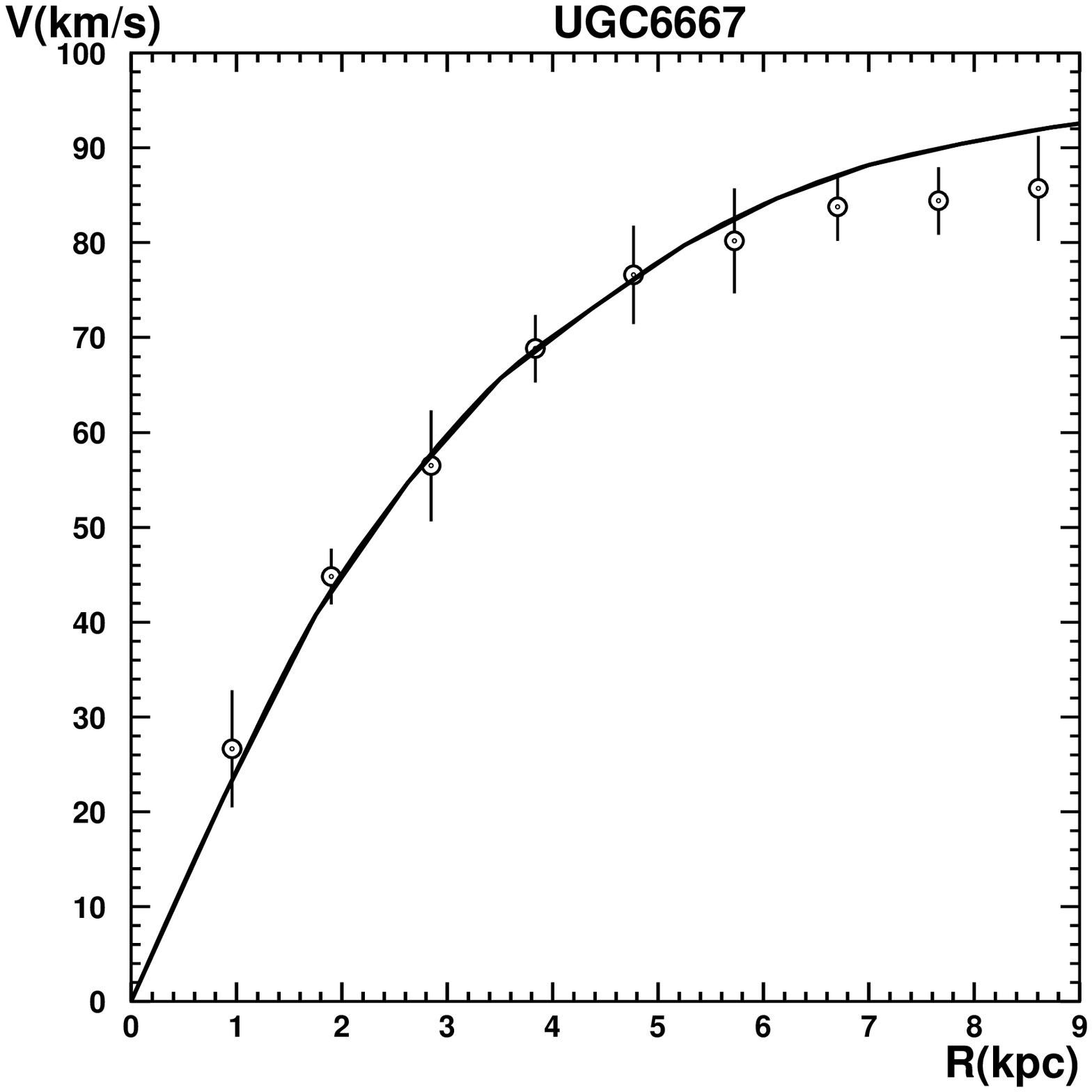} &
\includegraphics[width=60mm]{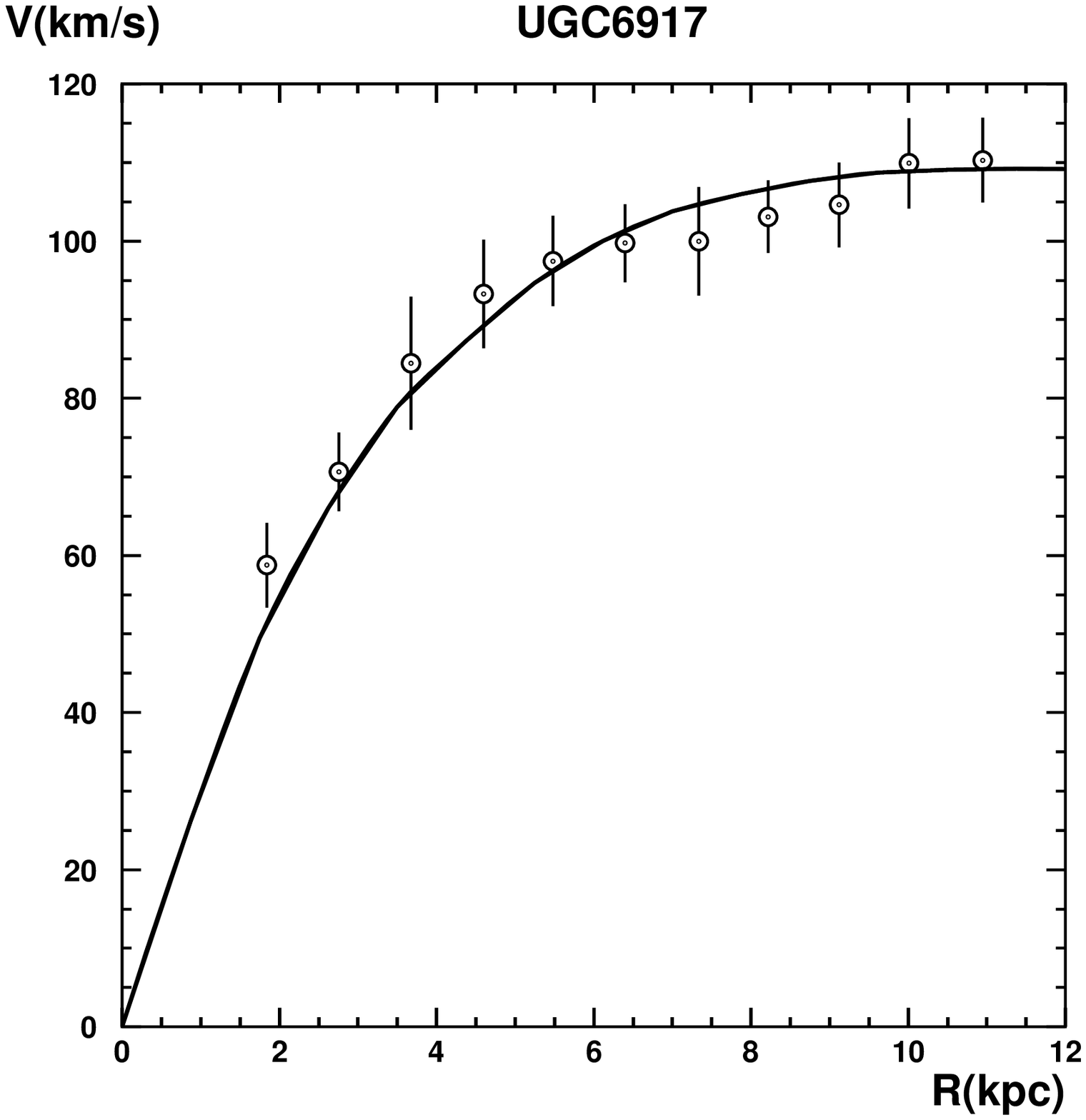}&
\includegraphics[width=60mm]{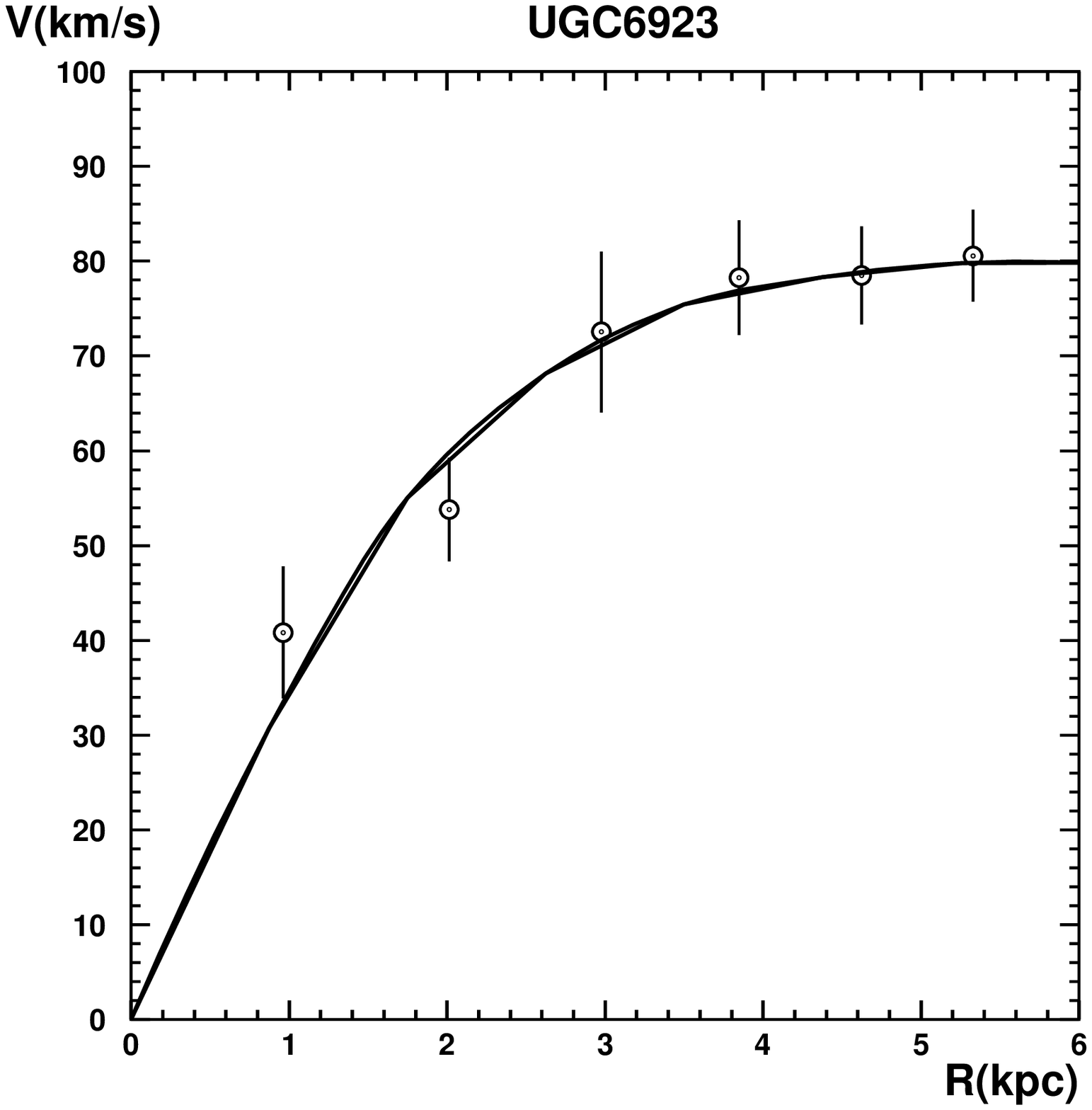}\\
\includegraphics[width=60mm]{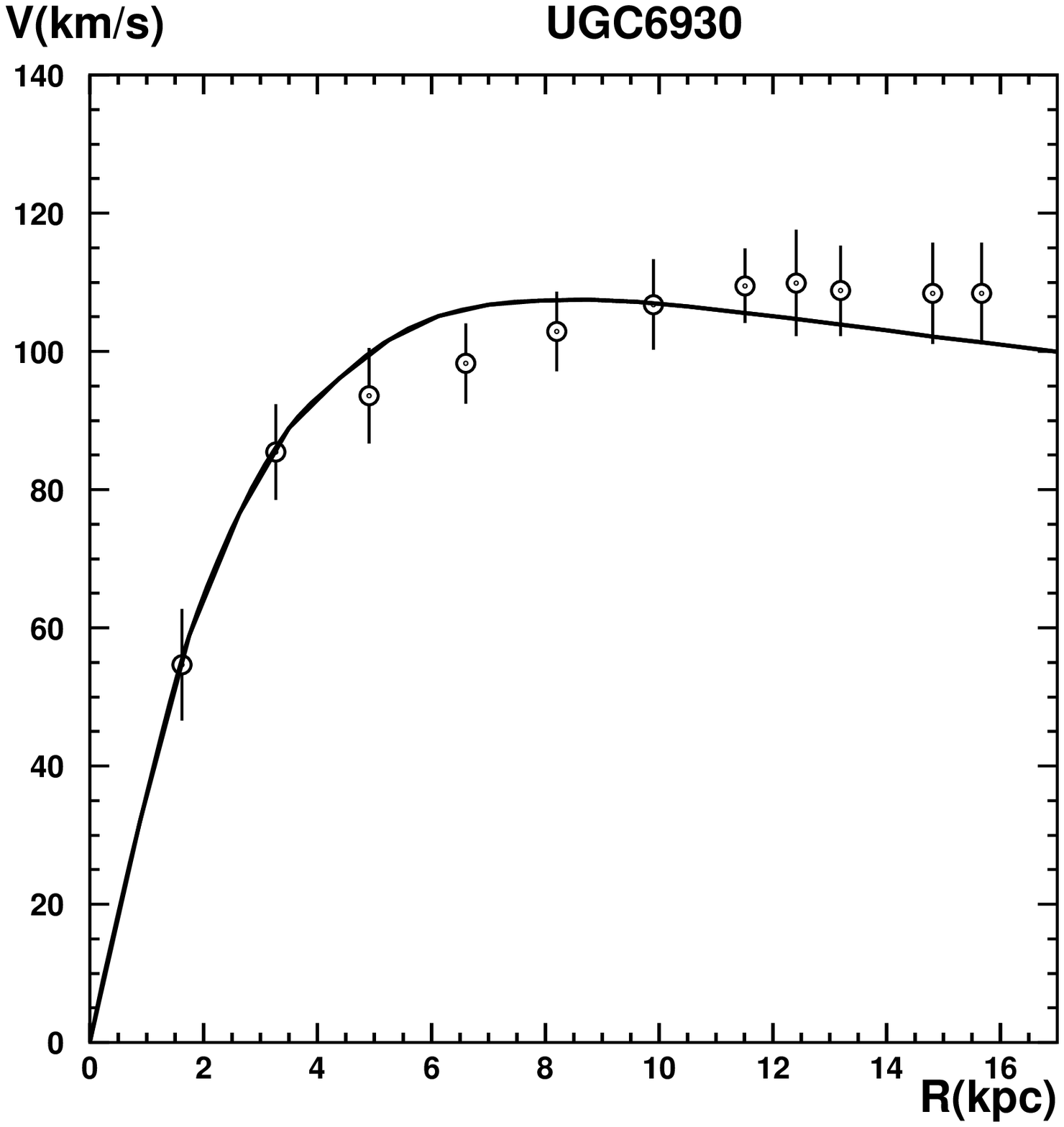}&
\includegraphics[width=60mm]{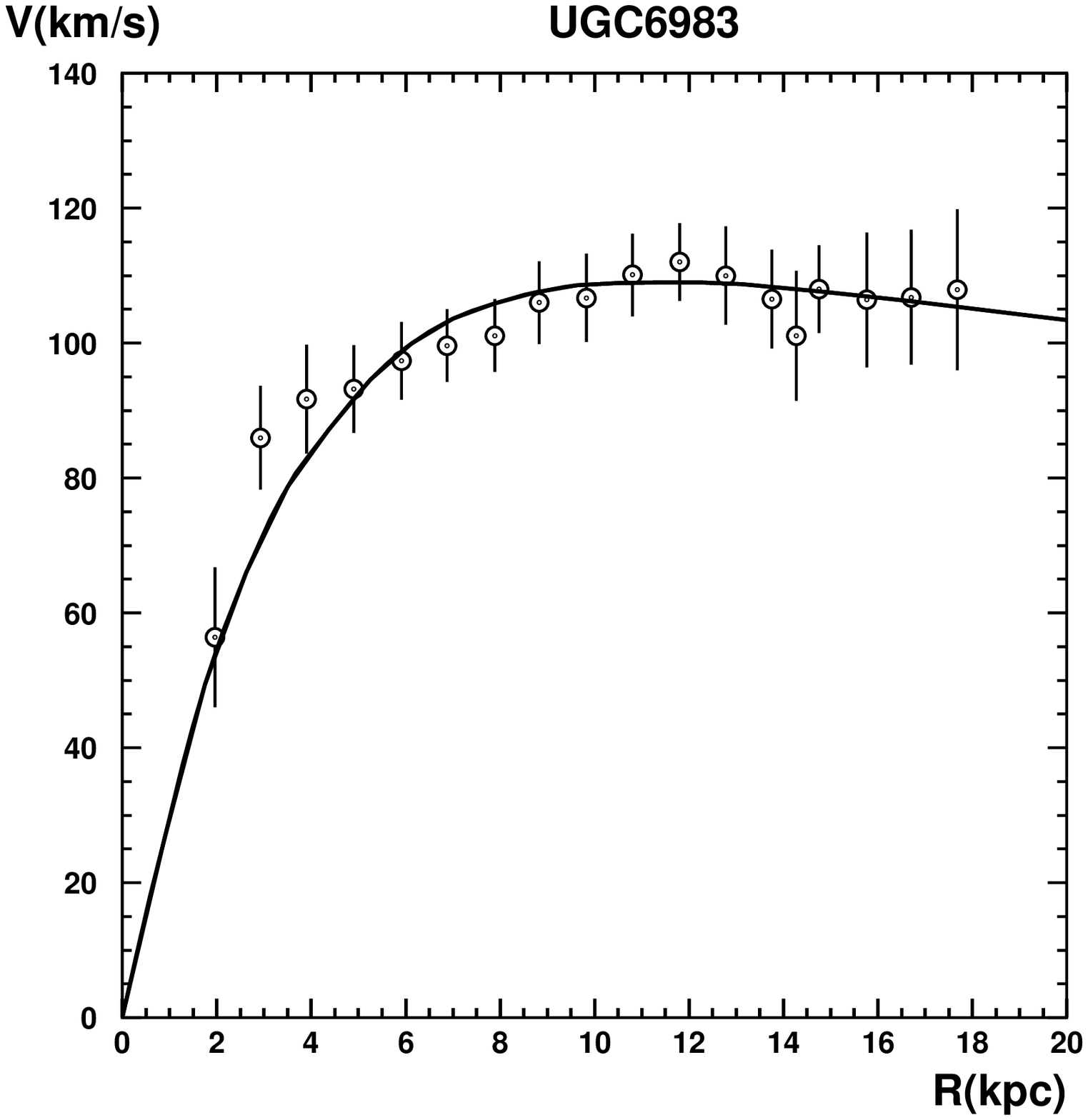}&
\includegraphics[width=60mm]{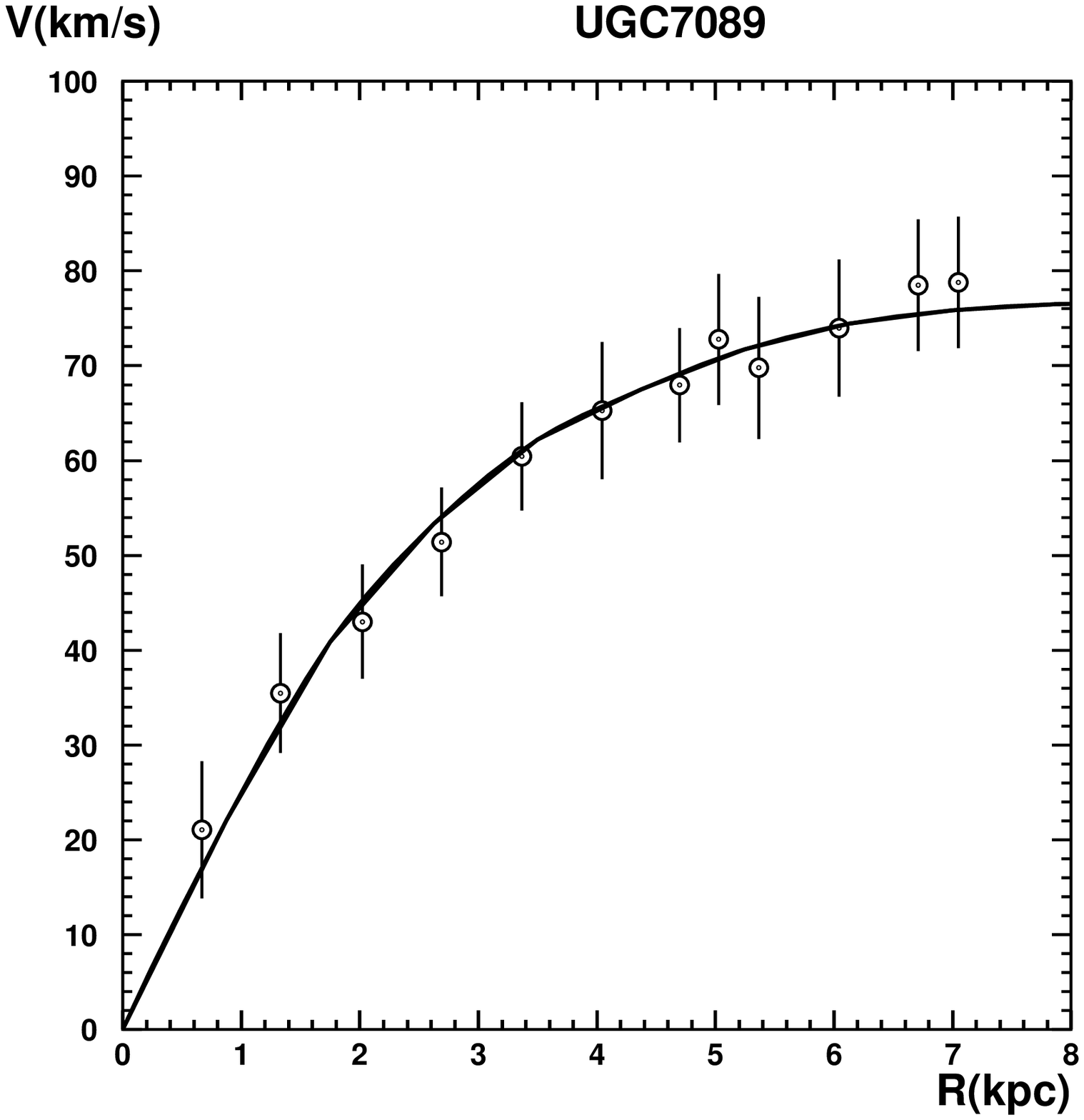}\\
\end{tabular}
\end{center}
\caption {Continued ...}
\newpage
\end{figure*}

\subsection{Ursa Major galaxies}

Next,  we choose twenty-seven galaxies from the Ursa Major cluster for testing nonlocal gravity (NLG) theory. The Ursa Major cluster of galaxies is a spiral-rich member of  the Virgo supercluster and is located at a distance of about 18.6 Mpc. We adopt as universal parameters the best values of $\alpha$ and $\mu$ obtained from fitting THINGS galaxies and displayed in Eq.~\eqref{parameters1+2}, namely, $\alpha = 10.94$ and $\mu = 0.059~{\rm kpc^{-1}}$, and let the stellar mass-to-light ratio $\Upsilon_\star$ of the galaxies be the only free parameter in our analysis. We find the best values of $\Upsilon_\star$ by fitting the observed rotation curves of galaxies with the NLG theory.

\setcounter{table}{1}
\begin{table*}
\begin{center}
%\begin{longtable}{|c|c|c|c|c|c|c|c|c|c|c|}
\caption{ Galaxies from the Ursa Major cluster~\cite{Ursa1,Ursa2,Ursa3} with the best-fitting stellar mass-to-light ratio $\Upsilon_\star$ for each galaxy. The columns contain (1)
the name of the galaxy, (2) the type of galaxy, (3) the distance of the galaxy,
 (4) the extinction-corrected absolute magnitude of the galaxy in the $B$ band, (5) the
characteristic length $\hat R$ of the galaxy, (6) the mass of neutral hydrogen, (7) the
overall mass of the galaxy calculated from $M_{disk} =L_B \Upsilon_{\star}+ \frac43 M_{\rm HI}$, (8) the reddening-corrected
color~\cite{sanders}, (9) the internal extinction of galaxy in the $B$
band, (10) the best-fit value for the stellar mass-to-light ratio
$\Upsilon_{\star}$, normalized to the solar value, and (11) the
reduced $\chi^2$ for the best fit to the data. \label{tab2}}
\begin{tabular}{|c|c|c|c|c|c|c|c|c|c|c|}
\hline\hline
       &      & Distance & $M_B$          & ${\hat R}$ & $M_{\rm HI}$          & $M_{disk}$     &$B-V$        &  $A_B$          &$\Upsilon_{\star}$     &   $\chi^2$    \\
Galaxy & Type & (Mpc)    & (mag) &(kpc)  & $(10^{10}M_\odot)$& $(10^{10}M_\odot)$  &(mag)       &  (mag)        & $(M_\odot/L_\odot)$  &   $(N_{\rm d.o.f.})$   \\
 (1)   & (2)  & (3)      & (4)            & (5)                       & (6)                &(7)          &(8)            & (9)                & (10)     & (11) \\
\hline
NGC 3726 &HSB & 17.4     & -20.76          &3.2    &0.60               &   1.71            &$0.45$      & 0.06          &$0.272^{+0.020}_{-0.020}$        & 1.70       \\
NGC 3769 &HSB & 15.5     & -19.32         & 1.5   &0.41               &   0.82              &  0.64      & 0.084           &$0.390^{+0.16}_{-0.06}$        & 0.93      \\
NGC 3877 &HSB &15.5      & -20.60         &2.4    &0.11               &   1.80             &$0.68$      & 0.084          &$0.85^{+0.05}_{-0.05}$        & 0.42       \\
NGC 3893 &HSB &18.1      & -20.55          &2.4    & 0.59              &   2.22              & 0.56       & 0.077           & $0.49^{+0.04}_{-0.04}$       & 2.94       \\
NGC 3917 &LSB &16.9      &\dotfill           &2.8    &0.17               &   1.13           &$0.60$      & 0.077           &$0.68^{+0.03}_{-0.03}$        &  0.37      \\
NGC 3949 &HSB &18.4      &-20.22           &1.7    & 0.35              &   1.44           &$0.39$      & 0.078          &$0.42^{+0.04}_{-0.03}$        & 0.87       \\
NGC 3953 &HSB &18.7     &-21.05           &3.9    &0.31               &   3.80             &$0.71$      & 0.109          &$0.80^{+0.03}_{-0.03}$        & 3.09        \\
NGC 3972 &HSB &18.6     &\dotfill            &2.0   &0.13                &   0.87              &$0.55$      & 0.051          &$0.72^{+0.04}_{-0.04}$        & 0.99       \\
NGC 4010 &LSB &18.4     &-20.51            &3.4  &0.29                 &   1.09            &  --        & 0.088          &$0.79^{+0.07}_{-0.07} $        &     0.48    \\
NGC 4013 &HSB &18.6     &-20.08            &2.1  &0.32                 &   2.13              & $0.83$     & 0.060          &$0.82^{+0.03}_{-0.01}$       & 2.94        \\
NGC 4051 &HSB &14.6     &-20.71            &2.3  &0.18                 &   1.42             &$0.62$      & 0.047          &$0.52^{+0.04}_{-0.05}$        & 2.12        \\
NGC 4085 &HSB &19.0     &-19.12            &1.6  &0.15                 &  0.87             & $0.47$     & 0.066          &$0.56^{+0.08}_{-0.08}$       & 0.24        \\
NGC 4088 &HSB &15.8     &-20.95            &2.8  &0.64                 &   1.94            & $0.51$     & 0.071          &$0.37^{+0.03}_{-0.03}$       & 1.16        \\
NGC 4100 &HSB &21.4     &-20.51            &2.9  &0.44                 &   2.18             &$0.63$      & 0.084          &$0.47^{+0.02}_{-0.02}$        & 0.82        \\
NGC 4138 &LSB &15.6     &-19.50            &1.2  &0.11                 &   1.59           &$0.81$      & 0.051          &$1.75^{+0.18}_{-0.18}$        & 0.74        \\
NGC 4217 &HSB &19.6     &-20.33            &3.1  &0.30                 &   2.46             &$0.77$      & 0.063          &$0.68^{+0.03}_{-0.03}$        & 1.60        \\
NGC 4157 &HSB &18.7     &-20.72            &2.6  &0.88                 &   2.39             &$0.66$      & 0.077          &$0.42^{+0.03}_{-0.03}$        & 0.65        \\
NGC 4183 &HSB &16.7     &-19.46            &2.9  &0.30                 &   0.82              &$0.39$      & 0.055           &$0.40^{+0.04}_{-0.04}$        & 0.64       \\
\hline
%\end{longtable}
\end{tabular}
\end{center}
\end{table*}

\setcounter{table}{1}
\begin{table*}
\begin{center}
\caption{Continued ...}
\begin{tabular}{|c|c|c|c|c|c|c|c|c|c|c|}
\hline\hline
       &      & Distance & $M_B$          & ${\hat R}$ & $M_{\rm HI}$          & $M_{disk}$         &$B - V$        &  $A_B$          &$\Upsilon_{\star}$     &   $\chi^2$    \\
Galaxy & Type & (Mpc)    & (mag) &(kpc)  & $(10^{10}M_\odot)$& $(10^{10}M_\odot)$ & (mag)       &  (mag)        & $(M_\odot/L_\odot)$  &   $(N_{\rm d.o.f.})$   \\
 (1)   & (2)  & (3)      & (4)            & (5)                       & (6)                &(7)          &(8)            & (9)                & (10)     & (11) \\
\hline
NGC 4389 &HSB &15.5     &\dotfill            &1.2  &0.04                 &   0.40              & --        & 0.053          &$0.58^{+0.08}_{-0.08}$         &      1.54   \\
UGC 6399 &LSB &18.7     &-17.56            &2.4  &0.07                 &   0.52             &--         & 0.061          &$1.49^{+0.12}_{-0.12}$         & 0.02        \\
UGC 6446 &LSB &15.9     &-18.08            &1.9  &0.24                 &   0.34             & 0.39      & 0.059             &$0.095^{+0.005}_{-0.075}$        & 0.61       \\
UGC 6667 &LSB &19.8     &-17.83            &3.1  &0.10                 &  0.53              & 0.65      & 0.058           &$0.94^{+0.09}_{-0.09}$         & 0.42        \\
UGC 6917 &LSB &18.9     &-18.63            &2.9  &0.22                 &   0.74              & 0.53      & 0.098              & $0.80^{+0.05}_{-0.07}$        & 0.48        \\
UGC 6923 &LSB &18.0     &\dotfill            &1.5  &0.08                 &  0.32              & $0.42$    & 0.096          &$0.71^{+0.10}_{-0.10}$      & 0.42       \\
UGC 6930 &LSB &17.0     &\dotfill            &2.2  &0.29                 &  0.66               & 0.59      & 0.108            & $0.45^{+0.06}_{-0.06}$         & 0.68        \\
UGC 6983 &LSB &20.2     &-18.58            &2.9  &0.37                 &   0.73               & 0.45      & 0.096              &$0.43^{+0.06}_{-0.06}$        & 0.44        \\
UGC 7089 &LSB &13.9     &\dotfill            &2.3  &0.07                 &   0.34                  & --     & 0.055              &$0.70^{+0.09}_{-0.09}$        & 0.12        \\
\hline
%\end{longtable}
\end{tabular}
\end{center}
\end{table*}

The twenty-seven Ursa Major galaxies that we consider here are listed in Table~\ref{tab2} with the best-fitting $\Upsilon_\star$ and the corresponding $\chi^2$ per degree of freedom for each galaxy. To fit the data, we use the  disk parameters  $\hat R$, $M_B$ and $M_{\rm H}$ from the observation of spiral galaxies~\cite{Ursa1,Ursa2,Ursa3},  given in Table~\ref{tab2}, and calculate the
rotation curves of these galaxies. Here, $M_B$ is the extinction-corrected absolute magnitude of the galaxy in the $B$ band. From the best value for the stellar mass to light ratio $\Upsilon_\star$, we can determine the overall mass of  the disk from
$M_{disk} = L_B\Upsilon_\star + \frac43 M_{\rm H}$, where $L_B$ is the intrinsic luminosity of the galaxy in the $B$ band. For the galaxies in this list, the average value of reduced $\chi^2$ for all the galaxies is $\overline{\chi^2} =1.015$. Figure~\ref{fig2} represents the observational data with the best fits to the rotation curves of the chosen galaxies.

Next, we must show that the approach that we have adopted in this section is consistent with the astrophysics of star formation as well as the Tully-Fisher relation. These issues will now be addressed in turn.

\subsection{Stellar mass-to-light ratio versus color for galaxies}

The stellar mass-to-light ratio is correlated with  the color of galaxies according to the theories of star formation~\cite{bell2001,bell2003}. Such a relationship is expected to  depend upon the initial mass function (IMF) as well as the way stars are actually formed. Uncertainties exist, however, due to the choice of the IMF and  the Stellar Population Synthesis (SPS) models. Furthermore,  galaxies may appear redder and fainter than they actually are due to the presence of dust in their interstellar media.

Assuming the Salpeter mass function~\cite{salpeter}, for instance, the relation between
the mass-to-light ratio $\Upsilon_\star^B$ in the $B$ band and color for
galaxies is given by~\cite{bell2003}
\begin{equation}
\log_{10}(\Upsilon_\star^B) = 1.74\,(B-V) - 0.94\,.  \label{emlc}
\end{equation}
In the case of Kroupa's IMF, the slope of Eq.~\eqref{emlc} does not change, but the mass-to-light ratio shifts by $-0.35$ dex.

\begin{figure*}
\begin{center}
\begin{tabular}{cc}
\includegraphics[width=80mm]{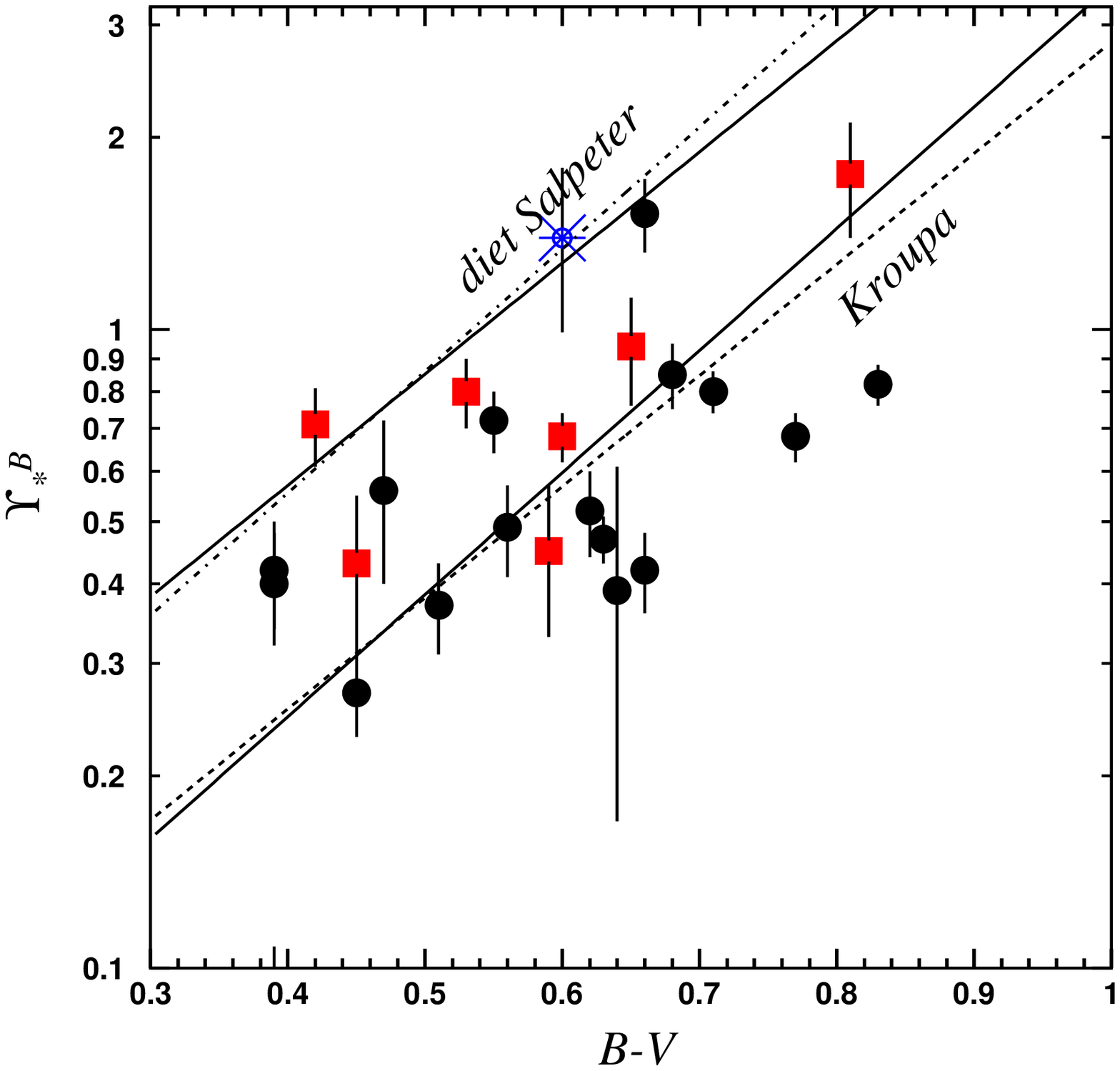} &
\includegraphics[width=80mm]{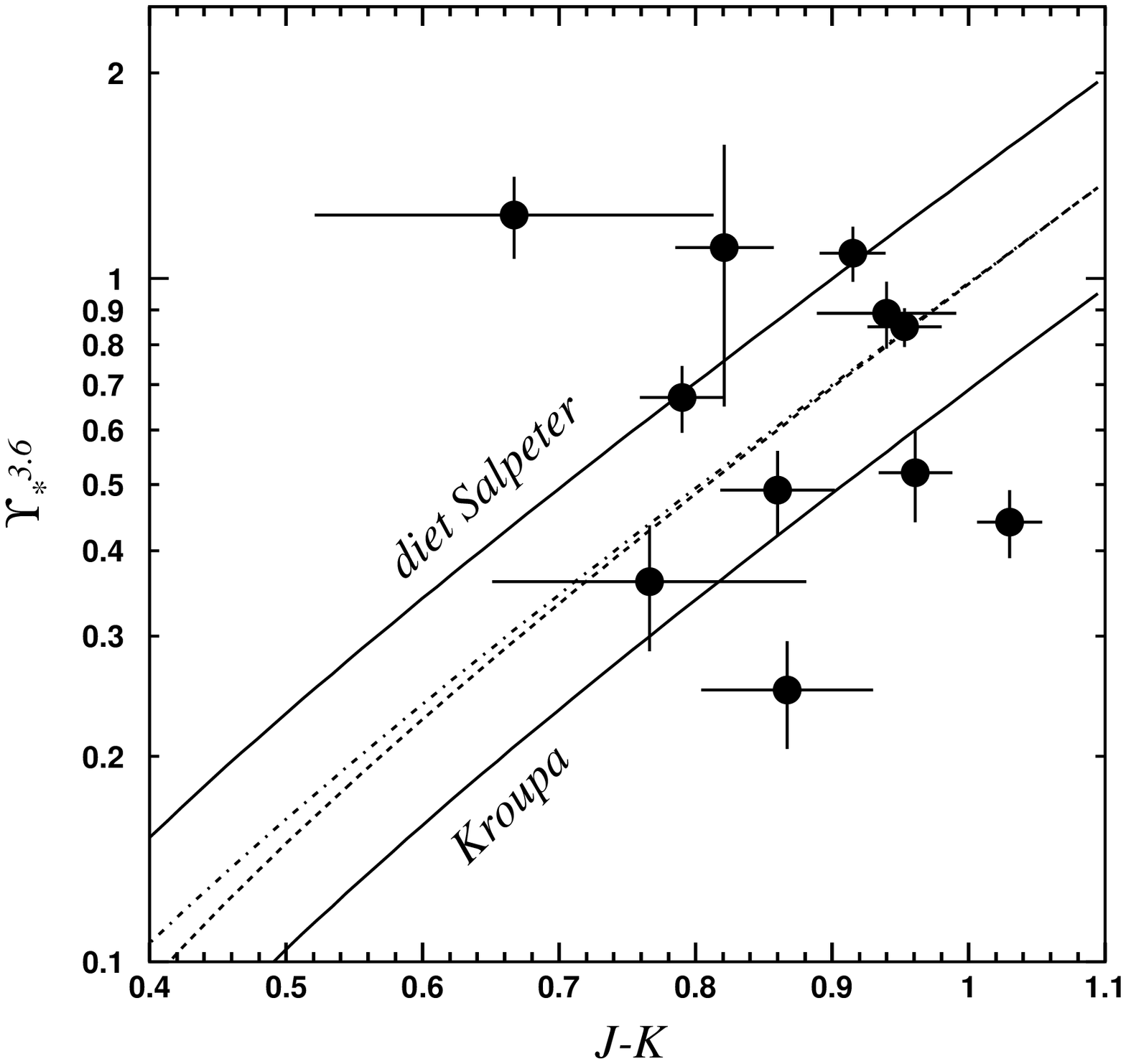}\\
\end{tabular}
\end{center}
\caption {Stellar mass-to-light ratio (in logarithmic scale) versus color for
galaxies in the Ursa Major catalog (left panel) and in
the THINGS catalog (right panel). Here $\Upsilon_\star$ is given in the $B$ band for the Ursa Major galaxies and in the $3.6~\mu$m band for the THINGS galaxies. We depict as solid lines
both the ``diet'' Salpeter and Kroupa IMF theoretical models with the margins to the models shown as dotted lines in both
panels. The black circles in the left panel represent HSB galaxies and
red squares represent LSB galaxies. For the sake of comparison, the blue star with coordinates (0.6, 1.4) represents the local value of $\Upsilon_\star$ in our Milky Way neighborhood~\cite{flynn}. All of the black circles in the right panel represent THINGS galaxies.
 \label{fig3} }
\end{figure*}

In longer wavelengths, the uncertainty in the relation between the stellar mass-to-light ratio and color for galaxies decreases significantly near the infrared (NIR). Therefore,  we adopt here the results of the analysis of
the magnitudes of galaxies in the $J$, $H$ and $K$ bands~\cite{bell2001}
as well as the observations in the $3.6~\mu$m band. On the basis of  the SPS
models, the relation between the mass-to-light ratio in the $K$ band
and color in the $J-K$ band is given by~\cite{bell2001}
\begin{equation}
\label{mlc}
\log_{10}(\Upsilon_\star^K) = 1.43\, (J-K)-1.38\,.
\end{equation}
Moreover, we can relate $\Upsilon_\star^{3.6}$ to the $J-K$ band from the relation between $\Upsilon_\star^K$ and $\Upsilon_\star^{3.6}$~\cite{oh}, namely,
\begin{equation}
\Upsilon_\star^{3.6} = 0.92\,\Upsilon_\star^K - 0.05\,. \label{kband}
\end{equation}
As before, in the case of Kroupa's IMF, the constant term in Eq.~\eqref{mlc}, $-1.38$, is reduced to $-1.53$.

We can now determine whether the mass-to-light ratio derived from the NLG theory  is in general agreement with the stellar synthesis models. To this end, we plot in Figure~\ref{fig3} the mass-to-light ratios both for the Ursa Major
galaxies in the $B$ band and the galaxies in the THINGS catalog in the $3.6~\mu$m band, and
compare the results with Eqs.~\eqref{emlc} and~\eqref{mlc}, respectively.  Let us
note that the colors and the magnitudes of galaxies in the
Ursa Major cluster are extinction corrected, and that we have divided the galaxies in
this plot into HSB and LSB galaxies in order to examine
their behaviors based on their types. Moreover, we compare our results with the average
stellar mass-to-light ratio of stars around the solar neighborhood using the \emph{Hipparcos} catalog~\cite{flynn}.

Though we have demonstrated the physical correlation between $\Upsilon_\star$ and the
color of galaxies, uncertainties still exist in the analytical relationship between these two parameters.
Finally, we remark that near-infrared observations provide more reliable
values for the stellar mass-to-light ratio than the visual-band
observations.

It is important to point out that on the basis of the results presented here, we can conclude that our approach is generally consistent with the theoretical astrophysical models of star formation.

\subsection{Tully-Fisher Relation}

We wish to determine here if the nonlocal gravity theory is in general agreement with the Tully-Fisher relation. In 1977, Tully and Fisher~\cite{TF} showed that there is an
empirical correlation between the intrinsic infrared luminosity of a spiral galaxy $L$ and the corresponding asymptotic rotation speed $v_c$, which may be roughly expressed as $L \propto v_c^4$. To test  this
empirical law using the nonlocal gravity model, we obtain the extinction-corrected absolute magnitude of a galaxy from observational data, namely, via the apparent magnitude, distance and extinction
factor. However, for the rotation velocity, we use the best-fitting parameters of our nonlocal gravity model, Eq.~\eqref{parameters1+2}, to calculate the corresponding rotation curve as in Figure 2. The result can be expressed as a plot of $M_B$, the absolute magnitude of a galaxy in the $B$ band, versus $V_{flat}$ in units of km s$^{-1}$, the rotation speed obtained from the flat part of the rotation curve.

\begin{figure}
\begin{center}
%\begin{tabular}
\includegraphics[width=80mm]{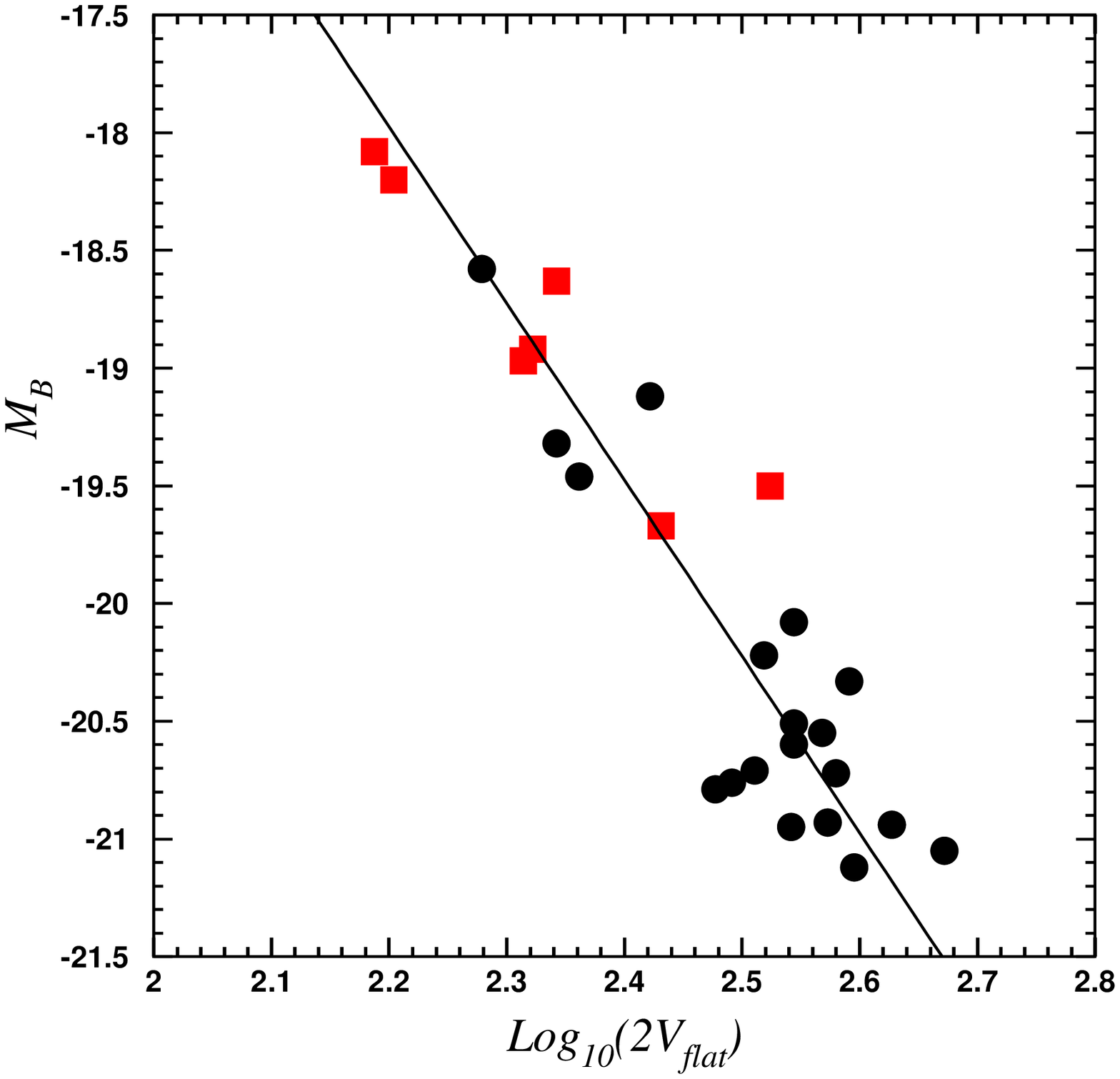}
%\end{tabular}
\end{center}
\caption{Plot of the extinction-corrected absolute magnitudes of galaxies in the subset of the Ursa Major
catalog versus the logarithm of $2V_{flat}$, where $V_{flat}$ is in units of km s$^{-1}$.
In this Tully-Fisher relation, the black circles
represent HSB galaxies and the red squares represent  LSB
galaxies. The flat rotation curves used to calculate $V_{flat}$ for this figure are determined on the basis of the
nonlocal gravity model by fitting to the observational data. The best fit to the data is given by $M_B =
-7.50\, \log_{10}(2V_{flat}) - 1.47$. \label{fig4}}
\end{figure}

To calculate $V_{flat}$ from the rotation curve, we follow here the
convention adopted in Ref.~\cite{ver2001}: (a) for a galaxy with a rising
rotation curve, $V_{flat}$ cannot be measured, (b) for a galaxy
with a ``flat" rotation curve, $V_{flat} = V_{max}$, where $V_{max}$ denotes the maximum of the rotation curve and (c) for a galaxy with
a declining rotation curve, $V_{flat}$ is calculated by averaging the
outer part of the rotation curve. Figure~\ref{fig4} displays the
distribution of galaxies in terms of absolute magnitude versus
the logarithm of $V_{flat}$ in the subset of the Ursa Major catalog. Here $V_{flat}$ is
calculated on the basis of the nonlocal gravity model. The best fit to the data points is given by
\begin{equation}
M_B = - 7.50\, \log_{10}(2V_{flat}) - 1.47\,, \label{tuequation}
\end{equation}
where $V_{flat} $ is in units of km s$^{-1}$. Our result for the Tully-Fisher relation in the $B$ band
 is compatible with the observational results given in
Ref.~\cite{ver2001}.

\section{Nonlocal Gravity: Solar System}

The parameters of nonlocal gravity model have been determined via the rotation curves of spiral galaxies in the previous section---see Eqs.~\eqref{parameters1+2} and~\eqref{parameter3}. With the model parameters $\alpha = 10.94\pm2.56$ and $\mu = 0.059\pm0.028~{\rm kpc^{-1}}$, we can now revisit the implications of nonlocal gravity for gravitational physics in the Solar System~\cite{NG3}.

In our nonlocal gravity model, the attractive Newtonian inverse-square force $F_N$ is modified by the addition of a repulsive Yukawa-type force as in Eq.~\eqref{II16} such that
\begin{equation}\label{IV1}
\frac{\delta F}{F_N} = \alpha\left[ 1 - (1+\frac12 \mu r)e^{-\mu r}\right]\,.
\end{equation}
At a radius of $r=10$ astronomical units, say, $\mu r\sim  10^{-9}$, so that
\begin{equation}\label{IV2}
\frac{\delta F}{F_N} \approx \frac{r}{\lambda_0} \sim 10^{-8}\,,
\end{equation}
where $\lambda_0 \approx 3\pm2$ kpc in our model in accordance with Eq.~\eqref{parameter3}. It follows from Eq.~\eqref{IV2} that within the Solar System, the force law reduces to the Tohline-Kuhn force given in Eq.~\eqref{II17}, since $0 < \mu r \ll 1$. The observational consequences of this perturbation due to nonlocal gravity have been investigated in detail in Ref.~\cite{NG3}. Nonlocal gravity is indeed consistent with current Solar System observations, since the new effects turn out to be rather small even compared to the main Einsteinian relativistic gravitational effects. To illustrate this point, we mention briefly here the extra bending of light rays by the Sun and the extra retrograde perihelion precession of Mercury.  Our discussion is based on the theoretical results presented near the end of section II. That is, assuming for the sake of simplicity that the Sun is a uniform sphere of radius $r_0$, then the gravitational force due to the Sun experienced by a test particle of unit mass located at $r>r_0$ from the center of the Sun is given by Eq.~\eqref{III13}, which, for our present purposes, may be approximated by the Tohline-Kuhn force~\eqref{II17}.

The deflection angle of light rays by an astronomical mass $M$ due to an additional attractive force of the form $G M m/(\lambda_0 r)$ is larger than the Einstein angle by a \emph{constant} amount given by $2\pi G M /(c^2 \lambda_0)=(M/M_\odot) \vartheta_\odot$, where
\begin{equation}\label{IV3}
\vartheta_\odot=\frac{2 \pi G M_\odot}{c^2 \lambda_0} \sim 10^{-16}\, \rm{rad}\,,
\end{equation}
which is nearly ten orders of magnitude smaller than the Einstein deflection angle.

The situation for the perihelion precession of Mercury is more hopeful, however~\cite{NG3, Tohl, K}. The extra force due to our nonlocal gravity model---see, for example, Eq.~\eqref{III13}---is essentially radial and conservative; therefore, its influence on planetary orbits is such that a Keplerian ellipse remains planar and keeps its shape on the average, but it has a retrograde precessional motion. The pericenter precession frequency is given by~\cite{NG3, Tohl, K}
\begin{equation}\label{IV4}
{\dot w}=-\frac{1}{2 \lambda_0} \Big(\frac{G M}{A}\Big)^{1/2} I(e)\,,
\end{equation}
where $A$ and $e$ are respectively the semimajor axis and eccentricity of the orbit. Here,
\begin{equation}\label{IV5}
I(e)= \frac{2}{e^2}\Big[(1-e^2)^{1/2}-(1-e^2)\Big]\,
\end{equation}
decreases monotonically from unity at $e=0$ to zero at $e=1$.

For the orbit of Mercury, $A \approx 6\times 10^{12}$ cm and $e\approx 0.2$, so that for Mercury,
 ${\dot w}\approx -0.2\pm0.1$ seconds of arc per century. We note that this is about $-4\times 10^{-3}$ of the observed  excess value for Mercury, which was successfully explained by Einstein on the basis of general relativity theory. In the Solar System, the rate of precession of perihelion is measured by optical methods with an uncertainty of about $0.4$ seconds of arc per century. On the other hand, the uncertainty can be reduced to about $0.2$ seconds of arc per century via radar measurements~\cite{Shap}. Since Shapiro's review~\cite{Shap}, the uncertainty in the radar observations of Mercury has improved by about an order of magnitude. Moreover, the Newtonian contribution to the perihelion precession of Mercury due to solar oblateness is about 0.03 seconds of arc per century~\cite{Will}, while the corresponding contribution of the asteroid belt is expected to be smaller by perhaps two orders of magnitude.  There are, however, problems in the interpretation of observational data~\cite{Shap}. Nevertheless, future improvements in the measurement of planetary perturbations and their interpretations may make it possible to detect the effect of nonlocal gravity in the Solar System.

\section{Observational Tests:  \emph{Chandra}  X-ray Clusters of Galaxies}

Clusters of galaxies are filled with hot ionized gas that is luminous in X-rays. The low-density gas contains $\sim 10^{-3}$ atoms/cm$^3$ and has a temperature of order $10^8$ K. Most of the electrons originate from hydrogen and helium atoms that are fully ionized. We assume, as usual, that in the plasma the mass density of protons is nearly $\frac{3}{4}\,\rho_g$ and the mass density of helium ions is nearly $\frac{1}{4}\,\rho_g$, where $\rho_g$ is the mass density of the cluster gas. The gas contains most of the baryonic mass of rich clusters and is roughly at the \emph{virial temperature} $T$ that is related to the radial (i.e., line-of-sight)  velocity dispersion $\sigma_r$ of the galaxies in the cluster, namely, $k_B\,T\approx \mu_p m_p \sigma_r^2$, where $\mu_p$ is the \emph{mean} atomic mass of the plasma (electrons and ions) in amu, $\mu_p\approx 0.6$, and $m_p$ is the proton mass.
In this section, we employ the gas density profile $\rho_g(r)$ of a cluster and the corresponding temperature profile $T(r)$, obtained from the observational data provided by the \emph{Chandra} X-ray telescope, to extract the magnitude of the gravitational acceleration $g(r)$ inside the cluster. To this end, we assume that the gas is in hydrostatic equilibrium, so that $dP/dr=-\rho_g~ g(r)$, where the gas pressure $P$ is given by $P/(k_B\,T)=\rho_g/(\mu_p m_p)$ in accordance with the ideal gas law~\cite{sarazin}. It follows that for a spherically symmetric system in hydrostatic equilibrium, the magnitude of acceleration  on a test  point particle is related to the observable parameters by~\cite{sarazin}
\begin{equation}
\frac{k_B T(r)}{\mu_p m_p\, r}\left(\frac{d\ln\rho_g(r)}{d\ln r} + \frac{d\ln T(r)}{d\ln
r}\right) = - g(r)\,. \label{halo}
\end{equation}
In order to have correct dynamics with only  baryonic matter and no actual dark matter, the left-hand side of this equation should be equal to the right-hand side given by Eq.~\eqref{II16} of our nonlocal gravity model. In Eq.~\eqref{II16}, for $\mu r \gg 1$, the force between two point particles reduces to the inverse-square force law augmented by the constant factor of $1+\alpha$. Thus in this case, a homogeneous spherical shell of matter exerts no force on a test particle in its interior, but attracts an exterior  test particle as though the mass of the shell were concentrated at its center. The radius of a cluster is of order
$1000$ kpc; therefore,  in the outer parts of the cluster and away from the central region that has the highest mass density, consider the determination of acceleration of gravity on a point mass at fixed radius $r$. Except for the mass interior to a sphere of radius $\sim\mu^{-1}\approx 17$ kpc about our fixed point mass , we can safely neglect the repulsive Yukawa force in Eq.~\eqref{II16} in comparison with the attractive Newtonian force. As the mass within the sphere of radius $\sim\mu^{-1}\approx 17$ kpc is very small compared to the mass of the cluster, we can simply estimate $g(r)$ using the Newtonian inverse-square force law augmented by $1+\alpha$, as discussed in detail in section II. It follows that at a radial distance $r$ well beyond the central regions of a cluster, we have the approximate expression $g(r)\approx (1+\alpha)GM_{\rm dyn}(r)/r^2$, and therefore Eq.~\eqref{halo} can be written \emph{approximately} as
\begin{equation}
M_{\rm dyn}(r) \approx - 3.68\times 10^{10}~\frac{r  T(r)} {1+\alpha}  \left(\frac{d\ln \rho_g(r)}{d\ln r}  +\frac{d\ln T(r)}{d\ln r} \right ) M_\odot\,,
\label{master}
\end{equation}
where $r$ is the radial distance from the center of the cluster in kpc, $M_{\rm dyn}(r)$ is  the dynamical  mass of the cluster within a sphere of radius $r$  and $T(r)$ is the temperature of the gas in keV.

We now proceed to the determination of the
gas density and temperature profiles for \emph{Chandra} X-ray clusters via the best fit to the observational data. In this way,  we compute the dynamical mass of the cluster within the radial distance $r$ from  Eq.~\eqref{master} and compare the result to the baryonic mass of the cluster, namely, the net mass of gas and stars, which we obtain by volume integration of $\rho_g$ from a radius of 100 kpc out to radius $r$ and the addition of mass of the stars up to radius $r$, without resorting to any actual dark matter. We start the integration of the gas density profile from $r=100$ kpc rather than from $r=0$, since there are high measurement uncertainties in the density of gas at the central part of the cluster; moreover, the contribution of the central region with $r<100$ kpc to the mass of the whole cluster with radius of $\sim 1000$ kpc is expected to be rather small. The mass of the cluster is dominated by its baryonic content and we expect that the baryonic mass and the dynamical mass would essentially agree if the gravitational force that balances  gas pressure within the cluster is correctly represented by nonlocal gravity.

\subsection{Gas density profile}

The temperature of the hot gas  in the cluster is of the order of  keV and the hot plasma mainly emits X-rays as a result of thermal bremsstrahlung via the free-free radiation process. There are line emissions by the ionized heavy elements as well. A detailed discussion of the various emission mechanisms is contained in Ref.~\cite{sarazin}. It turns out that the net amount of emitted radiation is proportional to the product of number densities of electrons $n_e$ and protons $n_p$. Moreover, this product is related to the gas density as well; that is,
\begin{equation}\label{rhog}
\rho_g(r) \approx 1.24\, m_p \Big[n_e(r)\,n_p(r)\Big]^{\frac{1}{2}}\,.
\end{equation}

In Ref.~\cite{chandra}, the properties of thirteen nearby relaxed galaxy clusters in the \emph{Chandra} catalog were studied by constructing three-dimensional radial profiles of gas density and temperature that were then projected along the line of sight and fit to observational data. We use the three-dimensional models provided in Ref.~\cite{chandra} to compute $M_{\rm dyn}(r)$ from Eq.~\eqref{master}.

To fit the observational data for the \emph{Chandra} X-ray clusters, the so-called $\beta$ model, which provides the standard expression for the density function $n_e(r)\,n_p(r)$ of the form~\cite{cav}
\begin{equation}
\frac{n_0^2}{(1+r^2/r_c^2)^{3\beta}}\,,
\label{nee}
\end{equation}
has been modified in Ref.~\cite{chandra}  in such a way that: (i) it has a cusp at the center, (ii) at large radii, X-ray brightness is steeper than that given by the $\beta$ model and (iii) a second $\beta$-model component is added with a smaller core radius in order to have more freedom near the center of the cluster.  Following Ref.~\cite{chandra}, we thus assume
\begin{equation}
n_e(r)\,n_p(r) =  \frac{ (r/r_c)^{-\alpha'}}{(1+r^2/r_c^2)^{3\beta-\alpha'/2}}\frac{n_0^2}{(1+r^\gamma/r_s^\gamma)^{\varepsilon/\gamma}} +\frac{n_{0}'{}^2}{(1+r^2/r_{c}'{}^2)^{3\beta^\prime}}\,,
\label{modne}
\end{equation}
where $\gamma=3$ throughout. Moreover, a limitation is placed on $\varepsilon$, namely, $\varepsilon \le 5$, in order to avoid unphysical radial variations in density~\cite{chandra}.

Table~\ref{tab3} represents the best fit to the density function~\eqref{modne} for ten clusters of galaxies in the \emph{Chandra} catalog~\cite{chandra}. Each cluster is considered to be a spherical configuration of matter with an effective radius of $r_{500}$; more precisely, $r_{500}$ is defined to be the cluster radius within which the average overdensity is 500 times the critical density of the universe at the redshift of the cluster in the dark matter model. The particular radial function~\eqref{modne} adopted in Ref.~\cite{chandra} to represent $\rho_g^2$ has nine unknown parameters and there are various degeneracies among them. The numerical values of these parameters have been specified in Ref.~\cite{chandra} and adopted here in Table~\ref{tab3}. No error estimates for the model parameters were given in Ref.~\cite{chandra}; hence, there are no error estimates for the nine parameters of the gas density profiles in Table~\ref{tab3}.

\begin{table*}
\caption{The parameters of the gas density profiles for the sample of ten clusters of galaxies observed by the \emph{Chandra} telescope and studied in Ref.~\cite{chandra}. This table has been constructed from the results given in Ref.~\cite{chandra}. The first column specifies the name of cluster and the second column specifies the radius of the outer boundary (i.e., $r_{500}$) where the cluster is observed in X-rays; that is, $r_{500}$ is a useful measure of the size of the cluster. Other columns represent the best-fit values of the parameters of the three-dimensional density profile~\eqref{modne}.}

\begin{ruledtabular}
\begin{tabular}{ccccccccccc}
Cluster &
$r_{500}$ &
{$n_0$} &
{$r_c$} &
{$r_s$} &
{$\alpha'$} &
{$\beta$} &
{$\varepsilon$} &
{$n_{0}'$} &
{$r_{c}'$} &
{$\beta'$}
%{$\beta_{eft,500}$}
\\
 & {(kpc)} & {($10^{-3}$~cm$^{-3}$)}
 & {(kpc)} & {(kpc)} & & & & {($10^{-1}$~cm$^{-3}$)}\\
\hline
A133          \dotfill & $1007\pm 41$&  4.705 & 94.6 & 1239.9 & 0.916 & 0.526 & 4.943 &   0.247 &    75.83 &   3.607     \\
%A262          \dotfill &  $650\pm 21$&  2.278 &  70.7 &  365.6 & 1.712 & 0.345 & 1.760 & -- &    -- & --   \\
A383          \dotfill &  $944 \pm32$ &   7.226 & 112.1 &  408.7 & 2.013 & 0.577 & 0.767 &   0.002 &     11.54 &   1.000 \\
A478          \dotfill & $1337\pm 58$ &   10.170 & 155.5 & 2928.9 & 1.254 & 0.704 & 5.000 &   0.762 &      23.84 &   1.000   \\
A907          \dotfill &$1096\pm 30$ &   6.257 & 136.9 & 1887.1 & 1.556 & 0.594 & 4.998 & -- &    -- & -- \\
A1413         \dotfill &$1299\pm 43$&   5.239 & 195.0 & 2153.7 & 1.247 & 0.661 & 5.000 & -- &    -- & --  \\
A1795         \dotfill & $1235\pm 36$&    31.175 &  38.2 & 682.5 & 0.195 & 0.491 & 2.606 & 5.695 &  3.00 & 1.000      \\
A1991         \dotfill &$732\pm 33$&   6.405 &  59.9 &  1064.7 & 1.735 & 0.515 & 5.000 &   0.007 &     5.00 &   0.517    \\
A2029         \dotfill & $1362\pm 43$&    15.721 &  84.2 &  908.9 & 1.164 & 0.545 & 1.669 &   3.510 &     5.00 &   1.000    \\
A2390         \dotfill & $1416\pm 48$&   3.605 & 308.2 & 1200.0 & 1.891 & 0.658 & 0.563 & -- &    -- & --  \\
%RXJ\,1159+5531\dotfill & $700\pm 57$&   0.191 & 591.9 &  640.7 & 1.828 & 0.838 & 4.869 &   0.457 &      11.99 &   1.000  \\
MKW4          \dotfill &  $634\pm 28$&   0.196 & 578.5 & 595.1 & 1.895 & 1.119 & 1.602 &   0.108 &      30.11 &   1.971    \\
%USGC~S152     \dotfill & \dotfill&    27.098 &   5.8 &  467.5 & 2.612 & 0.453 & 3.280 & -- &    -- & --  \\
\end{tabular}
\end{ruledtabular}
\label{tab3}
\end{table*}

\subsection{Temperature profile}

The temperature $T(r)$ that appears in Eq.~\eqref{master} is in fact $T_{3D}(r)$, the three-dimensional radial profile of the gas temperature, which has to be properly projected along the line of sight~\cite{vikh2006}, since it is the projected two-dimensional profile that can be compared directly with observational data. The plasma in the cluster can be divided into two regions: a cooling zone near the center and  the outer part of the cluster. The temperature profiles for these two regions have been modeled by two different functions~\cite{chandra}.
At the center of a cluster the temperature decreases due perhaps to radiative cooling in this region; hence, it is useful to define~\cite{allen}
\begin{equation}\label{in}
\Theta_{in}(r) = \frac{(x_0 + T_{min}/T_0)}{x_0+1}\,, \qquad x_0:= (\frac{r}{r_{cool}})^{a_{cool}}\,.
\end{equation}

For the outside of the central cooling zone, it is useful to define~\cite{chandra}
\begin{equation}\label{out}
\Theta_{out}(r) = \frac{(r/r_t)^{-a'}}{\big[1 + (\frac{r}{r_t})^b \big]^{c'/b}}\,,
\end{equation}
where $r_t$ represents the radial transition region.
The overall three-dimensional temperature profile of a cluster is then given by~\cite{chandra}
\begin{equation}
T_{3D}(r) = T_0\, \Theta_{in}(r) \Theta_{out}(r).
\label{Tprofile}
\end{equation}
Table~\ref{tab4} represents the best fits to the temperature profiles~\eqref{Tprofile} of the ten
clusters of galaxies in the \emph{Chandra} catalog~\cite{chandra}. The numerical values of the eight model parameters given in Table~\ref{tab4} are based on Ref.~\cite{chandra}. No error estimates for these parameters were provided in Ref.~\cite{chandra}.

\begin{table*}
\caption{The parameters of the three-dimensional temperature profiles for the sample of ten clusters of galaxies observed by the \emph{Chandra} telescope and studied in Ref.~\cite{chandra}.  This table has been constructed from the results given in Ref.~\cite{chandra}. The first column specifies the name of the cluster and the other columns specify the parameters of the temperature profile~\eqref{Tprofile}.}
\begin{ruledtabular}
\begin{tabular}{ccccccccc}
Cluster &
$T_0$ &
{$r_t$} &
{$a'$} &
{$b$} &
{$c'$} &
{$T_{min}/T_0$} &
$r_{cool}$&
$a_{cool}$\\
%$\beta_{t,500}$\\

 & {(keV)} & {(Mpc)} & & & &  & {(kpc)} &  \\
\hline
A133          \dotfill & 3.61 &      1.42 & 0.12 &  5.00 & 10.0 & 0.27   & 57     & 3.88       \\
%A262          \dotfill & 2.42 &      0.35 &  -0.02 & 5.00   & 1.1 & 0.64  & 19     & 5.25   &    --      \\
A383          \dotfill & 8.78 &      3.03 & -0.14 &  1.44 & 8.0  & 0.75   & 81     &  6.17  \\
A478          \dotfill & 11.06&     0.27 & 0.02 &  5.00 & 0.4    & 0.38  & 129    & 1.60      \\
A907          \dotfill & 10.19&     0.24 & 0.16 & 5.00 & 0.4     & 0.32  & 208    & 1.48   \\
A1413         \dotfill & 7.58 &     1.84 & 0.08 & 4.68 & 10.0   & 0.23  & 30      & 0.75   \\
A1795         \dotfill & 9.68 &     0.55 &  0.00 & 1.63 & 0.9    & 0.10  & 77      & 1.03        \\
A1991         \dotfill & 2.83 &     0.86 &  0.04 &  2.87 & 4.7   & 0.48  & 42      &  2.12       \\
A2029         \dotfill & 16.19&     3.04 &  -0.03 &  1.57 & 5.9 & 0.10  & 93      &  0.48      \\
A2390         \dotfill & 19.34&     2.46 &  -0.10 & 5.00 &10.0 & 0.12  & 214    & 0.08  \\
%RXJ\,1159+5531\dotfill & 3.74& 0.10 & 0.09 &  0.77 & 0.4  & 0.13 & 22       &  1.68 &    --\\
MKW4          \dotfill &  2.26&    0.10 & -0.07 & 5.00 & 0.5    & 0.85  & 16      &  9.62    \\
%USGC~S152     \dotfill &  3.24& 0.59 &  0.01 &  0.27 & 0.8 & 0.37  & 31        & 3.24 &   --\\
\end{tabular}
\end{ruledtabular}
\label{tab4}
\end{table*}

\setcounter{figure}{4}
\begin{figure*}
\begin{center}
\begin{tabular}{cc}
\includegraphics[width=65mm]{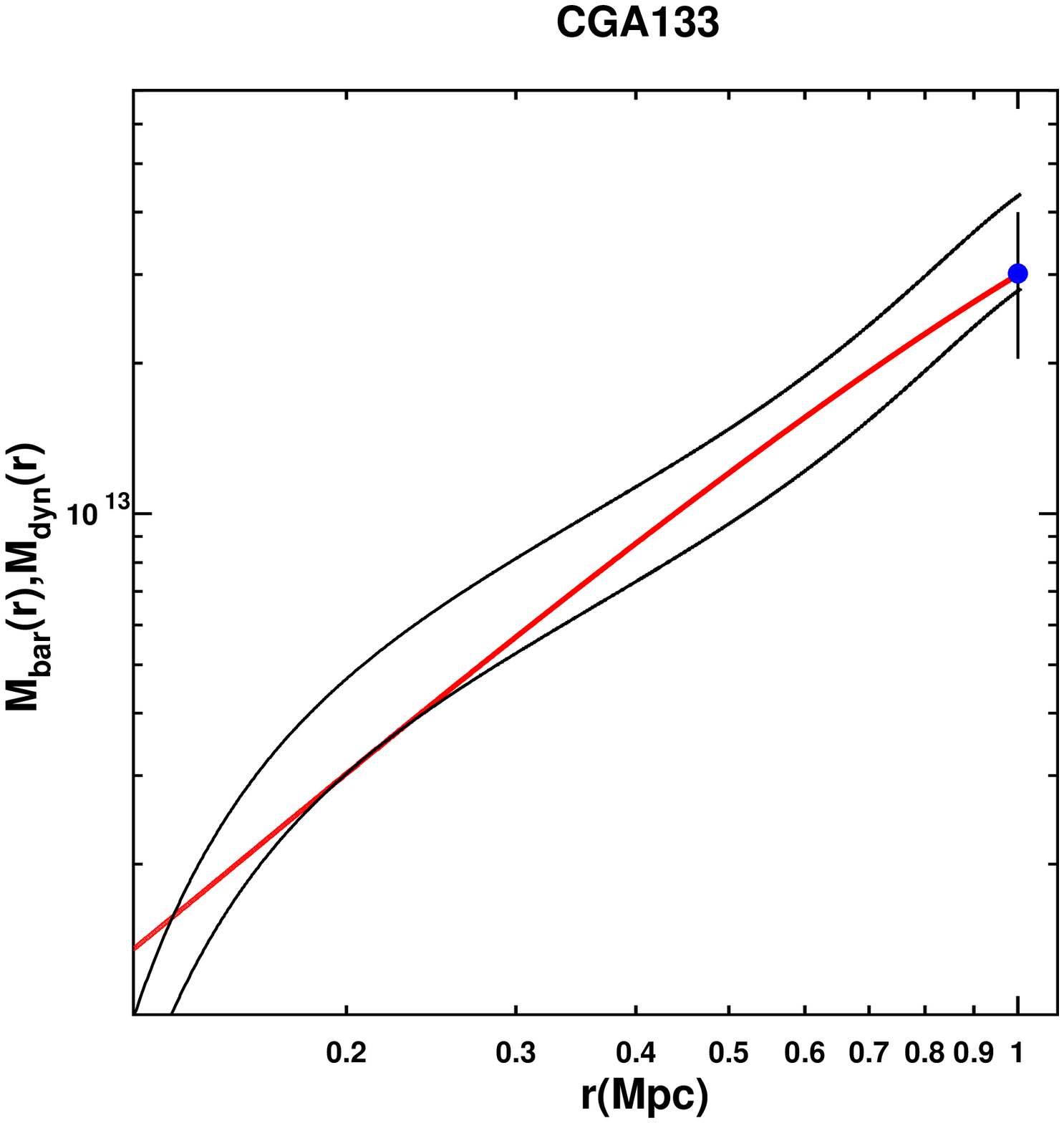}&
\includegraphics[width=65mm]{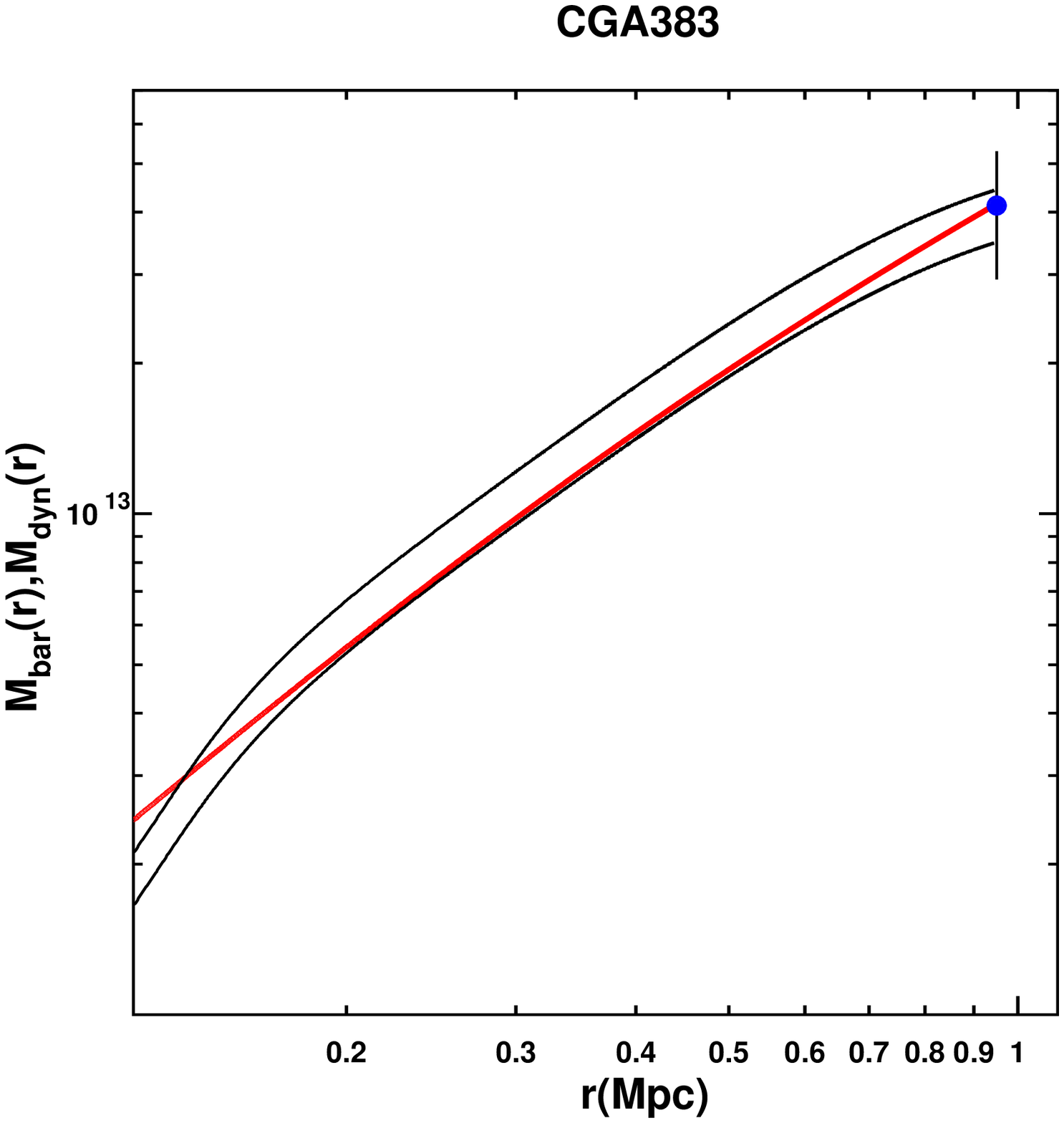} \\
\includegraphics[width=65mm]{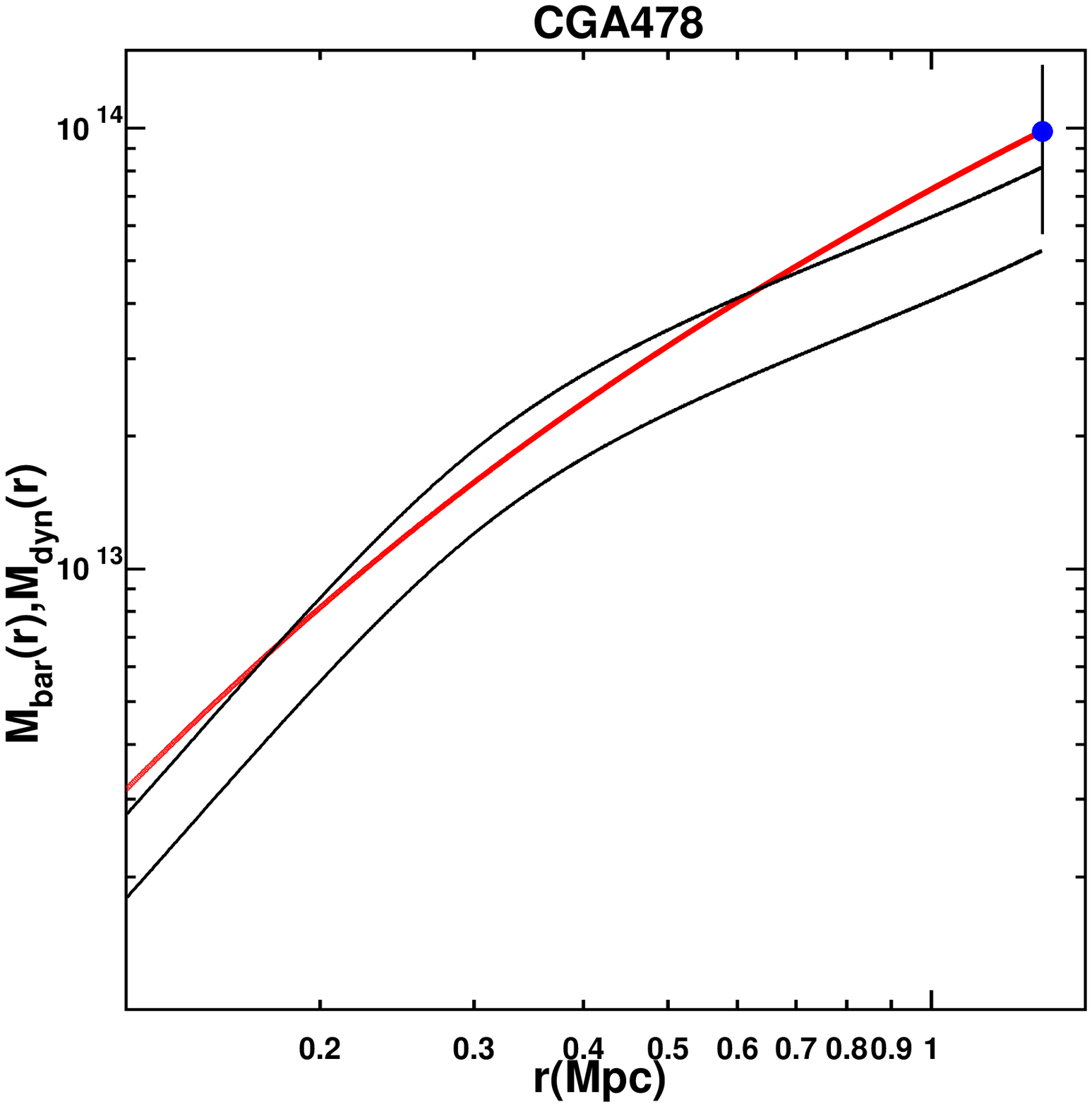} &
\includegraphics[width=65mm]{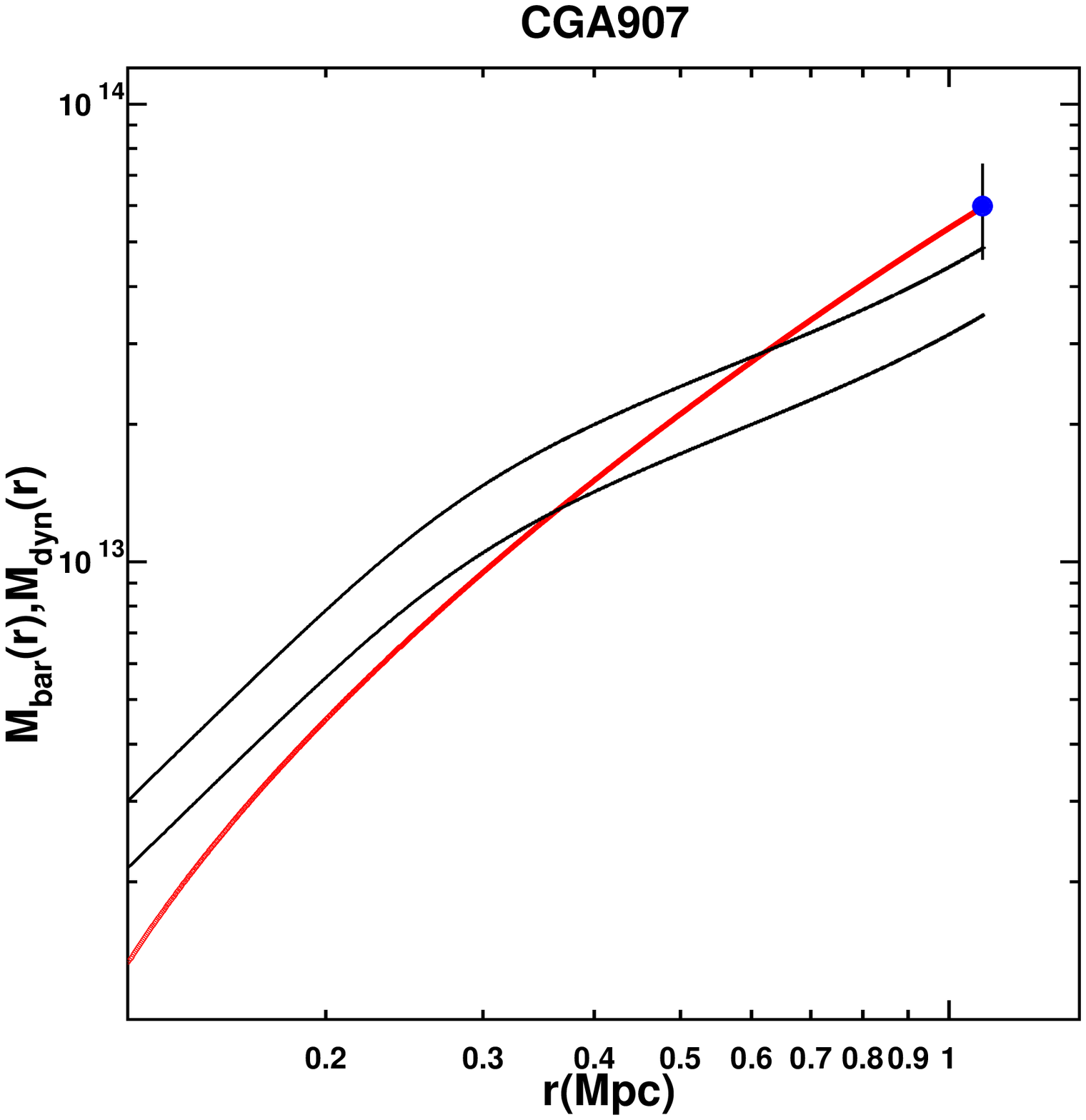}\\
\includegraphics[width=65mm]{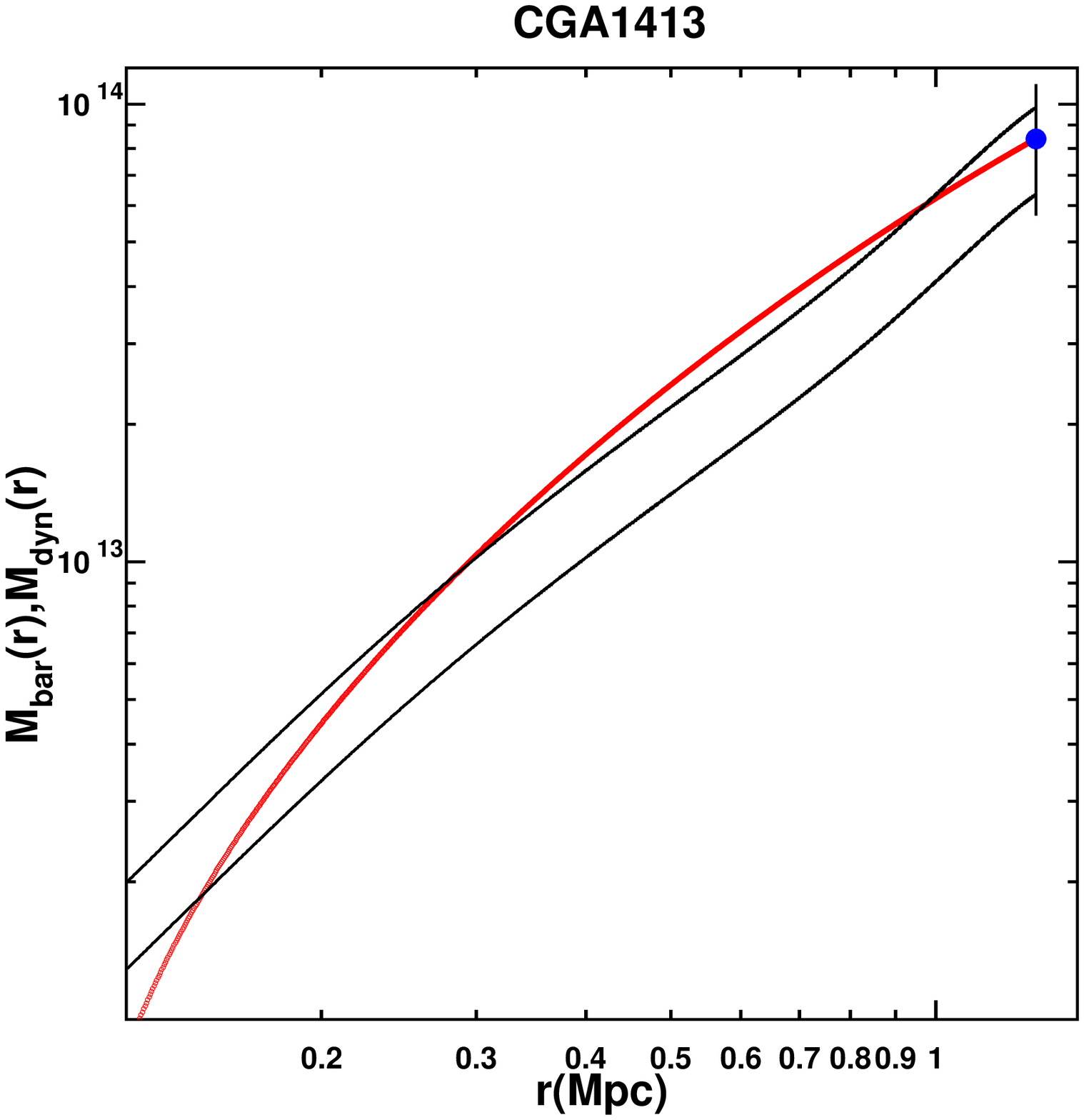}&
\includegraphics[width=65mm]{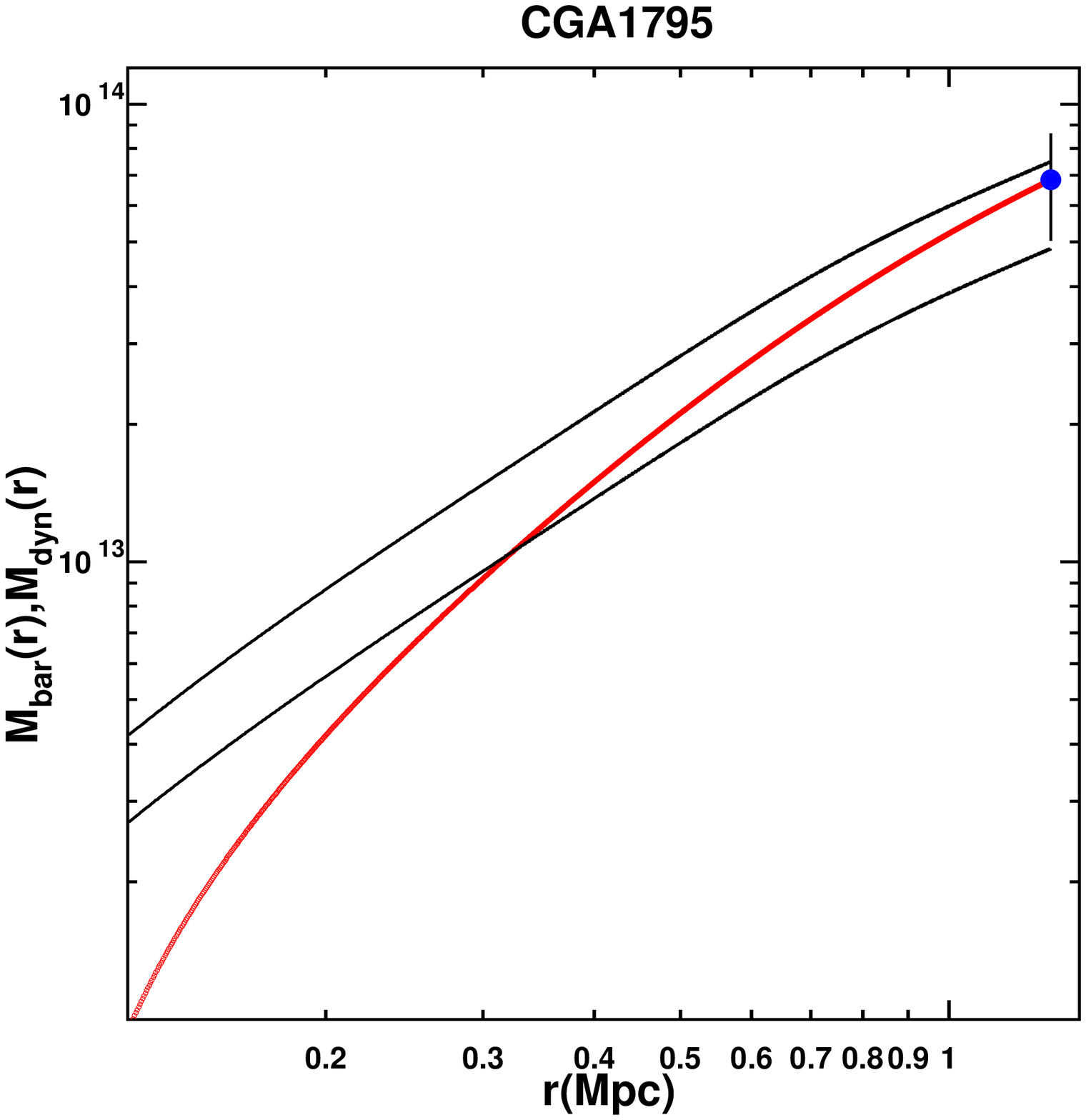}\\
\end{tabular}
\end{center}
\caption {Mass profile of a cluster is plotted as a function of distance from the center of cluster for the sample of ten \emph{Chandra} X-ray clusters studied in Ref.~\cite{chandra}. The thick red line results from the integration of observed distribution of gas plus stars, while the thin black lines represent the dynamical mass calculated from Eq.~\eqref{master} given by nonlocal gravity for $\alpha = 10.94\pm2.56\,$. The error bar at the end of the red line gives an indication of the measurement error along the red line.
\label{cluster} }
\newpage
\end{figure*}

\setcounter{figure}{4}
\begin{figure*}
\begin{center}
\begin{tabular}{ccc}
\includegraphics[width=65mm]{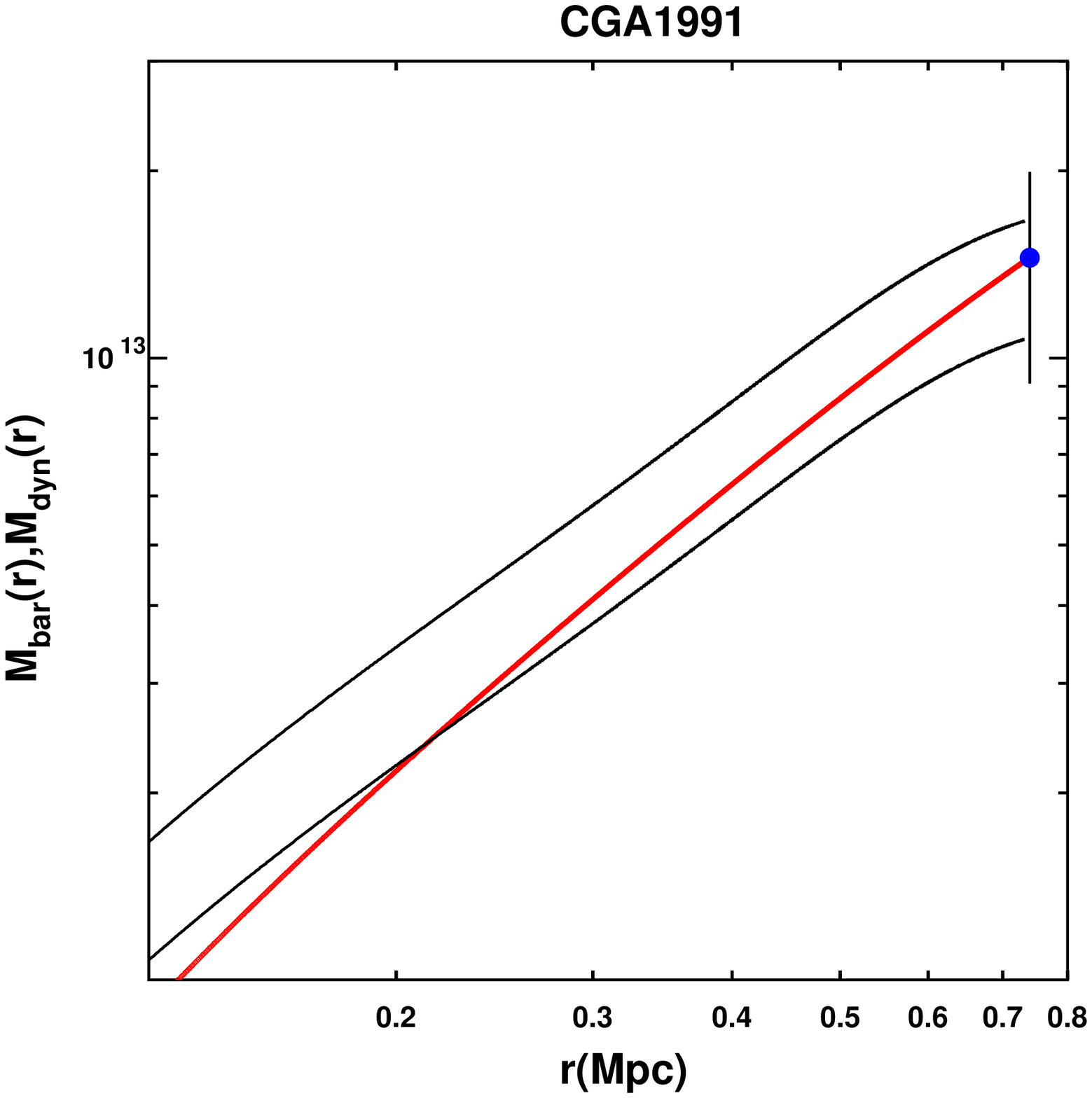} &
\includegraphics[width=65mm]{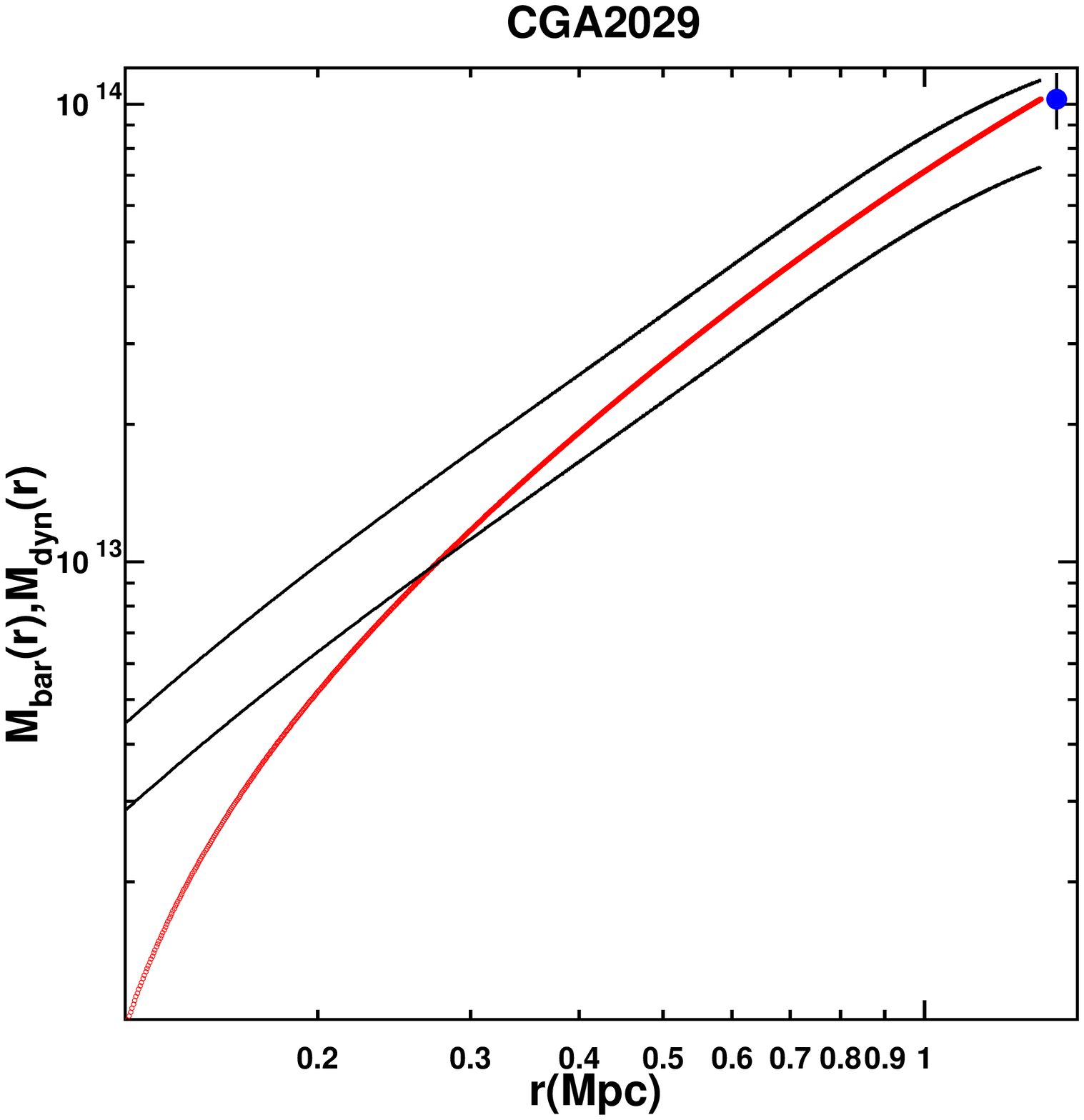}\\
\includegraphics[width=65mm]{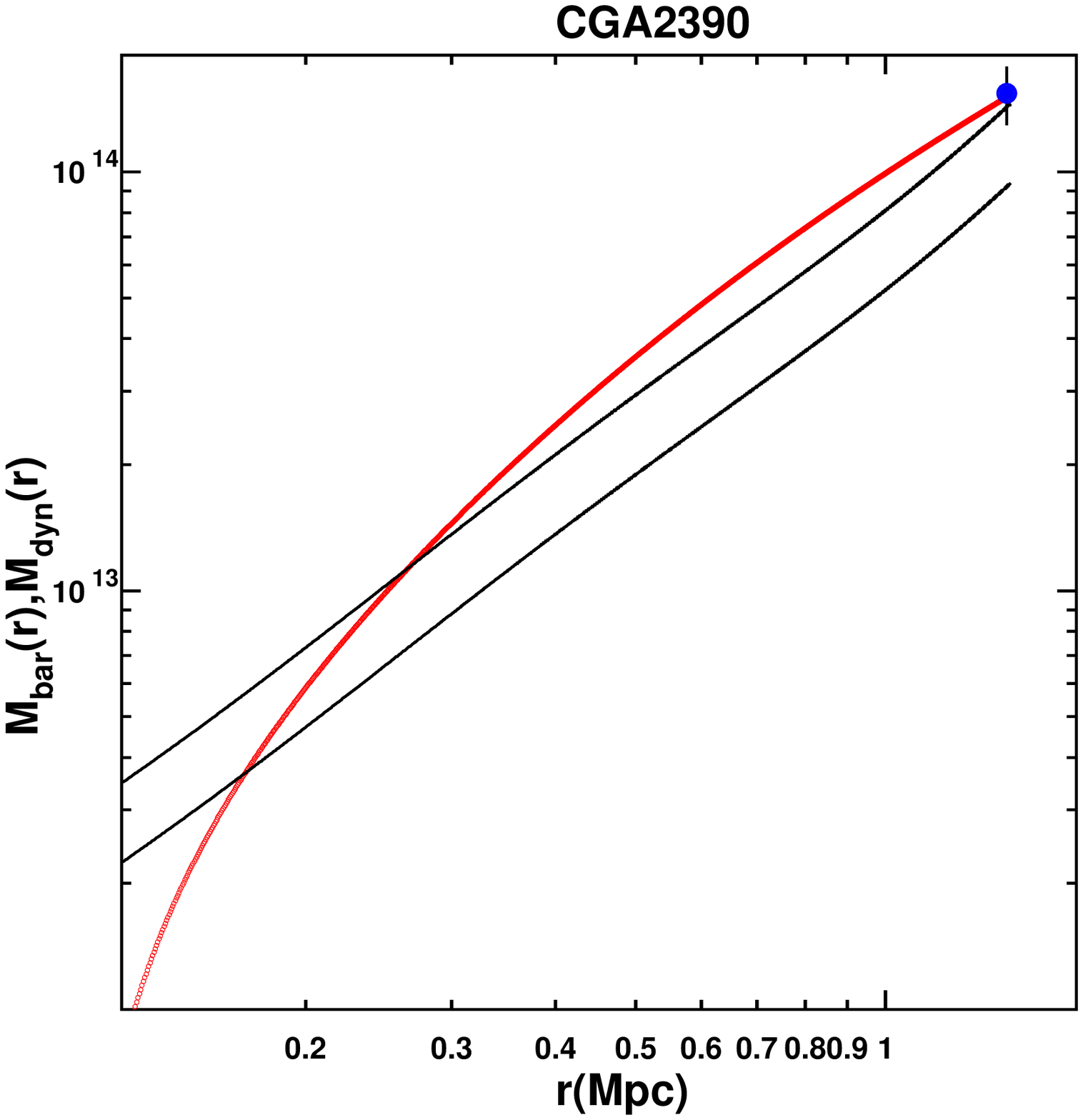} &
\includegraphics[width=65mm]{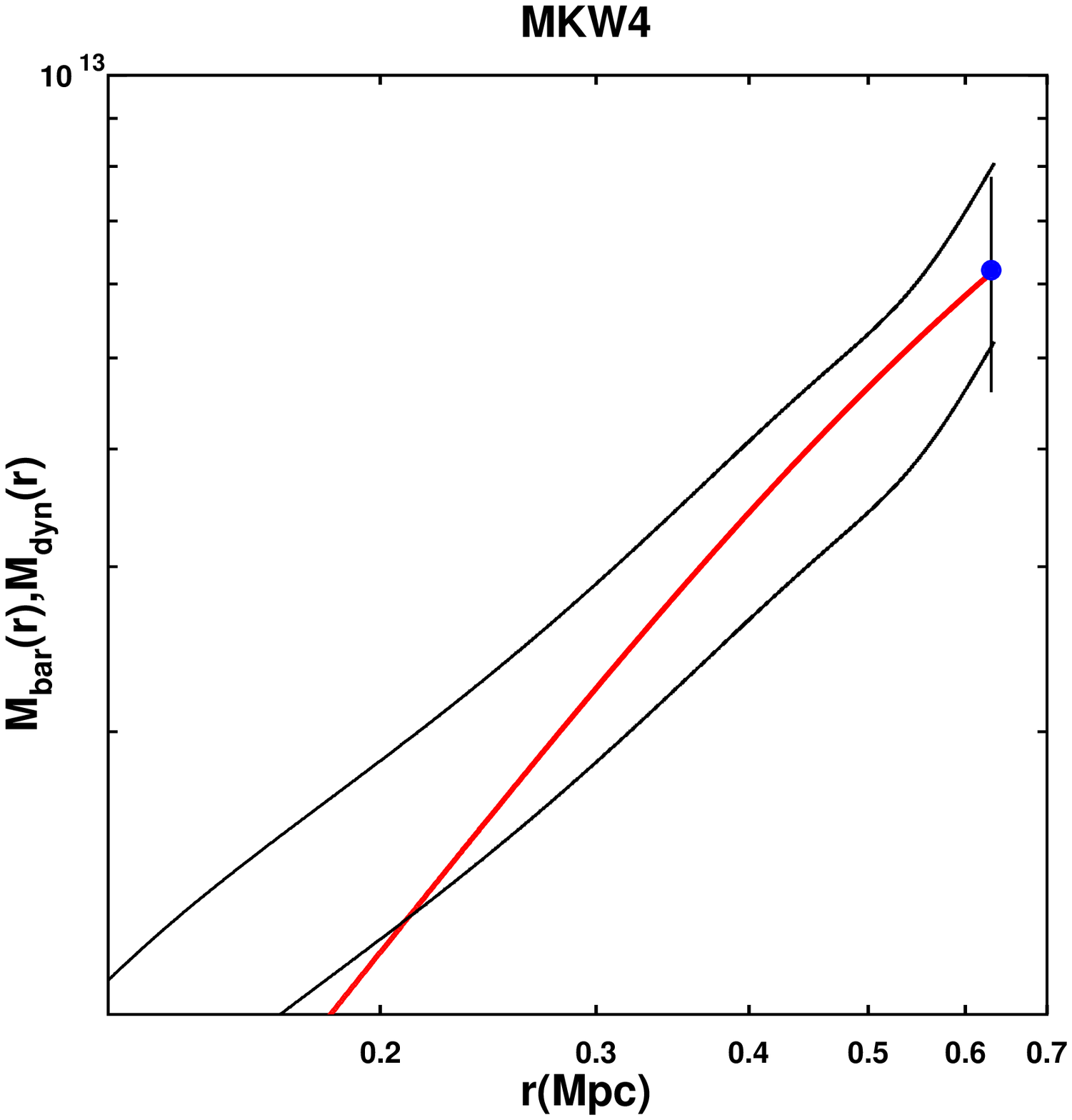} &
\end{tabular}
\end{center}
\caption {Continued ...}
\newpage
\end{figure*}

\begin{table*}
\caption{Dynamical mass in nonlocal gravity versus the observed baryonic mass of clusters. In this table, the dynamical masses of clusters in nonlocal gravity, based on Eq.~\eqref{master}, are compared with the observed baryonic masses of the clusters. The columns describe: (1)  the name of the cluster from the \emph{Chandra} catalog~\cite{chandra}, (2) the mass of the cluster in the dark matter model up to $r_{500}$; this quantity, given in Ref.~\cite{chandra}, is needed in Eq.~\eqref{mass} and its error estimate results from the uncertainty in the measurement of temperature and X-ray flux, (3) the mass of gas obtained by integrating the gas density over the volume of the cluster up to $r_{500}$; the error estimates are adapted from Ref.~\cite{chandra}, (4) the mass of stars from  the empirical relation~\eqref{mass}, (5) the overall baryonic mass; here, we neglect the error estimates associated with the contribution of stars, since it is small in comparison with the contribution of the gas to the total baryonic mass, and (6) the total dynamical mass inferred from Eq.~\eqref{master}.}
\begin{ruledtabular}
\begin{tabular}{cccccc}
(1) Cluster &
(2) $M_{500}$ &
(3) $M_{gas}$ &
(4) $M_{stars}$ &
(5) $M^{\rm c}_{\rm bar}$ &
(6) $M^{\rm c}_{\rm dyn}$ \\

 & $(10^{14} M_\odot)$  & $(10^{13} M_\odot)$  & $(10^{13} M_\odot)$   & $(10^{13} M_\odot)$ & $(10^{13} M_\odot)$\\

  % & $\times 10^{14} M_\odot$ &  &$\times 10^{13} M_\odot$ & $\times 10^{13} M_\odot$ & $\times 10^{13} M_\odot$ & $\times 10^{13} M_\odot$  \\
\hline
A133   \dotfill &$3.14 \pm 0.36$ & $2.61\pm0.54$  &  0.408 & $3.02\pm0.54$ &$ 3.41_{-0.60}^{+0.94}$  \\
%&$1.13\pm0.07$ &$0.067\pm0.002$ &$3.20\pm 0.49$  &   $2.82\pm0.54$   & $1.14\pm0.11$     & $0.75\pm0.07$  \\
%A262          \dotfill &   0.669  & 0.133 & 0.803  &0.767 \\
%   $0.34\pm0.05$  & $0.067\pm0.003$   &\dotfill                  &  \dotfill                    & $0.34\pm0.06$   & $0.22\pm0.04$     \\
A383          \dotfill &$3.10 \pm 0.32$ &$3.72\pm 0.59$  &  0.398 & $4.12\pm 0.59$  & $3.47_{-0.66}^{+0.94}$ \\
% $0.124\pm0.007$   &$1.64\pm0.14$  & $0.092\pm0.005$  &$3.09\pm 0.41$ &    $3.79\pm0.59$    & $1.65 \pm0.20$  & $1.50\pm0.21$     \\
A478          \dotfill &$7.83 \pm 1.04$ &$9.06 \pm 2.04$ & 0.765  & $9.82\pm 2.04$ & $6.43_{-1.13}^{+1.79}$ \\
% $4.12\pm0.26$  & $0.098\pm0.004$  &$7.76\pm1.28$  &      $9.21\pm2.04$  & $4.16 \pm0.40$  & $4.03\pm0.42$  \\
A907          \dotfill &$4.71 \pm 0.39$ &$ 5.46\pm0.72$   & 0.528 & $5.99\pm0.72$ &  $4.22_{-0.67}^{+1.06}$ \\
%  & $2.21\pm0.14$  & $0.091\pm0.003$&$4.61\pm0.53$     &   $5.65\pm0.72$  & $2.23 \pm0.15$  & $2.01\pm0.13$ \\
A1413         \dotfill &$7.78 \pm 0.83$ &$ 7.86\pm 1.34$  & 0.527  & $8.39\pm 1.34$ & $7.70_{-1.36}^{+2.20}$ \\
% & $3.01\pm0.18$  &  $0.094\pm0.003$   &$7.65\pm1.02$ &       $8.10\pm 1.34$ & $3.04\pm0.28$  & $2.83\pm0.26$ \\
A1795         \dotfill &$6.57 \pm 0.69$ & $6.18\pm0.90 $ &0.644  & $6.83\pm0.90$ & $5.87_{-1.06}^{+1.61}$ \\
%&$2.75\pm0.16$&$0.088\pm0.003$ &$6.09\pm0.73$&      $6.27\pm 0.90$  &$2.80\pm0.26$   & $2.42\pm0.22$ \\
A1991         \dotfill & $1.28 \pm 0.20$ &$1.25\pm0.27 $  &0.208 & $1.45\pm0.27$ & $1.30_{-0.23}^{+0.36}$  \\
% &$0.63\pm0.08$ & $0.069\pm0.004$  &$1.24\pm0.21$ &     $1.25 \pm 0.27$   &$0.63\pm0.10$ & $0.43\pm0.08$   \\
A2029         \dotfill &$8.29 \pm 0.79$ &$9.47\pm 1.46$  & 0.788& $10.26\pm 1.46$ & $8.86_{-1.57}^{+2.42}$ \\
% &$4.26\pm0.28$ & $0.091\pm0.003$ &$8.09 \pm 1.02$  &  $9.85\pm 1.46$    & $4.30\pm0.43$ & $4.68\pm0.46$ \\
A2390         \dotfill &$10.88 \pm 1.05$ &$14.48\pm2.46 $ &0.787 &$15.26\pm2.46$ & $11.42_{-2.03}^{+3.13}$ \\
%RXJ\,1159  \dotfill & \dotfill & &\dotfill  \\
%& $0.30\pm0.03$ &$0.042\pm0.002$   &\dotfill   & \dotfill   & $0.30\pm0.04$ & $0.13\pm0.02$ \\
MKW4          \dotfill & $0.74 \pm 0.09$ &$0.47\pm0.08$ & 0.149 &$0.62\pm0.08$ & $0.63_{-0.11}^{+0.18}$\\
% & $0.28\pm0.03$  & $0.045\pm0.002$  &$0.77\pm0.08$ &$0.47\pm0.08$ &  $0.28\pm0.04$   & $0.11\pm0.02$      \\
\end{tabular}
\end{ruledtabular}
\label{tab5}
\end{table*}

Appendix C contains explicit expressions for the logarithmic derivatives of the gas density and temperature profiles that appear in Eq.~\eqref{master}. We use these functions with the parameters given in Tables~\ref{tab3} and~\ref{tab4} for the clusters of
 galaxies observed by the \emph{Chandra} telescope and studied in Ref.~\cite{chandra} to calculate the \emph{dynamical} mass in the framework of the nonlocal gravity theory. We then compare the dynamical mass from Eq.~\eqref{master} with the \emph{baryonic} mass from the sum of the masses of stars and gas. It is important to note that the models given in Ref.~\cite{chandra} are expected to be reasonably reliable at intermediate cluster radii, but not at either very small or very large cluster radii~\cite{vikh_priv}.

For the calculation of the baryonic mass of the cluster within a sphere of radius $r$, we first integrate $\rho_g$ using Eqs.~\eqref{rhog} and~\eqref{modne} over the volume of a sphere of radius $r$. To this result, we must then add the mass of the stars within a sphere of radius $r$. We assume, for the sake of simplicity, that the stellar component of a cluster follows an isothermal distribution in the cluster; therefore, $\rho_\star \propto 1/r^2$, where $\rho_\star$ is the density of stars in the cluster. It follows that the mass of the stars within a sphere of radius $r$ increases linearly with $r$; hence, the result is $(r/r_{500})M_{stars}$, where $M_{stars}$ is the net mass of the stars within the cluster.

For the stars' net mass, we use a simplified version of an empirical relation given in Ref.~\cite{vikh2}, namely,
\begin{equation}
\frac{M_{stars}}{10^{12}M_{\odot}} \approx 1.8\, (\frac{M_{500}}{10^{14}M_{\odot}})^{0.71}\,,
\label{mass}
\end{equation}
noting that here $M_{500}$ is the mass of the cluster in the dark matter model and is given below in Table~\ref{tab5}.
Eq.~\eqref{mass} assumes a Hubble constant of $H_0=71$ km s$^{-1}$ Mpc$^{-1}$; moreover, there is a systematic uncertainty here regarding the stellar mass-to-light ratio, which we have simply ignored. Furthermore, it is mentioned
in Ref.~\cite{vikh2} that the original form of this empirical relation is expected to hold with a scatter of 31\%.

Figure~\ref{cluster} compares the dynamical mass within a sphere of radius $r$ according to nonlocal gravity with the corresponding observed baryonic mass
of the cluster. To provide an estimate of the measurement error for the observed baryonic mass, we use the uncertainty in the determination of the X-ray flux, which in turn results in the uncertainty in the gas fraction of the cluster given in Table 3 of Ref.~\cite{chandra}; hence, we deduce the error bars in Figure~\ref{cluster}. Based on our theoretical model, we expect general agreement between theory and observation in the outer mid-regions of the clusters and this is essentially confirmed by the results displayed in Figure~\ref{cluster}.

Finally, it is interesting to compare the \emph{total} dynamical mass of the cluster, $M^{\rm c}_{\rm dyn} := M_{\rm dyn}(r_{500})$, with the \emph{total} baryonic mass of the cluster, $M^{\rm c}_{\rm bar} :=M_{\rm bar} (r_{500})$, for the clusters under consideration here. The overall dynamical and baryonic masses of the clusters up to radius $r_{500}$ are given in Table~\ref{tab5}. Let ${\cal M} := M^{\rm c}_{\rm dyn}/M^{\rm c}_{\rm bar}$; then, we expect from NLG theory that ${\cal M}$ should be essentially unity. On the other hand, significant observational uncertainties exist in the determinations of $M^{\rm c}_{\rm bar}$ and $M^{\rm c}_{\rm dyn}$. As in Ref.~\cite{rah13}, we simply compute the best fit to the linear relation $M^{\rm c}_{\rm dyn}={\cal M}\, M^{\rm c}_{\rm bar}$. The result is given in Figure 6. The best-fitting ratio is given by ${\cal M}=0.84 \pm 0.04$, which is in general agreement with the NLG theory in view of the presence of various uncertainties in our estimates of the dynamical and baryonic masses of the clusters.

\begin{figure}
\begin{center}
%\begin{tabular}
\includegraphics[width=80mm]{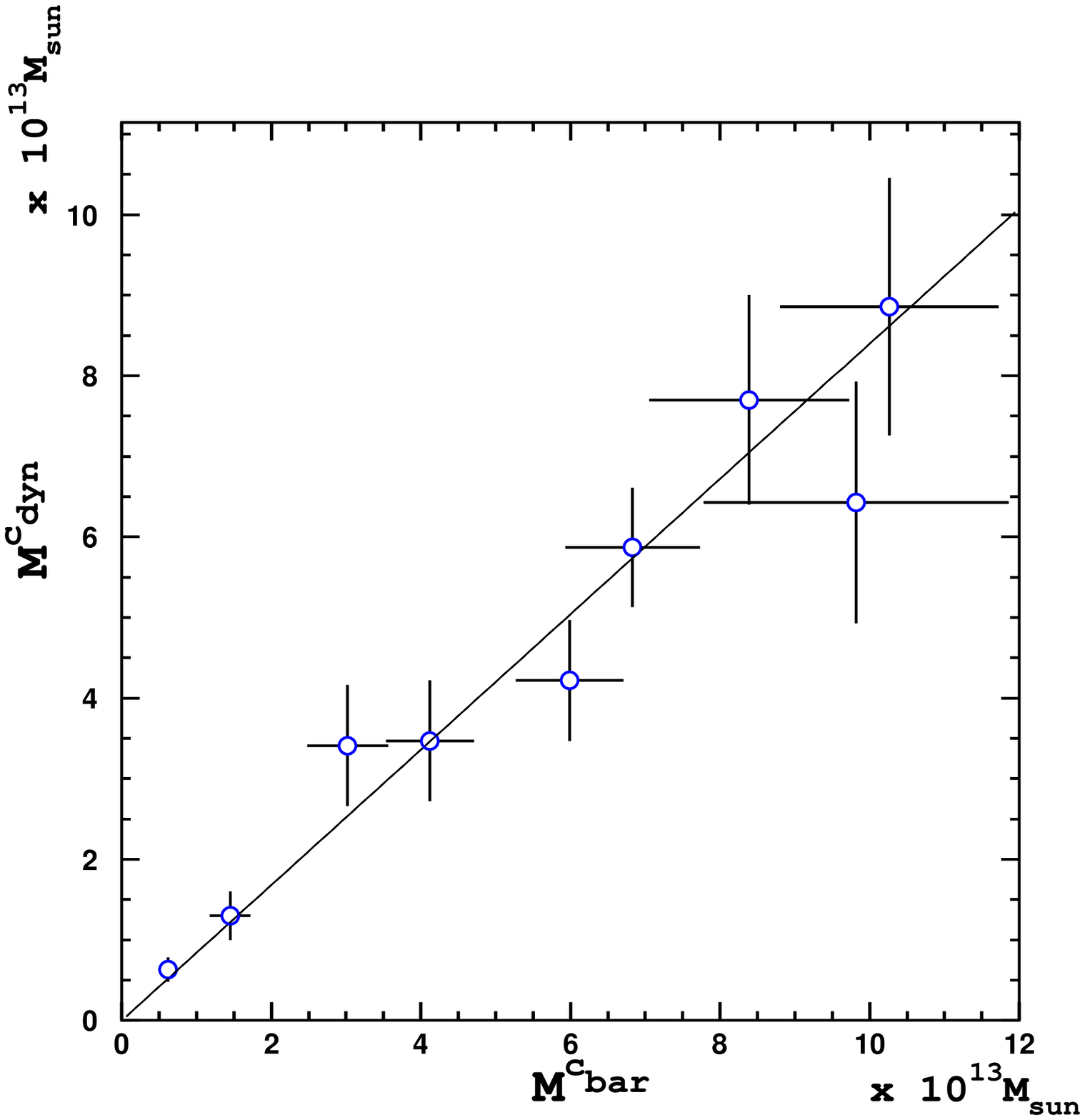}
\includegraphics[width=80mm]{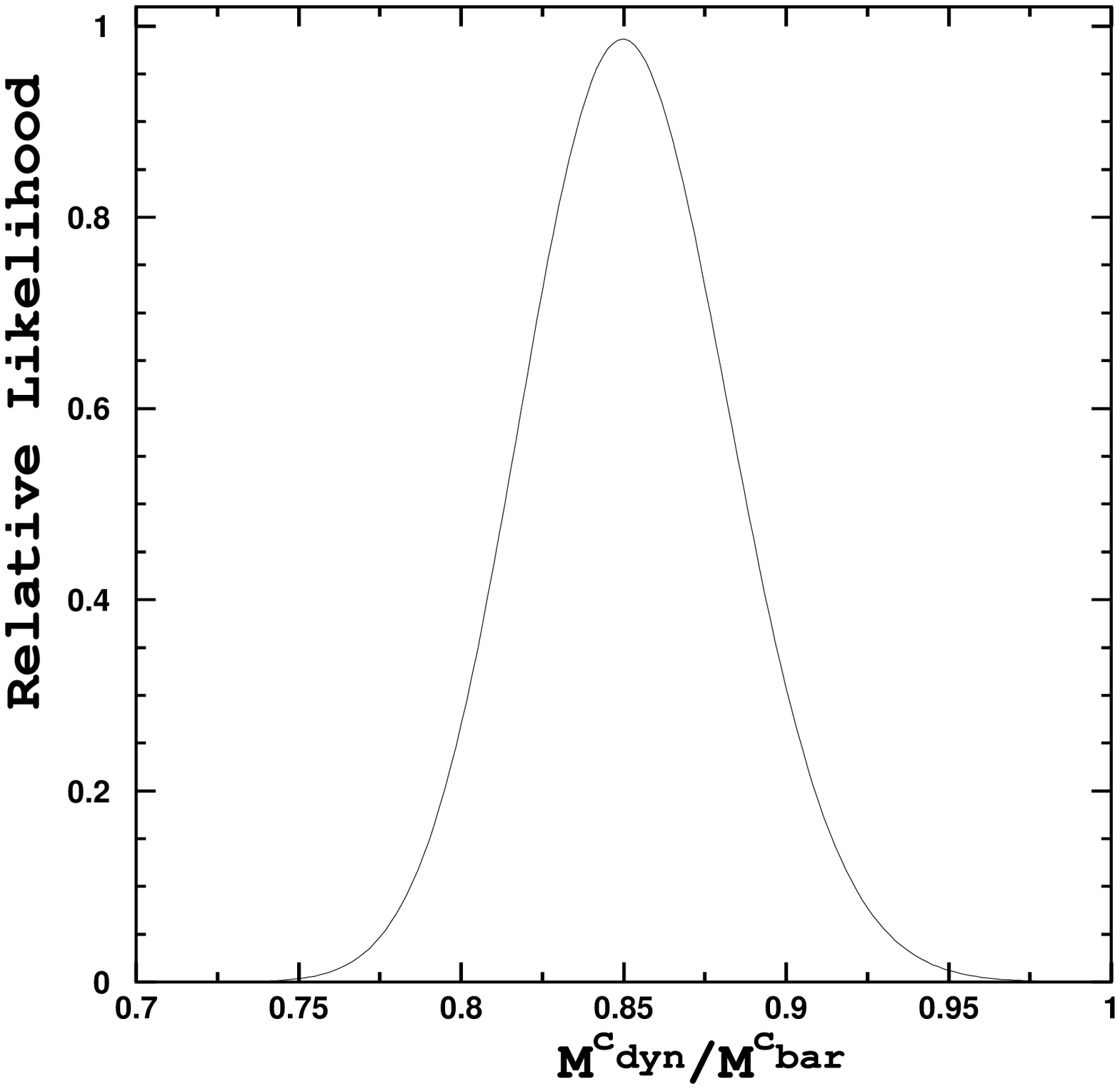}
%\end{tabular}
\end{center}
\caption{\label{comp} Left panel: The best linear fit to the relation between the dynamical masses according to nonlocal gravity and the observed
baryonic masses of the ten \emph{Chandra} X-ray clusters of Table~\ref{tab5}. Here the best-fitting slope is $M^{\rm c}_{\rm dyn}/M^{\rm c}_{\rm bar} =  0.84 \pm 0.04$. Right panel: The likelihood
function for parameter ${\cal M}=M^{\rm c}_{\rm dyn}/M^{\rm c}_{\rm bar}$.}
\end{figure}

\section{Discussion}

The recent classical nonlocal generalization of Einstein's general relativity involves a scalar causal kernel that must be determined via observation. The situation here is somewhat similar to the electrodynamics of media, where Maxwell's original equations remain formally unchanged, but new physics is contained in the constitutive relations that are in general nonlocal. In nonlocal general relativity, Einstein's equations expressed in their equivalent teleparallel form remain formally unchanged as well; however, the nonlocal constitutive relation introduces new aspects of the gravitational interaction via the scalar kernel. Indeed, nonlocality is here significant on galactic scales and can simulate dark matter. That is, in this theory there is no actual dark matter and what appears as dark matter in astrophysics is essentially a manifestation of the nonlocal aspect of the gravitational interaction.

The implications of nonlocal gravity theory have thus far been investigated in the linear domain, which, just as in general relativity, involves linearized gravitational waves as well as the Newtonian regime of the theory. It has been shown in  recent investigations that for  gravitational radiation, the situation in nonlocal gravity is essentially the same as in general relativity~\cite{NG6, NG7}. The present paper therefore deals with the Newtonian regime of nonlocal gravity, where we assume a simple kernel with two parameters $\alpha$ and $\mu$ and find a modified force law, where the Newtonian inverse-square attraction is combined with a Yukawa-type repulsion, which decays with radial distance $r$ as $\exp{(-\mu r)}$. From the new force law, we determine the rotation curves of spiral galaxies and compare the theory with observational data in order to fix the parameters of our model. We find that for  the best value of $\mu$, $\mu^{-1} \approx 17$ kpc,  and  for $r\gg \mu^{-1}$, the force of gravity is Newtonian except that Newton's gravitational constant $G$ must be replaced by $G(1+\alpha)$, where the best  value of $\alpha$ is $ \approx 11$. Regarding our confidence in the values of these parameters, it is important to point out that in our comparison of the theory with observation in section III, we always find one global minimum in the $(\alpha, \mu, \Upsilon_\star^{3.6})$ space; furthermore, our work on the clusters of galaxies in section V crucially depends on the value of $\alpha$. We then demonstrate that our approach is consistent with the known astrophysical correlation between the stellar mass-to-light ratio and the color of galaxies. Moreover, our results are consistent with the Tully-Fisher relation for spiral galaxies. Extending our nonlocal gravity theory to clusters of galaxies, we show that cluster dynamics is consistent with the measured baryonic content of galaxy clusters.

Nonlocal gravity theory introduces a modification of the Newtonian gravitational force that accounts for gravitational physics from the scale of the Solar System to that of a galaxy cluster without any recourse to dark matter. It remains to study gravitational lensing as well as nonlocal cosmological models in order to have a more complete confrontation of nonlocal general relativity with experiment.

\begin{acknowledgments}
We are grateful to Friedrich Hehl, Jeffrey Kuhn, Roy Maartens and Haojing Yan for valuable discussions. We acknowledge using THINGS, ``The HI Nearby Galaxy Survey"~\cite{things}.
\end{acknowledgments}

\appendix{}

\section{Kernel Parameter $a_0$}\label{appA}

The purpose of this appendix is to show that neglecting $a_0$, $0<a_0/\lambda_0 \ll 1$ and $0<a_0 \mu \ll 1$, in the reciprocal kernel of nonlocal gravity theory in the Newtonian regime has a negligible influence on the conclusions of this paper.

Solving the modified Poisson equation with either kernel $q_1$ or $q_2$, given respectively by Eq.~\eqref{II6} or Eq.~\eqref{II7}, results in the modification of the Newtonian inverse-square force law given by
\begin{equation}\label{A1}
 F(r)=\frac{Gm_1m_2}{r^2}\left \{1-{\cal E}(r)+\alpha \Big[1-(1+\frac{1}{2}\mu r)e^{-\mu r}\Big]\right \}\,
\end{equation}
instead of Eq.~\eqref{II16}. Here ${\cal E}(r)$ is either ${\cal E}_1(r)$ or ${\cal E}_2(r)$ given by
\begin{equation}\label{A2}
 {\cal E}_i(r)=4 \pi \int_0^r [q(\rho)-q_i(\rho)]\,\rho^2\,d\rho\,, \qquad i=1,2\,,
\end{equation}
where $q(r)$ is our adopted reciprocal kernel given by Eq.~\eqref{II8} that is obtained from either $q_1$ or $q_2$ by ignoring parameter $a_0$. We find that
\begin{equation}\label{A3}
 {\cal E}_1(r)=\frac{a_0}{\lambda_0}\left \{-\frac{r}{r+a_0}e^{-\mu r}+2e^{\mu a_0} \Big[E_1(\mu a_0)-E_1(\mu a_0+\mu r) \Big] \right \}\,
\end{equation}
and
\begin{equation}\label{A4}
 {\cal E}_2(r)=\frac{a_0}{\lambda_0}e^{\mu a_0}\Big[E_1(\mu a_0)-E_1(\mu a_0+\mu r)\Big]\,,
\end{equation}
where $E_1(u)$ is the exponential integral function defined in Eq.~\eqref{II21}.

It turns out that ${\cal E}_1(r)$ and ${\cal E}_2(r)$ are positive, monotonically increasing functions that start from zero at $r=0$ and asymptotically approach ${\cal E}_1(\infty)=2{\cal E}_{\infty}$  and ${\cal E}_2(\infty)={\cal E}_{\infty}$, respectively, where
\begin{equation}\label{A5}
 {\cal E}_{\infty}=\frac{a_0}{\lambda_0}e^{\mu a_0}E_1(\mu a_0)\,.
\end{equation}
Here $0<a_0/\lambda_0 \ll 1$ and $0< \mu a_0 \ll 1$; hence, it follows from Eq.~\eqref{II23} that $0<  {\cal E}_{\infty} \ll 1$ for sufficiently small $a_0/\lambda_0$. For instance, with $a_0/\lambda_0=10^{-3}$ and the parameters of our nonlocal gravity model as in Eq.~\eqref{parameters1+2}, we have ${\cal E}_{\infty}\approx 0.008$. We conclude that $a_0/\lambda_0$ can always be chosen to be  so small  that ${\cal E}(r)$ is such that $0 \le {\cal E}(r) \ll 1$ and can therefore be neglected in comparison to unity in Eq.~\eqref{A1} for the considerations of this paper.

\section{Model Fitting}\label{appB}

For the rotation curve of each spiral galaxy under consideration in this paper, let there be ${\cal N}$ observational data points such as $v_i\pm \sigma_i$ at radial distance $r_i$ for $i=1,2,3, ..., {\cal N}$. The errors in the measurement data are assumed to be Gaussian. The nonlocal gravity model predicts instead a rotation curve given by $V(r; p)$, where $p=(p_1, p_2, ..., p_n)$ represents the set of $n$ free model parameters that should be determined from a comparison of the model with the data. For instance, for each THINGS galaxy in the present work, $n=3$ for $\alpha$, $\mu$ and $\Upsilon_\star^{3.6}$. Assuming that the data points are independent of each other, the goodness-of-fit of the data to the model is measured via the chi-squared statistic, namely,
\begin{equation}\label{B1}
 \chi^2=\sum_{i=1}^{{\cal N}}\Big(\frac{v_i-V_i}{\sigma_i}\Big)^2\,,
\end{equation}
where $V_i :=V(r_i; p)$ and $\chi^2$ is thus a function of the parameters of the model. The number of degrees of freedom, $N_{d.o.f.}$, is defined to be
\begin{equation}\label{B2}
 N_{d.o.f.}={\cal N} - n\,.
\end{equation}
The chi-squared probability distribution is given by
\begin{equation}\label{B3}
{\cal P}(\chi^2)= \frac{1}{2^\nu\, \Gamma(\nu)}\,(\chi^2)^{\nu-1}e^{-\frac{1}{2}\chi^2}\,,
\end{equation}
where
\begin{equation}\label{B4}
2 \nu := N_{d.o.f.}\,.
\end{equation}
The mean value of $\chi^2$ according to Eq.~\eqref{B3} is $N_{d.o.f.}$ and its variance is $2 N_{d.o.f.}$. Therefore, if the model is correct, we expect that  the value of $\chi^2$ with the best-fitting parameters for the model is near its mean, so that the \emph{reduced chi-squared} defined by $\chi^2 / N_{d.o.f.}$ is near unity.

To find the best-fitting parameters of the model from the minimum of the chi-squared statistic, we calculate the values of $\chi^2$ over a large grid of parameters. For Gaussian variables, the corresponding \emph{likelihood} is proportional to $\exp{(-\chi^2 / 2)}$; therefore, a normalized probability distribution can be determined in this way for $(p_1, p_2, ..., p_n)$ over the grid. The \emph{marginalized likelihood} distribution for one parameter is obtained from the grid probability distribution by summing over the values of  the other parameters.

Finally, for a set of galaxies, we obtain the best-fitting parameters $\alpha$ and $\mu$ of our model by the net $\chi^2$ of the set, which we obtain by summing the $\chi^2$'s of the different galaxies, as they are assumed to be independent. For instance, for the twelve THINGS galaxies in section III, the \emph{combined likelihood} distribution is assumed to be proportional to
\begin{equation}\label{B5}
\exp{\Big(-\frac{1}{2}\,\sum_{j=1}^{12}\chi_j^2\,\Big)}\,,
\end{equation}
from which we obtain the best-fitting parameters of our nonlocal gravity model given in Eq.~\eqref{parameters1+2}.

\section{Formulas for Calculating $M_{\rm dyn}$}\label{appC}

In Eq.~\eqref{master}, we need the radial derivatives of the three-dimensional gas density and  temperature profiles. The required terms are given by
\begin{eqnarray}\label{C1}
\nonumber \frac{d\ln\rho_g(r)}{d\ln r} = \frac{1}{H(r)} \Bigg[-{{n_0}^2}\alpha'
\left(1+\frac{r^2}{{r_c}^2}\right)^{0.5 \alpha' -3 \beta}
\left(\frac{r}{{r_c}}\right)^{-\alpha' } \left(1+r^{\gamma}
{r_s}^{-\gamma }\right)^{-\frac{\varepsilon }{\gamma }} \\
 \nonumber  { {n_0}^2}( \alpha' -6 \beta )
\left(1+\frac{r^2}{{r_c}^2}\right)^{-1+0.5 \alpha' -3 \beta }
\left(\frac{r}{{r_c}}\right)^{-\alpha' +2} \left(1+r^{\gamma }
{r_s}^{-\gamma }\right)^{-\frac{\varepsilon }{\gamma }}\\
\nonumber - {6 {n_{0}'}^2\beta'} \left(\frac{r}{{r_{c}'}}\right)^2
\left(1+\frac{r^2}{{r_{c}'}^2}\right)^{-1-3 \beta'}\\
 -{n_0}^2  \varepsilon\left(\frac{r}{r_c}\right)^{- \alpha'} \left(\frac{r}{r_s}\right)^\gamma
\left(1+\frac{r^2}{{r_c}^2}\right)^{0.5 \alpha' -3 \beta }
\left(1+r^{\gamma } {r_s}^{-\gamma }\right)^{-1-\frac{\varepsilon
}{\gamma }}  \Bigg]\,,
\end{eqnarray}
where
\begin{equation}\label{C2}
H(r) = 2 \left[{n_{0}'}^2
\left(1+\frac{r^2}{{r_{c}'}^2}\right)^{-3 \beta'}+{n_0}^2 \left(1+\frac{r^2}{{r_c}^2}\right)^{0.5 \alpha'
-3 \beta } \left(\frac{r}{{r_c}}\right)^{-\alpha'}
\left(1+r^{\gamma } {r_s}^{-\gamma }\right)^{-\frac{\varepsilon
}{\gamma }}\right]
\end{equation}

and

\begin{eqnarray}\label{C3}
\frac{d\ln T(r)}{d\ln r}& = & a_{cool}\left(\frac{r}{{r_{cool}}}\right)^{a_{cool}}\frac{T_0-T_{min}}{\Big [1+\left(\frac{r}{r_{cool}}\right)^{a_{cool}}\Big]\Big [T_{min}+T_0\left(\frac{r}{r_{cool}}\right)^{a_{cool}} \Big ]
}\nonumber\\
&-&{a'} -\frac{c' \left(\frac{r}{{r_t}}\right)^b}{1+\left(\frac{r}{{r_t}}\right)^b}\,.
\end{eqnarray}

\end{document}